\documentclass[amsmath,amssymb,aps,twocolumn,nofootinbib]{revtex4-2}
\usepackage[toc,page]{appendix}
\usepackage[utf8]{inputenc}
\usepackage{graphicx}

\usepackage{apptools}
\AtAppendix{
    \numberwithin{theorem}{section}
	\numberwithin{figure}{section}
	\numberwithin{table}{section}
    \numberwithin{equation}{section}
    
}

\makeatletter 
    
\renewcommand\onecolumngrid{
\do@columngrid{one}{\@ne}%
\def\set@footnotewidth{\onecolumngrid}
\def\footnoterule{\kern-6pt\hrule width 1.5in\kern6pt}%
}

\renewcommand\twocolumngrid{
        \def\footnoterule{
        \dimen@\skip\footins\divide\dimen@\thr@@
        \kern-\dimen@\hrule width.5in\kern\dimen@}
        \do@columngrid{mlt}{\tw@}
}%

\makeatother    

\newtheorem{theorem}{Theorem}

\newtheorem{lemma}[theorem]{Lemma}

\newcommand{\ket}[1]{{\left| {#1} \right\rangle}}
\newcommand{\bra}[1]{{\left\langle {#1} \right|}}

\newcommand{\WignerThreej}[6]{{\left( \begin{array}{ccc} {#1} & {#2} & {#3}  \\  {#4} & {#5} & {#6} \end{array} \right)}}
\newcommand{\WignerSixj}[6]{{\left\{ \begin{array}{ccc} {#1} & {#2} & {#3}  \\  {#4} & {#5} & {#6} \end{array} \right\}}}

\newcommand{\boldgreek}[1]{{\mbox{\boldmath$ {#1} $}}}
\newcommand{\boldgreeksmall}[1]{{\mbox{\scriptsize\boldmath$ {#1} $\normalsize}}}
\newcommand{\shrinkify}[1]{\textstyle {#1} \displaystyle}

\newcommand{\ii}{\mathchoice
  {\textsl{i}\;\!}
  {\textsl{i}\;\!}
  {\textsl{i}\:\!}
  {\textsl{i}\:\!}
}

\newcommand{\ee}{\textsl{e}}

\usepackage[greek,english]{babel}
\newcommand{\pp}{\mathchoice
  {\textsl{\greektext p}\;\!}
  {\textsl{\greektext p}\;\!}
  {\textsl{\greektext p}\:\!}
  {\textsl{\greektext p}\:\!}
}

\makeatletter
\newcommand*{\bigcdot}{}
\DeclareRobustCommand*{\bigcdot}{%
  \mathbin{\mathpalette\bigcdot@{}}%
}
\newcommand*{\bigcdot@scalefactor}{.5}
\newcommand*{\bigcdot@widthfactor}{1.15}
\newcommand*{\bigcdot@}[2]{%
  \sbox0{$#1\vcenter{}$}
  \sbox2{$#1\cdot\m@th$}%
  \hbox to \bigcdot@widthfactor\wd2{%
    \hfil
    \raise\ht0\hbox{%
      \scalebox{\bigcdot@scalefactor}{%
        \lower\ht0\hbox{$#1\bullet\m@th$}%
      }%
    }%
    \hfil
  }%
}
\makeatother

\begin{document}

\title{Angular Geometry of Atomic Multipole Transitions}

\author{Wesley C. Campbell}
\email{wes@physics.ucla.edu}
\affiliation{Department of Physics and Astronomy, University of California Los Angeles, Los Angeles CA, 90095 USA}

\date{\today}

\begin{abstract}
A simple way to calculate Rabi frequencies is outlined for interactions of atomic or nuclear multipole moments with laser fields that focuses on their relative geometry.  The resulting expression takes the form of a dot product between the laser polarization and a vector spherical harmonic, thereby naturally connecting to the multipole's far-field spontaneous-emission pattern and providing a way to visualize the interaction.  Since the vector spherical harmonics are not yet a standard tool in quantum science, their relevant properties are reviewed.  This approach is illustrated in the calculation of a variety of beam effects, yielding both perturbative corrections and some nontrivial cases with non-vanishing coupling.

\end{abstract}

\maketitle

\section{Introduction}\vspace{-12pt}
The spherical multipole expansion of the quantum-mechanical interaction Hamiltonian of an atom (or molecule, or nucleus\footnote{The results I present are just as valid for nuclei or gas-phase molecules interacting with light as they are for atoms; the choice to refer only to atoms is intended only to simplify the wording.}) with propagating electromagnetic radiation partitions it into terms involving well-defined angular-momentum quantum numbers.  This has been studied at length by many authors, and treatments are available with sophisticated microscopic descriptions utilizing irreducible spherical tensor products, spherical Bessel functions, and Hansen multipoles (see, e.g., \cite{RoseMultipole,ShoreAndMenzel,HamiltonNuclear,WeissbluthAtoms,LandauQED,BiedenhornAndLouck,BrinkAndSatchler,WalterRJohnson}). Often, the focus of such studies is at least partly the calculation of the atomic transition moments themselves.

Advances in trapped-atom clocks and quantum processors have demonstrated systems in which multipole transitions between angular momentum eigenstates are driven by lasers, well beyond the applicability of the dipole approximation.  For example, the laser-driven electric octupole ($E3$) transition from ${}^2\mathrm{S}_{1/2}$ to ${}^2\mathrm{F}_{7/2}^o$ in $\mathrm{Yb}^+$ \cite{Roberts1997Observation} is used to stabilize state-of-the-art atomic clocks \cite{Huntemann2012HighAccuracy,Tofful2024Optical}.  Electric quadrupole ($E2$) transitions are driven in a variety of atoms for quantum logic operations, precision measurement, and optical clockwork \cite{Ludlow2015Optical}, and a recent experiment highlighted the role of transverse gradients in that coupling \cite{Schmiegelow2016Transfer}.  We seek an adroit understanding of how the geometry of the laser beam and the atom's angular momentum components interact to dictate selection rules and resonant Rabi frequencies.

While methods for calculating geometric contributions to atom-light interactions have been known for many years, the complexity involved in calculating high-rank tensor products can sometimes obscure the basic physics at play, particularly for scientists who may not be deeply familiar with irreducible spherical tensor analysis.  Further, the inclusion of near-field descriptions of the interaction tends to introduce mathematical complexity (such as spherical cylinder functions) that is not necessary for calculating coupling to laser fields if the reduced atomic transition moment is known (for example, through measurement).  These aspects of the existing literature on this topic can make it difficult to concisely describe all but a couple of the most-likely experimental geometries, such as the application of linearly-polarized light with $\mathbf{k}$ either perpendicular or parallel to the quantization axis.

Here, I outline a simple method to calculate multipole-transition resonant Rabi frequencies that highlights the relationship between the applied light and the multipole's native far-field emission pattern.  I assume that the reduced transition moment is either provided or to-be-measured to focus on the coupling strength's dependence on geometric features of the laser field and the atom's angular momentum.  The method can be applied with minimal knowledge of spherical-tensor analysis,\footnote{Readers interested in getting to the results without the spherical tensor formalism are encouraged to skip directly to \S \ref{sec:RabiFreq}.} and can be framed in terms of experimentally-convenient quantities like the Jones vector and $\mathbf{k}$ vector of the laser light without needing to explicitly calculate their components in the quantization basis.  Further, I illustrate with examples how this form naturally admits visualization, and generalization to coherent, multi-beam geometries; off-axis beams; Hermite-Gauss and Laguerre-Gauss modes; and vector-mode beams.

This paper starts with an operational introduction to calculating laser-atom coupling strengths, followed by examples, with the majority of the derivation details and background material in appendices.  As an illustration of the application of this method to quantum science, I introduce beam corrections and coherent, multi-beam geometries that allow more-efficient coupling to individual multipole components than is possible with a single, plane-wave beam using the same total optical power.

\section{Multipole expansion}\label{sec:MultipoleExpansion}
We will begin by taking the interaction between an atom and an externally-applied electric field\footnote{I focus in this paper on electric multipole transitions, but the description of the field applies just as well to magnetic fields, and most of these results can be applied to magnetic multipole transitions with the appropriate definition of the atomic operator (Appendix \ref{app:Spon}).} $\mathbf{E}$ in free space as a multipole expansion of the electron charge distribution about the nucleus in spatial derivatives of the field, $H = \sum_K H_{EK}$ with
\begin{widetext}
\begin{align}
    H_{EK} = &\,\, -\frac{1}{K!}
     \sqrt{\frac{K!\,(2K+1)}{(2K+1)!!}} \sum_p (-1)^p \,\,\left( \sqrt{\frac{4 \pp}{2K + 1}}\, \sum_i q^{}_i \,r_i^K \, Y_{K,p}(\mathbf{\hat{r}}_i) \right)\,\, T^{(K)}_{-p}[\;T^{(K-1)}[\underbrace{\boldsymbol{\nabla},\boldsymbol{\nabla}, \ldots , \boldsymbol{\nabla}}_{K-1}],\mathbf{E}]. \label{eq:HEKComplex}
\end{align}
\end{widetext}
Here, the sum over $i$ extends over the electrons, and the term in parentheses is the standard definition of the $p$ component of the electric $2^K$-pole moment,
\begin{align}
    T^{(K)}_p[Q^{(E)}] \equiv \sqrt{\frac{4 \pp}{2K + 1}}\, \sum_i q^{}_i \,r_i^K \, Y_{K,p}(\mathbf{\hat{r}}_i), \label{eq:2KPoleMoment}
\end{align}
where $Y_{K,p}(\mathbf{\hat{r}}_i)$ is a spherical harmonic of the angular position ($\mathbf{\hat{r}}_i \doteq (\vartheta_i,\varphi_i)$) of the $i$th electron (where $\doteq$ stands for ``is represented by'' \cite{Sakurai}).
For example, if $K=1$, $T^{(1)}[Q^{(E)}] = \sum_i q_i\, r_i\,\frac{4 \pp}{3}\,\, Y_{(1)}(\mathbf{\hat{r}}_i) = \sum_i q_i\, \mathbf{r}_i = \mathbf{d}$ is the electric dipole moment operator.\footnote{I use the notation $Y_{(K)}(\mathbf{\hat{r}})$ to denote the rank-$K$ irreducible tensor whose $p$ component is the spherical harmonic $Y_{K,p}(\mathbf{\hat{r}})$.}

The rightmost term in (\ref{eq:HEKComplex}), which I will shortly simplify to a more tractable form, is an irreducible spherical tensor product that is essentially the $(K-1)$th spatial derivative of the electric field (the \emph{field-derivative tensor}).  As suggested by the leading factor of $1/K!$, $H_{EK}$ is essentially just the $K$th term in a Taylor expansion of the interaction Hamiltonian between a charge distribution and the electric field \cite{Jackson}.

Next, since we are concerned primarily with atom-laser interactions, we will specify that the externally-applied field's spatial distribution consists of monochromatic, transverse vector plane waves given by
\begin{align}
    \mathbf{E}(\mathbf{r},t) = \frac{\mathcal{E}_0}{2}(\boldgreek{\hat{\epsilon}}\, \ee^{\ii (\mathbf{k} \boldsymbol{\cdot} \mathbf{r} - \omega t)} + \mathrm{c.c.})
\end{align}
where $\boldgreek{\hat{\epsilon}}$ is the unit vector describing the polarization and $\mathrm{c.c.}$ denotes the complex conjugate of the preceding expression.  Since the polarization is independent of $\mathbf{r}$, we see that $\boldsymbol{\nabla}\,\boldgreek{\hat{\epsilon}}\, \ee^{\ii (\mathbf{k} \boldsymbol{\cdot} \mathbf{r} - \omega t)} = (\ii \hspace{-0.6ex} k)\, \mathbf{\hat{k}}\, \boldgreek{\hat{\epsilon}} \,\ee^{\ii (\mathbf{k} \boldsymbol{\cdot} \mathbf{r} - \omega t)}$ and we can utilize the replacement
\begin{align}
    \boldsymbol{\nabla} \rightarrow (\pm \ii \hspace{-0.1ex} k)\, \mathbf{\hat{k}}
\end{align}
where the upper sign is for the $\boldgreek{\hat{\epsilon}}\, \ee^{\ii (\mathbf{k} \boldsymbol{\cdot} \mathbf{r} - \omega t)}$ term and the lower sign is for its complex conjugate.

\begin{figure}[t]
\centering
\includegraphics[scale=1.0]{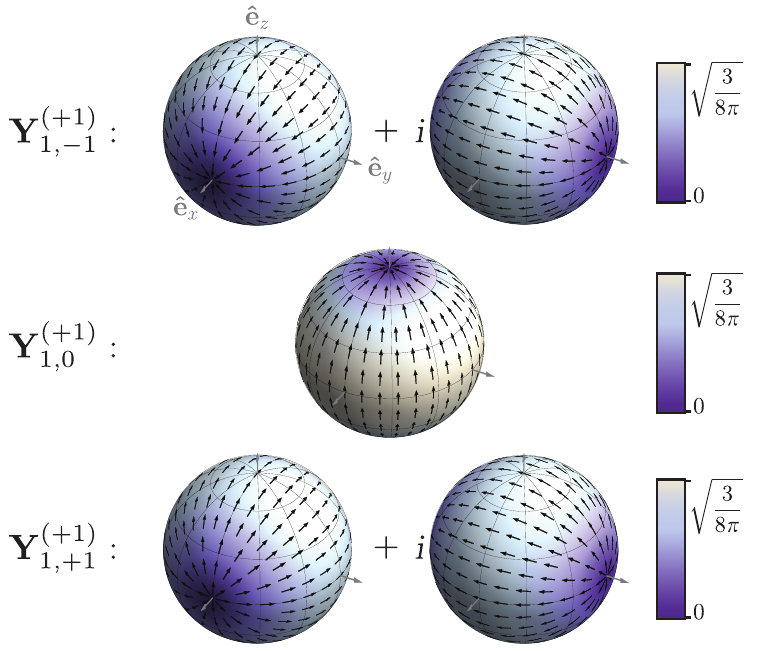}
\caption{Rank-1 vector spherical harmonics of type $\lambda = +1$, depicted on the unit sphere.  The color scale depicts the vector magnitude of each part, while the arrows, which are all the same length, only indicate direction.  It is recommended to disable the \texttt{Enhance thin lines} option in \texttt{Preferences} $\rightarrow$ \texttt{Page Display} to see the details in these plots.}\label{fig:Y1s}
\end{figure}

\begin{figure}[t]
\centering
\includegraphics[scale=1.0]{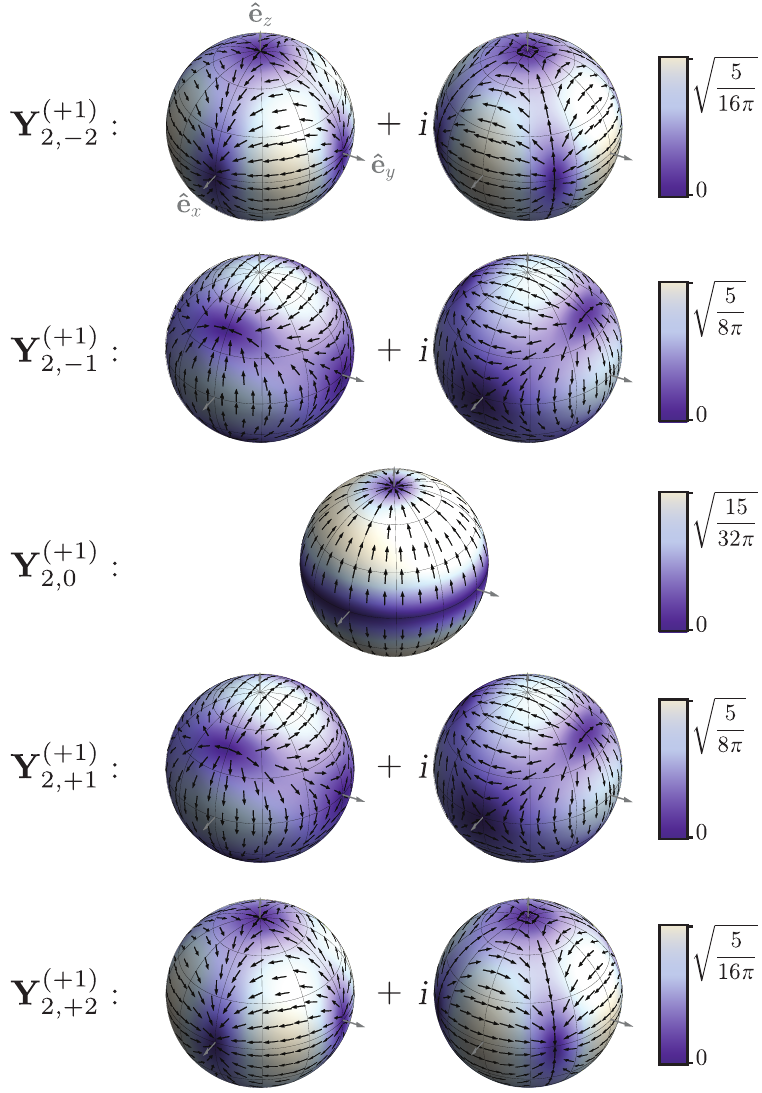}
\caption{Rank-2 vector spherical harmonics of type $\lambda = +1$.  The rank-3 vector spherical harmonics are plotted in Appendix \ref{app:VectorSphericalHarmonics}.}\label{fig:Y2s}
\end{figure}

Next, we can simplify the expression for the field-derivative tensor by introducing the following polarization identity for vectors $\mathbf{\hat{k}} \perp \boldgreek{\hat{\epsilon}}$ (Appendix \ref{app:PolIdty}):
\begin{widetext}
\begin{align}
    T^{(K)}[T^{(K-1)}[\underbrace{\mathbf{\hat{k}},\mathbf{\hat{k}}, \ldots, \mathbf{\hat{k}}}_{K-1}], \boldgreek{\hat{\epsilon}}] = &\,\, \sqrt{\frac{(K-1)!(K+1)\,\, 4 \pp}{(2K+1)!!}}\,\left( \boldgreek{\hat{\epsilon}} \bigcdot \mathbf{Y}^{(+1)}_{(K)}(\mathbf{\hat{k}})\right).\label{eq:PolIdty}
\end{align}
\end{widetext}
Here, $\mathbf{Y}^{(+1)}_{(K)}(\mathbf{\hat{k}})$ is a \emph{vector spherical harmonic}, which is a rank-$K$ irreducible spherical tensor whose $p$ component (denoted by $\mathbf{Y}^{(+1)}_{K,p}(\vartheta_k,\varphi_k) = k \boldgreek{\nabla}_\mathbf{k}\,Y_{K,p}(\vartheta_k,\varphi_k)/\sqrt{K(K+1)}$) is a 3-space vector that is perpendicular to $\mathbf{\hat{k}}$ \cite{Varshalovich1988Quantum}.  An introduction to vector spherical harmonics is given in Appendix \ref{app:VectorSphericalHarmonics}, along with explicit formulas for the first few, which are also shown in Figs.\ \ref{fig:Y1s}, \ref{fig:Y2s}, and \ref{fig:Y3s}.  The numerical factor in front of the dot product gives $\sqrt{8 \pp/3}$, $\sqrt{4\pp/5}$, and $\sqrt{32 \pp/105}$, for $K=1$, $2$, and $3$.

This allows us to write the interaction Hamiltonian between the atom's electric $2^K$-pole moment and externally-applied plane waves in the form
\begin{widetext}
\begin{align}
    H_{EK} = & \,\, -\frac{\mathcal{E}_0}{2}\, (\ii k)^{(K-1)}\, \sqrt{\frac{(K+1)\, 4 \pp}{K\, (2K+1)!!\, (2K-1)!!}}\,\,\sum_p (-1)^p \,\,T^{(K)}_p[Q^{(E)}] \, \left( \boldgreek{\hat{\epsilon}}\bigcdot \mathbf{Y}^{(+1)}_{K,-p}(\mathbf{\hat{k}}) \right)\, \ee^{\ii (\mathbf{k} \boldsymbol{\cdot} \mathbf{R}_\mathrm{a} - \omega t)} + \mathrm{c.c.} \label{eq:HEK}
\end{align}
\end{widetext}
\onecolumngrid

\twocolumngrid
\onecolumngrid

\twocolumngrid
\noindent where I have added the dependence on $\mathbf{R}_\mathrm{a}$, the location of the atom's center of mass.  The numerical factor with the square root is  $\sqrt{8 \pp/3}$, $\sqrt{2 \pp/15}$, and $\sqrt{16 \pp/4725}$ for $K=1$, $2$, and $3$.

Writing the Hamiltonian in the form (\ref{eq:HEK}) effects a few desirable outcomes.  When paired with the definition of the electric $2^K$-pole moment (\ref{eq:2KPoleMoment}), this form takes advantage of the irreducible spherical tensor formalism, but expresses the result in terms of simple (and potentially more familiar) quantities like spherical harmonics and a vector dot product (cf.\ (\ref{eq:HEKComplex})).  Further, as will be explained in more detail in the next section, it keeps the atomic operator conveniently in a form that is easy to evaluate in the quantization frame while expressing the field-derivative tensor in a form that can be easily evaluated in the frame that is likely most convenient for experiments -- the \emph{helicity frame} (see Appendix \ref{app:HelicityFrame} and Fig.\ \ref{fig:BasisVectors}).  In effect, the polarization identity (\ref{eq:PolIdty}) can automatically handle the transformation of laser field parameters into the quantization frame without requiring that transformation be done explicitly before putting parameters into the Hamiltonian, a simplification that may be useful even for rank-$1$ interactions. 

Last, as discussed below, the vector dot product (along with depictions such as figures \ref{fig:Y1s} and \ref{fig:Y2s}) invites an intuitive interpretation and visualization of the interaction that can be difficult to glean from a Hamiltonian steeped in spherical tensor products.  Despite having arrived at (\ref{eq:HEK}) by considering an atom interacting with an ``incoming'' laser field, the polarization identity (\ref{eq:PolIdty}) allows us to identify $\mathbf{Y}^{(+1)}_{K,-p}(\mathbf{\hat{k}})$ as the normalized vector emission distribution for far-field spontaneous emission, an insight that applies for arbitrarily-high-rank multipole processes.

\section{Calculation of the Rabi Frequency}\label{sec:RabiFreq}
In the experimental context, the ``strength'' of a transition moment can be described by its Einstein $A$ coefficient.  For a two-level system, the Einstein $A$ coefficient is related simply to the spontaneous emission lifetime $\tau$ via $A = 1/\tau$, where the probability that an atom in field-free space prepared in the excited state at $t=0$ remains in the excited state at time $t>0$ is $\ee^{-t/\tau}$.

The Einstein $A$ coefficient is convenient to use as a proxy for the strength of a transition because it is often straightforward to measure, has the same dimensions no matter the nature (magnetic vs.\ electric) or rank (dipole, quandrupole, octupole, etc.) of the transition moment(s) involved, conveniently contains all of the most-difficult-to-calculate atomic properties, and describes the energy scale associated with the phenomenon of saturation.  As detailed in Appendix \ref{app:Spon}, the Hamiltonian (\ref{eq:HEK}) can be used to derive the general expression for the Einstein $A$ coefficient of an electric $2^K$-pole transition:
\begin{align}
    A_{EK} = &\,\, \frac{2(K+1)}{K\,(2K-1)!!\, (2K+1)!!} \, \left( \frac{\omega}{c} \right)^{2K+1} \nonumber \\
    & \hspace{5mm} \times \alpha \,c \,\, \frac{\big|\bra{J_\mathrm{e}}\! | \, T^{(K)}_{\cdot}[Q^{(E)}]\, |\!\ket{J_\mathrm{g}}/e \big|^2}{2J_\mathrm{e} + 1},\label{eq:AEK}
\end{align}
where the numerical factor preceding $(\omega/c)^{2K+1}$ is $4/3$, $1/15$, and $8/4725$ for $K=1$, $2$, and $3$.  By inverting (\ref{eq:AEK}), we can obtain a formula for the reduced matrix element\footnote{I adopt the convention of Racah for defining the reduced matrix element, in which the Wigner-Eckart theorem takes the form \begin{align*} \langle J'\;M'| T^{(K)}_p | J\;M\rangle = & \,\, (-)^{J'-M'} \WignerThreej{J'}{K}{J}{-M'}{p}{M} \langle J'| \!| T^{(K)}_\cdot |\! |J\rangle. \end{align*}} in terms of the Einstein $A$ coefficient, to within a sign.

To write an expression for the resonant Rabi frequency for driving a particular $2^K$-pole-moment component $p$, I adopt the definition of $\Omega$ such that a resonant population inversion would be driven in time $t_{\pp}$ if $\Omega\, t_{\pp} = \pp$.  We can apply the Wigner-Eckart theorem to write
\begin{widetext}
\begin{align}
    \boxed{\Omega_\mathrm{eg} =  s_J\,(-1)^{J_\mathrm{e}-M_\mathrm{g}}\frac{e\mathcal{E}_0}{\hbar}\, \sqrt{\frac{2 \pp \,A_{EK}}{\alpha \, c} \, \,(2J_\mathrm{e}+1) \left( \frac{c}{\omega}\right)^{3}}\,\, \WignerThreej{J_\mathrm{e}}{K}{J_\mathrm{g}}{-M_\mathrm{e}}{p}{M_\mathrm{g}}  \,\, \left( \boldgreek{\hat{\epsilon}} \bigcdot \mathbf{Y}^{(+1)}_{K,-p}(\mathbf{\hat{k}}) \right) }\label{eq:RabiFreq}
\end{align}
\end{widetext}
where I define
\begin{align*}
    s_J \equiv &\,\,(\ii)^{K-1} \,\ee^{\ii \mathbf{k} \boldsymbol{\cdot} \mathbf{R}_\mathrm{a}}\times \mathrm{sign}\left[-\frac{1}{e} \bra{J_\mathrm{e}}\! | \, T^{(K)}_{\cdot}[Q^{(E)}]\, |\!\ket{J_\mathrm{g}} \right].
\end{align*}
For many applications involving only one transition moment and one laser beam, the phase and sign of $\Omega_\mathrm{eg}$ can be ignored.

Equation (\ref{eq:RabiFreq}) is the central result of this paper; it provides an easy-to-use form for calculating Rabi frequencies for arbitrary-rank transitions as a function of experimentally-convenient parameters of the applied field.  The calculation of $\mathbf{Y}^{(+1)}_{K,-p}(\mathbf{\hat{k}})$ is outlined in Appendix \ref{app:VectorSphericalHarmonics}, with the most-common examples given explicitly, and readers unfamiliar Wigner $3j$ symbols (such as the term in large parentheses) are recommended to a student-friendly introduction in Zare \cite{Zare}.
This form can be augmented to include sub-structure (such as hyperfine structure) as described in Appendix \ref{app:Hyperfine}, and is consistent, for $K=2$, with the popular $E2$ treatment of James \cite{James1998Quantum}.  The magnetic multipole expression can be obtained through the replacement $\mathcal{E}_0  \sqrt{A_{EK}} \, \boldgreek{\hat{\epsilon}} \rightarrow c\, \mathcal{B}_0 \sqrt{A_{MK}}\, \boldgreek{\hat{\beta}} $ with $\boldgreek{\hat{\beta}} \equiv \mathbf{\hat{k}} \boldsymbol{\times} \boldgreek{\hat{\epsilon}}$.   

\section{Applications}
In some cases, the use of the form (\ref{eq:RabiFreq}) for writing the resonant Rabi frequency for a laser-driven transition allows intuition, ease of calculation, or both.  In this section, I start with simple example applications and gradually increase the complexity to include new results that illustrate the use of this formalism for modeling laser-atom interactions in nontrivial geometries.
\subsection{Elementary examples}

\subsubsection{$E1$ transitions}
I begin by noting a conceptual simplification that occurs for the $K=1$ case, which is that the field-derivative tensor (\ref{eq:PolIdty}) takes the form
\begin{align}
    \sqrt{\frac{8 \pp}{3}}\,\boldgreek{\hat{\epsilon}} \bigcdot \mathbf{Y}^{(+1)}_{1,-p}(\mathbf{\hat{k}}) = & \,\, \left( \boldgreek{\hat{\epsilon}}\right)_{-p}.\label{eq:E1Simple}
\end{align}
The vector dot product essentially extracts the appropriate component of the laser's polarization in the complex spherical basis of the quantization frame.  Using Appendix \ref{app:VectorSphericalHarmonics}, we can write the rank-$1$ vector spherical harmonics for $\lambda=+1$ in the basis of circular-polarization states as
\begin{align}
     \mathbf{Y}^{(+1)}_{1,0}(\mathbf{\hat{k}}) = & \,\, \sqrt{\frac{3}{8 \pp}}\, \sin(\vartheta_k)\left[ \frac{1}{\sqrt{2}}(\mathbf{\hat{e}}'_{+1} - \mathbf{\hat{e}}'_{-1})\right] \\
     \mathbf{Y}^{(+1)}_{1,\pm 1}(\mathbf{\hat{k}}) = & \,\,\ee^{\pm \ii \varphi_k} \sqrt{\frac{3}{8 \pp}} \left[ \cos^2\big(\shrinkify{\frac{\vartheta_k}{2}}\big)\mathbf{\hat{e}}'_{\pm 1} + \sin^2\big(\shrinkify{\frac{\vartheta_k}{2}}\big)\mathbf{\hat{e}}'_{\mp 1}\right],
\end{align}
or in the basis of linearly polarized states with polarization along $\boldgreek{\hat{\vartheta}}$ and $\boldgreek{\hat{\varphi}}$ as
\begin{align}
     \mathbf{Y}^{(+1)}_{1,0}(\mathbf{\hat{k}}) = & \,\, -\sqrt{\frac{3}{8 \pp}}\, \sin(\vartheta_k)\, \boldgreek{\hat{\vartheta}} \label{eq:Y10lin}\\
     \mathbf{Y}^{(+1)}_{1,\pm 1}(\mathbf{\hat{k}}) = & \,\, \mp \ee^{\pm \ii \varphi_k} \sqrt{\frac{3}{16 \pp}} \left[ \cos(\vartheta_k)\boldgreek{\hat{\vartheta}} \pm \ii \boldgreek{\hat{\varphi}}\right].
\end{align}
In the absence of hyperfine structure, this can be used directly with Eq.\ (\ref{eq:RabiFreq}) to calculate the resonant Rabi frequency.  The inclusion of hyperfine structure is described in Appendix \ref{app:Hyperfine}.

We can calculate the power-optimal geometry for driving a particular $\Delta M \equiv M_\mathrm{e} - M_\mathrm{g}$ transition by finding the conditions that maximize $|\boldgreek{\hat{\epsilon}} \bigcdot \mathbf{Y}^{(+1)}_{1,-\Delta M}|$, which are $\vartheta_k= \frac{\pp}{2}$ and $\boldgreek{\hat{\epsilon}} \propto \boldgreek{\hat{\vartheta}}$ for $\Delta M = 0$, $\vartheta_k=\frac{\pp}{2} \pm \frac{\pp}{2}$ and $\boldgreek{\hat{\epsilon}} \propto \boldgreek{\hat{\vartheta}} \mp \ii \boldgreek{\hat{\varphi}}$ for $\Delta M = +1$, and $\vartheta_k=\frac{\pp}{2} \pm \frac{\pp}{2}$ and $\boldgreek{\hat{\epsilon}} \propto \boldgreek{\hat{\vartheta}} \pm \ii \boldgreek{\hat{\varphi}}$ for $\Delta M = -1$.  We see that in this case, the power-optimal geometry for a given choice of $\Delta M$ sets the resonant Rabi frequency for all other choices of $\Delta M$ to zero.

As a concrete example, we can start with the most general geometry, in which the quantization direction $\mathbf{\hat{e}}_{z}$ makes an angle $\vartheta_k$ with an applied field's $\mathbf{k}$ vector.  For this example, I will specify that the field is initially linearly polarized along $\boldgreek{\hat{\varphi}}$ before passing through a quarter-wave plate whose fast axis has been tilted via a $+\mathbf{\hat{k}}$ rotation though angle $\theta_\mathrm{q}$ from $\boldgreek{\hat{\varphi}}$.  Our task is to calculate $|\Omega_\mathrm{eg}|$ for a $\Delta M = 0$ $E1$ transition as a function of $\theta_\mathrm{q}$.

Using Jones calculus (Appendix \ref{app:JonesCalculus}), we can write the polarization after the quarter-wave plate as
\begin{align*}
    \boldgreek{\hat{\epsilon}} = \shrinkify{\frac{1}{\sqrt{2}}}\,\ee^{\ii \pp/4}\sin(2\theta_\mathrm{q}) \,\boldgreek{\hat{\vartheta}} + \left[\sin^2(\theta_\mathrm{q})\! -\! \ii \cos^2(\theta_\mathrm{q}) \right] \!\boldgreek{\hat{\varphi}},
\end{align*}
and so the vector dot product with the $-p=0$ component of $\mathbf{Y}^{(+1)}_{(1)}(\mathbf{\hat{k}})$ gives us
\begin{align*}
    \boldgreek{\hat{\epsilon}} \bigcdot \mathbf{Y}^{(+1)}_{1,0}(\mathbf{\hat{k}}) = -\sqrt{\frac{3}{16 \pp}}\,\ee^{\ii \pp/4}\sin(2\theta_\mathrm{q})\sin(\vartheta_k).
\end{align*}
We therefore obtain
\begin{align*}
    |\Omega_\mathrm{eg}| = &\,\,\frac{e \mathcal{E}_0}{\hbar}\sqrt{\frac{3 A_{E1}}{8\alpha \, c} \, \,(2J_\mathrm{e}+1) \left( \frac{c}{\omega}\right)^{3}}\nonumber \\
    & \,\, \times \WignerThreej{J_\mathrm{e}}{1}{J_\mathrm{g}}{-M_\mathrm{e}}{0}{M_\mathrm{g}}\sin(2\theta_\mathrm{q})\sin(\vartheta_k).
\end{align*}
Since $\Delta M = 0$, this is a ``$\pi$ transition,'' which should be maximized for the maximum magnitude of the projection of the polarization on $\mathbf{\hat{e}}_z$.  In accordance, (\ref{eq:Y10lin}) shows that the emission pattern is $\boldgreek{\hat{\vartheta}}$ polarized (see Fig.\ \ref{fig:Y1s}), and the magnitude of the resonant Rabi frequency we calculate is minimized when the quarter-wave plate does nothing ($\theta_\mathrm{q} = n\frac{\pp}{2},\,\,n \in \mathbb{Z}$, yielding $\boldgreek{\hat{\epsilon}} = \boldgreek{\hat{\varphi}}$) and maximized when it circularizes the polarization ($\theta_\mathrm{q} = (2n+1)\frac{\pp}{4}$, which gives the largest fraction of $\pi$-polarized light under these constraints).

\subsubsection{$E2$ transitions}\label{sec:E2Transitions}
The electric quadrupole case does not admit a simplification of the type in Eq.\ (\ref{eq:E1Simple}), so visualization of $\mathbf{Y}_{2,-p}(\mathbf{\hat{k}})$ (see Fig.\ \ref{fig:Y2s}) can be used to gain intuition for the dot product in (\ref{eq:RabiFreq}).  In the absence of hyperfine structure, little needs to be added (other than the form of $\mathbf{Y}_{2,-p}(\mathbf{\hat{k}})$, which is outlined in Appendix \ref{app:VectorSphericalHarmonics}) to calculate the resonant Rabi frequency directly from (\ref{eq:RabiFreq}); hyperfine structure can be added as outlined in Appendix \ref{app:Hyperfine}.

One result that becomes apparent with this formalism is that, unlike the $E1$ case, for an $E2$ transition there is not guaranteed to be a way to isolate only a particular $\Delta M$ component of the spectrum (assuming they are degenerate) by choosing only $\mathbf{\hat{k}}$ and $\boldgreek{\hat{\epsilon}}$ for a single source of infinite plane waves.  If we seek to suppress all but one particular $\Delta M$, we require a geometry for which $|\boldgreek{\hat{\epsilon}} \bigcdot \mathbf{Y}^{(+1)}_{K,-\Delta M}(\mathbf{\hat{k}})|$ is finite at a zero of the function
\begin{align*}
    S_{2,\Delta M}(\boldgreek{\hat{\epsilon}},\mathbf{\hat{k}}) \equiv \sum_{p \neq \Delta M} \left| \boldgreek{\hat{\epsilon}} \bigcdot \mathbf{Y}^{(+1)}_{2,-p}(\mathbf{\hat{k}})\right|,
\end{align*}
which only exists in a few, special cases.

For example, to maximize the Rabi frequency for a $\Delta M = 2$ transition, the power distribution ($W(\vartheta_k)$, see Appendix \ref{app:VectorSphericalHarmonics}) shows that $\vartheta_k = \frac{\pp}{2}$ has the largest coupling per unit solid angle, and the form of $\mathbf{Y}^{(+1)}_{2,-2}(\shrinkify{\frac{\pp}{2}},\varphi_k)$ (see Fig.\ \ref{fig:Y2s}) shows that $\boldgreek{\hat{\epsilon}} \propto \boldgreek{\hat{\varphi}}$ will achieve the maximum coupling.  However, in this geometry ($\mathbf{\hat{k}},\boldgreek{\hat{\epsilon}} \perp \mathbf{\hat{e}}_z$), there is also finite coupling to the $\Delta M = -2$ component of the quadrupole transition moment, a potentially undesired effect that will be discussed in the next section.

\subsection{Single-photon transitions with multiple, phase-coherent plane waves}
In the examples discussed so far, the phase of $\Omega_\mathrm{eg}$ has not played a role, but this changes if we move on to consider simultaneous driving by multiple sources of infinite plane waves, all at the same frequency and with a well-defined phase relationship between them.  This scenario admits the possibility of interference, both between terms driving different multipoles and between terms driving the same multipole moment.  Focusing on the latter, we can write the resonant Rabi frequency for multiple incident plane waves  $j$ with the same frequency through the replacement
\begin{align*}
    \mathcal{E}_0\,\left( \boldgreek{\hat{\epsilon}} \bigcdot \mathbf{Y}^{(+1)}_{K,-p}(\mathbf{\hat{k}})\right)\, &\ee^{\ii (\mathbf{k} \boldsymbol{\cdot} \mathbf{R}_\mathrm{a} - \omega t)} \nonumber \\
    & \hspace{-15mm} \rightarrow \sum_j \mathcal{E}_{0,j}\, \left(\boldgreek{\hat{\epsilon}}_j \bigcdot \mathbf{Y}^{(+1)}_{K,-p}(\mathbf{\hat{k}}_j)\right)\, \ee^{\ii (\mathbf{k}_j \boldsymbol{\cdot} \mathbf{R}_\mathrm{a} - \omega t + \phi_j)}
\end{align*}
in Eq.\ (\ref{eq:HEK}).

An elementary example of this for $K=1$ may be familiar to the reader: if an $E1$ transition is driven simultaneously by two, equal-strength, linearly-polarized fields with $\boldgreek{\hat{\epsilon}}_1 = \mathbf{\hat{e}}_x$ and $\boldgreek{\hat{\epsilon}}_2 = \mathbf{\hat{e}}_y$, the relative strength of the $\sigma^+$ (meaning $\Delta M = +1$) and $\sigma^-$ ($\Delta M = -1$) transitions can be tuned by controlling the relative phase (say, $\phi_2$) of these two waves.  Generalizing this slightly using the formalism presented here to non-orthogonal plane waves, if we take $\vartheta_{k,j} = \frac{\pp}{2}$ and $\boldgreek{\hat{\epsilon}}_j = \boldgreek{\hat{\varphi}}(\mathbf{\hat{k}}_j)$ (meaning both $\mathbf{k}$ vectors and their polarizations are contained in the $xy$ plane, but their azimuthal angles from $\mathbf{\hat{e}}_x$ are $\varphi_{k,j}$, see Fig.\ \ref{fig:BasisVectors}), the interference appears in
\begin{align*}
    \Omega_\mathrm{eg} \propto &\,\, \boldgreek{\hat{\epsilon}}_1 \!\bigcdot\! \mathbf{Y}^{(+1)}_{1,\pm 1}(\mathbf{\hat{k}}_1)\, +\, \ee^{\ii \phi_2} \,\boldgreek{\hat{\epsilon}}_2\! \bigcdot\! \mathbf{Y}^{(+1)}_{1,\pm 1}(\mathbf{\hat{k}}_2) \\
    =&\,\, - \ii \ee^{\pm \ii \varphi_{k,1}} \sqrt{\frac{3}{16 \pp}}\, \left( 1 + \ee^{\ii \phi_2} \ee^{\pm \ii (\varphi_{k,2} - \varphi_{k,1})}\right).
\end{align*}
We see that the $\sigma^+$ transitions can be completely suppressed by choosing $\phi_2 = \pp + \varphi_{k,2} - \varphi_{k,1}$ while the $\sigma^-$ transitions vanish for $\phi_2 = \pp + \varphi_{k,1} - \varphi_{k,2}$.  In these scenarios, as long as $\mathbf{\hat{k}}_1$ and $\mathbf{\hat{k}}_2$ aren't collinear, the resonant Rabi frequency for the unsuppressed-$\sigma$ transition remains finite at this condition.  This may enable applications in constrained geometries, such as optical pumping to produce orientation or stretch-transition cycling, using only transverse ($\mathbf{\hat{k}} \perp \mathbf{\hat{e}}_z$) illumination.

A similar analysis can be applied for rank-$2$ transitions, where multi-plane-wave geometries can eliminate undesired transitions that do not vanish for single-wave illumination.  Taking the example from the end of \S \ref{sec:E2Transitions}, single-wave illumination cannot drive $\Delta M = \pm 2$ transitions while eliminating coupling to $\Delta M = \mp 2$ transitions through geometry alone.  However, adopting the same geometry as in the $K=1$ case above, we find that a two-wave geometry can furnish this ability.  The relevant interference term takes the form
\begin{align*}
    \Omega_\mathrm{eg} \propto &\,\, \boldgreek{\hat{\epsilon}}_1 \!\bigcdot\! \mathbf{Y}^{(+1)}_{2,\pm 2}(\mathbf{\hat{k}}_1)\, +\, \ee^{\ii \phi_2} \,\boldgreek{\hat{\epsilon}}_2\! \bigcdot\! \mathbf{Y}^{(+1)}_{2,\pm 2}(\mathbf{\hat{k}}_2) \\
    \propto &\,\, \left( 1 + \ee^{\ii \phi_2} \ee^{\pm \ii 2 (\varphi_{k,2} - \varphi_{k,1})}\right)
\end{align*}
and the $\Delta M = \mp 2$ transitions vanish for $\phi_2 =\pp \mp 2(\varphi_{k,2} - \varphi_{k,1})$.  In particular, two linearly-polarized waves that cross at an angle $\varphi_{k,2}-\varphi_{k,1} = \frac{\pp}{4}$ allow for maximum coupling to the desired $|\Delta M | = 2$ transition with complete suppression of the undesired one through control over their relative phase.

\subsection{Beam effects}
While the solutions provided so far have assumed infinite plane waves, they can be used as a basis for evaluating the coupling strength to a variety of \emph{beams}, by which I mean fields with finite transverse extent (Appendix \ref{app:Beams} provides a review of paraxial beams).  Transverse structure in the applied field (both its strength and polarization) can modify the coupling because it alters the distribution of the field in $\mathbf{k}$ space.  Working in the paraxial approximation ($|\mathbf{k}_\perp| \ll |\mathbf{k}|$), I will model the effect of transverse structure as a perturbative correction to the infinite-plane-wave solution when calculating the resonant Rabi frequency.   Our approach is based on an integral in Fourier space that a practitioner would likely use numerical integration methods to evaluate.  Here, I provide the first nontrivial terms in a series expansion in $1/(k w_0)$ for a variety of scenarios to illustrate the relative sizes of these effects.

\subsubsection{Centered Gaussian beams}\label{sec:CenteredGaussian}
A paraxial Gaussian beam with $1/\ee^2$ intensity radius $w_0$ produces an electric field in the plane of its minimum waist given by
\begin{align*}
    \mathbf{E}(\mathbf{r},t) = & \,\, \frac{\mathcal{E}_0}{2}\left(\boldgreek{\hat{\epsilon}}\, \exp \left(-\frac{\boldgreek{\rho}^2}{w_0^2} \right)\ee^{\ii(\mathbf{k} \boldsymbol{\cdot} \mathbf{r} - \omega t)} + \text{c.c.}\right)
\end{align*}
where $\boldgreek{\rho} \perp \mathbf{k}$, which could be written in the helicity basis (Appendix \ref{app:HelicityFrame}) as $\boldgreek{\rho} = x'\, \mathbf{\hat{e}}'_{x'} + y' \,\mathbf{\hat{e}}'_{y'}$.  Despite the fact that the wave fronts at the plane of the minimum waist of a Gaussian beam are all flat, the finite transverse extent of this beam leads to its having a spread in $\mathbf{k}$ vectors.  This is given quantitatively by the Fourier transform of the transverse beam profile,
\begin{align}
    \tilde{u}(\mathbf{k}_\perp) \equiv & \,\, \frac{1}{2 \pp} \int \mathrm{d}^2 \!\rho \,\ee^{-\ii \mathbf{k}_\perp \boldsymbol{\cdot} \boldgreeksmall{\rho}} \exp \left(-\frac{\boldgreek{\rho}^2}{w_0^2} \right) \nonumber \\
    = & \,\, \frac{w_0^2}{2}\exp\left(-\frac{\mathbf{k}_\perp^2 w_0^2}{4} \right). \label{eq:utilde}
\end{align}
where $\mathbf{k}_\perp \equiv k_{x'}\, \mathbf{\hat{e}}'_{x'} + k_{y'}\, \mathbf{\hat{e}}'_{y'}$ and I adopt the unitary transform convention with the sign of the Fourier-kernel exponent as defined by Sakurai \cite{Sakurai}.
The resonant Rabi frequency can be calculated using (\ref{eq:RabiFreq}) by replacing the dot product with the \emph{beam-coupling integral}
\begin{align}
    \left( \boldgreek{\hat{\epsilon}} \bigcdot \mathbf{Y}^{(+1)}_{K,-p}\right)_\mathbf{\hat{k}} \equiv &\,\,\boldgreek{\hat{\epsilon}}(\mathbf{\hat{k}}) \bigcdot \mathbf{Y}^{(+1)}_{K,-p}(\mathbf{\hat{k}})  \nonumber \\
    & \hspace{-17mm} \rightarrow   \frac{1}{2 \pp}\int \mathrm{d}^2 k_\perp \, \left( \boldgreek{\hat{\epsilon}} \bigcdot \mathbf{Y}^{(+1)}_{K,-p}\right)_{\boldsymbol{\hat{\ell}}} \,  \tilde{u}(\mathbf{k}_\perp) \equiv N_{K,-p}(\boldgreek{\hat{\epsilon}},\vartheta_k) \label{eq:IntegratePol}
\end{align}
where $\boldsymbol{\ell} \equiv \mathbf{k} \cos(\vartheta'_k) + \mathbf{k}_\perp \sin(\vartheta'_k)$ and $\boldsymbol{\hat{\ell}} \doteq (\vartheta_\ell, \varphi_\ell)$.  In the infinite-plane-wave limit, $\tilde{u}(\mathbf{k}_\perp) \rightarrow 2 \pp \delta^{(2)}(\mathbf{k}_\perp) = \delta(\vartheta'_k)/\sin(\vartheta'_k)$ and the beam-coupling integral $N$ converges to $\left( \boldgreek{\hat{\epsilon}} \bigcdot \mathbf{Y}^{(+1)}_{K,-p}\right)_\mathbf{\hat{k}}$.

As an example, we can calculate the correction due to the Gaussian shape of the beam on a particular $E1$ transition with $\Delta M = 0$.  I will consider that the beam's average $\mathbf{k}$ vector is along $\mathbf{\hat{e}}_x$, and that the beam is linearly polarized in the $\boldgreek{\hat{\vartheta}}$ direction.  As can be seen in Fig.\ \ref{fig:Y1s}, this would nominally drive this $\Delta M=0$ transition with maximum strength, and the infinite-plane-wave limit would give $\boldgreek{\hat{\epsilon}} \bigcdot \mathbf{Y}^{(+1)}_{1,0}(\pp/2,\varphi_k) = -\sqrt{\frac{3}{8 \pp}}$.  For the beam-coupling integral, we obtain
\begin{align*}
    N_{1,0}(\boldgreek{\hat{\vartheta}},\shrinkify{\frac{\pp}{2}}) \approx -\sqrt{\frac{3}{8 \pp}} \left(1- \frac{5}{(k w_0)^2} + \mathcal{O}\left[ \frac{1}{(k w_0)^4}\right] \right),
\end{align*}
and we see that the leading-order Gaussian-beam correction to the resonant Rabi frequency is a factor of $-5(kw_0)^{-2}$.  While the direction of $\mathbf{\hat{k}}$ and the beam polarization will determine the sign and precise size of the correction in other configurations, the scale is $\mathcal{O}[(k w_0)^{-2}]$ for $E1$ transition geometries where that term doesn't vanish due to symmetry.

\subsubsection{Off-center Gaussian beams}
Still considering an atom in the plane of minimum waist of a paraxial Gaussian beam, if the beam center is displaced from the atom's position (assumed to be the origin) in the transverse direction by $\boldgreek{\rho}_\mathrm{offs}$, $\tilde{u}(\mathbf{k}_\perp)$ picks up a (transverse spatial) frequency-dependent phase factor,
\begin{align*}
    \tilde{u}(\mathbf{k}_\perp,\boldgreek{\rho}_\mathrm{offs}) = & \,\, \frac{w_0^2}{2}  \ee^{-\mathbf{k}_\perp^2 w_0^2/4}\,\,\ee^{-\ii \mathbf{k}_\perp \! \boldsymbol{\cdot} \boldgreeksmall{\rho}_\mathrm{offs}}.
\end{align*}

On the one hand, a transverse displacement of a Gaussian beam will reduce the intensity at the atom, which will suppress the resonant Rabi frequency in most geometries.  However, the $\mathbf{k}_\perp$-dependent phase can also give rise to the dominant effect if the coupling to uniform intensity is zero.  For example, if $\mathbf{k} \parallel \mathbf{\hat{e}}_z$, an $E1$ $\Delta M = 0$ transition cannot be driven by either infinite plane waves or even a centered Gaussian beam (\S\ \ref{sec:CenteredGaussian}, also apparent in Fig.\ \ref{fig:Y1s}).  However, transversely displacing such a linearly-polarized Gaussian beam in the direction of its polarization by $\rho_\mathrm{offs}$ gives a nonzero coupling characterized by
\begin{align*}
     \left( \boldgreek{\hat{\epsilon}} \bigcdot  \mathbf{Y}^{(+1)}_{1,0}\right)_{\mathbf{\hat{e}}_z} &\\
     & \hspace{-20mm}\rightarrow  \ii \sqrt{\frac{3}{8\pp}}\,\,\frac{\rho_\mathrm{offs}}{w_0} \ee^{- \rho_\mathrm{offs}^2/w_0^2} \left( \frac{2\sqrt{2}}{kw_0} + \mathcal{O}\left[ \frac{1}{(k w_0)^3}\right]\right),
\end{align*}
which is maximized in magnitude at $\rho_\mathrm{offs} = \pm w_0/\sqrt{2}$.

As another example, we can use this formalism to visit a case demonstrated experimentally by Schmiegelow \textit{et al.} \cite{Schmiegelow2016Transfer}, in which a $|\Delta M| = 1$, $E2$ transition was driven by a linearly-polarized Gaussian beam with $\vartheta_k = \frac{\pp}{4}$ as a function of its displacement along the $\mathbf{\hat{e}}'_{x'} = \boldgreek{\hat{\vartheta}}$ direction (see Fig.\ \ref{fig:Y2s}).  In this geometry, $\boldgreek{\hat{\varphi}}$ polarization yields a coupling integral whose leading-order contribution is
\begin{align*}
    \left( \boldgreek{\hat{\epsilon}}_\boldgreeksmall{\hat{\varphi}} \bigcdot  \mathbf{Y}^{(+1)}_{2,\pm 1}\right)_{\mathbf{\hat{k}}} \rightarrow -\ii \sqrt{\frac{5}{32 \pp}}\,\ee^{-\rho_\mathrm{offs}^2/w_0^2}
\end{align*}
In contrast, it can been seen from Fig.\ \ref{fig:Y2s} that $\boldgreek{\hat{\vartheta}}$ polarization will not couple to $\mathbf{Y}^{(+1)}_{2,\pm 1}$ at $\vartheta_k = \frac{\pp}{4}$, and so the Rabi frequency will vanish for a centered Gaussian (at least to order $1/(kw_0)$, see \S \ref{sec:CenteredGaussian}).  For finite $\boldgreek{\rho}_\mathrm{offs} \parallel \boldgreek{\hat{\vartheta}}$, however, we find
\begin{align*}
    \left( \boldgreek{\hat{\epsilon}}_\boldgreeksmall{\hat{\vartheta}} \bigcdot  \mathbf{Y}^{(+1)}_{2,\pm 1}\right)_{\mathbf{\hat{k}}} \rightarrow -\ii \sqrt{\frac{5}{\pp}}\,\frac{1}{kw_0}\frac{\rho_\mathrm{offs}}{w_0}\,\ee^{-\rho_\mathrm{offs}^2/w_0^2}.
\end{align*}
The connection to the transverse field gradient in this case is particularly clear since $\boldgreek{\hat{\epsilon}}_\boldgreeksmall{\hat{\vartheta}} \bigcdot  \mathbf{Y}^{(+1)}_{2,\pm 1} \approx k_{x'}/k$, and the Fourier transform of the product of $\tilde{u}(k_{x'})$ and $k_{x'}$ is proportional to $\partial u(x')/\partial x'$.  The beam-coupling integral can therefore be viewed in this case as computing the transverse derivative of the field.

\subsubsection{Gouy phase}
Up to this point, I have neglected the corrections to the axial field gradient due to the beam shape, the leading-order of which (at the plane of the minimum waist) is the Gouy phase,
\begin{align*}
    \phi_{\mathrm{G},\mu}(z') = & \,\, \mu \,\;\mathrm{arctan}\shrinkify{\left( \frac{2z'}{k w_0^2}\right)} 
\end{align*}
where $\mu$ is a positive integer given by
\begin{align*}
    \mu \equiv &\,\, \left\{ \begin{array}{ll} m+n+1 & \mathrm{HG}_{mn} \\ 2n+|\ell|+1 & \mathrm{LG}_{n,\ell} \end{array} \right.
\end{align*}
for Hermite-Gauss (\S \ref{sec:HG}, upper) and Laguerre-Gauss (\S \ref{sec:LG}, lower) paraxial beams (see Appendix \ref{app:Beams}).
Equation (\ref{eq:HEK}) can be corrected for this finite-beam effect through the replacement
\begin{align*}
    (\ii k)^{(K-1)} \!\rightarrow \!(\ii k)^{(K-1)}\left( 1\! -\! \frac{2 \mu(K-1)}{(k w_0)^2} + \mathcal{O}\left[ \frac{1}{(k w_0)^4}\right] \right)
\end{align*}
and we see that the leading-order fractional Gouy-phase correction to the resonant Rabi frequency is $-2 \mu (K-1)(k w_0)^{-2}$.

\subsubsection{Hermite-Gauss beams}\label{sec:HG}
Hermite-Gauss modes (Appendix \ref{app:Beams}) introduce the possibility that the instantaneous field direction of even a paraxial beam flips sign across its transverse profile, which can lead to nontrivial corrections to the resonant Rabi frequency.  For example, even for a centered Gaussian beam, a $\Delta M = 0$ transition cannot be driven by an $E2$ moment if $\mathbf{\hat{k}} \perp \mathbf{\hat{e}}_z$ for any polarization (Fig.\ \ref{fig:Y2s}).  However, an $\mathrm{HG}_{10}$ Hermite-Gauss mode\footnote{I adopt the notation $\mathrm{HG}_{mn}$ for the Hermite-Gauss mode with $m$ local nulls in along $x'$ and $n$ local nulls along $y'$.} will flip its phase by $\pp$ across the $y'$ axis, allowing for simultaneous $+\boldgreek{\hat{\vartheta}}$ and $-\boldgreek{\hat{\vartheta}}$ fields in the same beam.  This can lead to a finite resonant Rabi frequency with $\mathbf{\hat{k}} \perp \mathbf{\hat{e}}_z$, where the infinite-plane-wave coupling is forbidden.

Quantitatively, the field in this case can be written as
\begin{align*}
    \mathbf{E}(\mathbf{r},t) = & \,\, \frac{\mathcal{E}_0}{2}\left(\boldgreek{\hat{\epsilon}}\,H_1\!\shrinkify{\left[ \frac{\sqrt{2}x'}{w_0}\right]} \ee^{-\boldgreeksmall{\rho}^2\!/w_0^2} \,\,\ee^{\ii(\mathbf{k} \boldsymbol{\cdot} \mathbf{r} - \omega t)} + \text{c.c.}\right)
\end{align*}
where $H_1[x] = 2x$ is a Hermite polynomial in the physics convention.  As we did for the Gaussian beam ($\mathrm{HG}_{00}$), we will work in Fourier space by introducing the Fourier transform of the transverse profile,
\begin{align*}
    \tilde{u}_{10}(\mathbf{k}_\perp) \equiv & \,\, \frac{1}{2 \pp} \int \mathrm{d}^2 \!\rho \,\ee^{-\ii \mathbf{k}_\perp \boldsymbol{\cdot} \boldgreeksmall{\rho}} \,H_1\!\shrinkify{\left[ \frac{\sqrt{2}x'}{w_0}\right]} \ee^{-\boldgreeksmall{\rho}^2\!/w_0^2} \\
    = & \,\, -\ii\frac{w_0^2}{2}H_1\!\shrinkify{\left[ \frac{k w_0 \tan(\vartheta'_k)\cos(\varphi'_k)}{\sqrt{2}}\right]}\, \ee^{-\mathbf{k}_\perp^2 w_0^2/4}.
\end{align*}
From here, the procedure for calculating the resonant Rabi frequency is the same as was used for Gaussian beams.  Taking $\boldgreek{\hat{\epsilon}} = \boldgreek{\hat{\vartheta}}$, we find that the resonant Rabi frequency for a $\Delta M = 0$, $E2$ transition with a $\mathrm{HG}_{10}$ beam propagating with $\mathbf{\hat{k}} = \mathbf{\hat{e}}_x$ (see Fig.\ \ref{fig:Y2s}) is given by (\ref{eq:RabiFreq}) with the replacement
\begin{align*}
    \left( \boldgreek{\hat{\epsilon}} \bigcdot \mathbf{Y}^{(+1)}_{2,0}\right)_\mathbf{\hat{k}} \rightarrow & \,\, \ii \sqrt{\frac{15}{8\pp}}\,\, \left( \frac{2\sqrt{2}}{kw_0} + \mathcal{O}\left[ \frac{1}{(k w_0)^3}\right]\right),
\end{align*}
from which we see that the Hermite-Gauss mode renders the resonant Rabi frequency finite in a geometry where infinite plane waves won't couple to the transition moment.

\subsubsection{Helical Laguerre-Gauss beams}\label{sec:LG}
Helical Laguerre-Gauss modes (Appendix \ref{app:Beams}) furnish a rotationally-symmetric basis for paraxial beams, and I will refer to $\mathrm{LG}_{n,\ell}$ as the mode whose field in the plane of minimum waist is given by
\begin{align*}
    \mathbf{E}(\mathbf{r},t)& \\
    & \hspace{-10mm} =  \frac{\mathcal{E}_0}{2}\left(\boldgreek{\hat{\epsilon}} \,\shrinkify{\left( \frac{\sqrt{2} \rho}{w_0}\right)^{|\ell|}} \! L_n^{(|\ell|)}\!\shrinkify{\left[ \frac{2\rho^2}{w_0^2}\right]} \ee^{-\boldgreeksmall{\rho}^2\!/w_0^2} \ee^{\ii \ell \varphi_\rho}\ee^{\ii(\mathbf{k} \boldsymbol{\cdot} \mathbf{r} - \omega t)} \!+\! \text{c.c.}\right),
\end{align*}
where $L^{(\alpha)}_k[x]$ is a generalized Laguerre polynomial and $\varphi_\rho \equiv \mathrm{arctan}(y'/x')$.  Here, the phase winding described by $\ee^{\ii \ell \varphi_\rho}$ can give rise to both small corrections and nontrivial couplings, and is the reason these are sometimes referred to as \emph{twisted} modes.  In their experiment reported in 2016, Schmiegelow and co-workers showed that an $E2$ transition in an atom could be driven by coupling the transverse gradient of a helical Laguerre-Gauss mode \cite{Schmiegelow2016Transfer}.

For example, in a $\mathbf{\hat{k}} = \mathbf{\hat{e}}_z$ geometry, the $\Delta M = \pm 2$ components of an $E2$ transition will vanish for infinite plane waves and centered Gaussian beams (Fig.\ \ref{fig:Y2s}).  For a centered $\mathrm{LG}_{0,+1}$ mode, the angular distribution is given by
\begin{align*}
    \tilde{u}(\mathbf{k}_\perp) = - \ii \frac{w_0^2}{2} \frac{k_\perp w_0}{\sqrt{2}}\, \ee^{\ii \varphi'_k}\ee^{- \mathbf{k}_\perp^2 w_0^2/4}.
\end{align*}
The coupling integral for the resonant Rabi frequency from an $\mathrm{LG}_{0,+1}$-mode beam with $\mathbf{\hat{k}} = \mathbf{\hat{e}}_z$ and left-handed circular polarization (LCP, see Appendix \ref{app:JonesCalculus}) on an $E2$ transition with $\Delta M = +2$ is
\begin{align*}
    \left( \boldgreek{\hat{\epsilon}} \bigcdot \mathbf{Y}^{(+1)}_{2,-2}\right)_\mathbf{\hat{k}} \rightarrow & \,\, \ii \sqrt{\frac{5}{8 \pp}} \left(\frac{2\sqrt{2}}{kw_0} + \mathcal{O}\left[ \frac{1}{(k w_0)^2}\right]\right)
\end{align*}
while the coupling to the $\Delta M = -2 $ component vanishes.  To pick out only the $\Delta M = -2 $ component in this geometry, the $\mathrm{LG}_{0,-1}$ mode with right-handed circular polarization (RCP) can be used, for which the coupling to $\Delta M = +2$ vanishes.

\subsubsection{Classical, nonseparable (vector-mode) beams}
Last, I consider the use of this formalism to analyze multipole transitions driven by single beams with (propagation-invariant) nonuniform polarization across their transverse profile, so-called \emph{vector modes} \cite{Rosales-Guzman2018Review}.  Since the polarization of the field in such cases depends upon $\boldgreek{\rho}$, the polarization degree of freedom is not separable from the transverse position coordinate, and the transverse spatial mode and polarization (``orbit'' and ``spin'') are sometimes said to be \emph{classically entangled} \cite{Spreeuw1998Classical}.  Classical (sometimes called \emph{local} or \emph{single-particle}) entanglement has been investigated in a number of systems and can retain many of the features of quantum entanglement, but lacks the nonlocality of quantum entanglement \cite{Hasegawa2003Violation,Souza2007Topological,Luis2009Coherence,Chen2010SinglePhoton,Borges2010BellLike,Qian2011Coherence,Goldin2010Simulating,Karimi2015Classical}.

Here, we can use the present formalism to compare the resonant Rabi frequency that can be achieved by illuminating an atom with a classical beam exhibiting such spin-orbit nonseparability to that driven by a beam with separable polarization and spatial mode (which we could call a \emph{scalar-mode} beam).

Our starting point is to consider a $\Delta M = 0$ $E1$ transition driven by a laser beam with $\mathbf{\hat{k}} = \mathbf{\hat{e}}_z$ (see Fig.\ \ref{fig:Y1s}).  Like many of the examples above, this transition is forbidden for infinite plane waves and centered Gaussian beams, but can be driven by centered beams with higher-order Hermite-Gauss or Laguerre-Gauss modes.  In particular, I will consider only the $\mathrm{HG}_{10}$ and $\mathrm{HG}_{01}$ Hermite-Gauss spatial modes (furnishing a $2$-dimensional ``orbit'' subspace) and linear polarization (a $2$-dimensional ``spin'' subspace) of the beam, which is treated classically as a continuous, un-quantized field using Maxwell's equations.

The nonseparable state of interest is a radially-polarized ``donut-beam'' state \cite{Toppel2014Classical} made of the superposition of $\mathbf{\hat{e}}_x$-polarized $\mathrm{HG}_{10}$ and $\mathbf{\hat{e}}_y$-polarized $\mathrm{HG}_{01}$ modes given by
\begin{align*}
    \mathbf{E}_\mathrm{vm}(\mathbf{r},t)& \\
    & \hspace{-15mm} = \shrinkify{\frac{\mathcal{E}_0}{2\sqrt{2}}} \left( \mathbf{\hat{e}}_x H_1\!\shrinkify{\left[ \frac{\sqrt{2}x}{w_0}\right]} \!+\! \mathbf{\hat{e}}_y H_1\!\shrinkify{\left[ \frac{\sqrt{2}y}{w_0}\right]} \right)\ee^{-\frac{x^2 + y^2}{w_0^2}}\,\ee^{\ii(kz - \omega t)}\! +\! \mathrm{c.c.}
\end{align*}
The coupling integral in this case can be broken into a sum of two integrals (one for each spatial mode involved), and yields
\begin{align*}
    \left( \boldgreek{\hat{\epsilon}} \bigcdot \mathbf{Y}^{(+1)}_{1,0}\right)_\mathbf{\hat{k}} \rightarrow &\,\, \ii \sqrt{\frac{3}{8 \pp}}\, \left(\frac{4}{kw_0} + \mathcal{O}\left[ \frac{1}{(k w_0)^3}\right]\right).
\end{align*}

In contrast, the transverse mode of a \emph{separable} beam in the basis spanned by $\mathrm{HG}_{10}$ and $\mathrm{HG}_{01}$ must be well-defined, as does its polarization, which we can express in the $x$--$y$ basis. If the polarization is linear and orthogonal to the direction in which the spatial mode has its oscillation, the coupling integral evaluates to 0 (at least to order $1/(kw_0)$).  Using this as a complete basis, we see that the largest-magnitude resonant Rabi frequency that can be achieved by a separable beam in this basis is $1/\sqrt{2}$ times the above result with a vector-mode beam.  This could be called ``nonseparability advantage'' in the sense that both cases appear to use the same resource costs (beam power, propagation direction, numerical aperture, maximum spatial frequency, etc.) but the beam with ``classical entanglement'' is more efficient at driving the transition.

\section{Discussion}
The novelty of what has been presented here is likely to be found somewhat ``in the eye of the beholder.''  Some of the more-advanced treatments (in particular, those of Shore and Menzel \cite{ShoreAndMenzel} and Johnson \cite{WalterRJohnson}) can be reduced to something resembling this form in certain limits, and the polarization identity (\ref{eq:PolIdty}) can be reconstructed through appropriate combinations of identities found in Varshalovich, Moskalev, and Khersonskii \cite{Varshalovich1988Quantum} as described in Appendix \ref{app:PolIdty}.  There is also nothing new about the idea that the laser-atom coupling should be proportional to the dot product of polarization with the atom's vector emission pattern.  Some of the conclusions in the applications section have appeared elsewhere before (and I have tried to cite those that I am aware of).  Nonetheless, it may prove useful to have the essential geometric dependence packaged in a compact (and general) form such as Eq.\ (\ref{eq:RabiFreq}), and to gain familiarity with the vector spherical harmonics as a means both to calculate and to visualize the essential geometrical aspects of the physics.

\section{Summary}
Experimentalists driving multipole transitions in the laboratory are unlikely to be occupied in trying to calculate atomic transition moments from electronic structure.  Far more commonly, they have access to knowledge of (or means to measure) the reduced transition matrix element, and understanding the dependence of the coupling on experimental parameters such as laser-beam directions and polarizations is of more practical importance.  This paper was written to provide just such a description.  While the main body of the paper outlines how the vector spherical harmonics can be used to gain intuition and  perform quantitative calculations that are unified across arbitrary-rank multipole transitions, there are appendices that follow that clearly set out the conventions I have used in the definitions and can guide readers (particularly students) through the rudiments of some of the basic concepts at work.

\begin{acknowledgments}
The author acknowledges Patrick M\"uller, Amar Vutha, Zachary Wall, Nils Huntemann, Nick Hutzler, and Jaideep Singh for helpful discussions.  This work was supported by ARO W911NF-24-S-0004 and NSF PHY-2207985 and OMA-2016245.
\end{acknowledgments}

\bibliography{Multipole}

\providecommand{\noopsort}[1]{}\providecommand{\singleletter}[1]{#1}%
\begin{thebibliography}{34}%
\makeatletter
\providecommand \@ifxundefined [1]{%
 \@ifx{#1\undefined}
}%
\providecommand \@ifnum [1]{%
 \ifnum #1\expandafter \@firstoftwo
 \else \expandafter \@secondoftwo
 \fi
}%
\providecommand \@ifx [1]{%
 \ifx #1\expandafter \@firstoftwo
 \else \expandafter \@secondoftwo
 \fi
}%
\providecommand \natexlab [1]{#1}%
\providecommand \enquote  [1]{``#1''}%
\providecommand \bibnamefont  [1]{#1}%
\providecommand \bibfnamefont [1]{#1}%
\providecommand \citenamefont [1]{#1}%
\providecommand \href@noop [0]{\@secondoftwo}%
\providecommand \href [0]{\begingroup \@sanitize@url \@href}%
\providecommand \@href[1]{\@@startlink{#1}\@@href}%
\providecommand \@@href[1]{\endgroup#1\@@endlink}%
\providecommand \@sanitize@url [0]{\catcode `\\12\catcode `\$12\catcode `\&12\catcode `\#12\catcode `\^12\catcode `\_12\catcode `\%12\relax}%
\providecommand \@@startlink[1]{}%
\providecommand \@@endlink[0]{}%
\providecommand \url  [0]{\begingroup\@sanitize@url \@url }%
\providecommand \@url [1]{\endgroup\@href {#1}{\urlprefix }}%
\providecommand \urlprefix  [0]{URL }%
\providecommand \Eprint [0]{\href }%
\providecommand \doibase [0]{https://doi.org/}%
\providecommand \selectlanguage [0]{\@gobble}%
\providecommand \bibinfo  [0]{\@secondoftwo}%
\providecommand \bibfield  [0]{\@secondoftwo}%
\providecommand \translation [1]{[#1]}%
\providecommand \BibitemOpen [0]{}%
\providecommand \bibitemStop [0]{}%
\providecommand \bibitemNoStop [0]{.\EOS\space}%
\providecommand \EOS [0]{\spacefactor3000\relax}%
\providecommand \BibitemShut  [1]{\csname bibitem#1\endcsname}%
\let\auto@bib@innerbib\@empty
\bibitem [{\citenamefont {Rose}(1955)}]{RoseMultipole}%
  \BibitemOpen
  \bibfield  {author} {\bibinfo {author} {\bibfnamefont {M.~E.}\ \bibnamefont {Rose}},\ }\href@noop {} {\emph {\bibinfo {title} {Multipole Fields}}}\ (\bibinfo  {publisher} {Wiley},\ \bibinfo {year} {1955})\BibitemShut {NoStop}%
\bibitem [{\citenamefont {Shore}\ and\ \citenamefont {Menzel}(1968)}]{ShoreAndMenzel}%
  \BibitemOpen
  \bibfield  {author} {\bibinfo {author} {\bibfnamefont {B.~W.}\ \bibnamefont {Shore}}\ and\ \bibinfo {author} {\bibfnamefont {D.~H.}\ \bibnamefont {Menzel}},\ }\href@noop {} {\emph {\bibinfo {title} {Principles of Atomic Spectra}}}\ (\bibinfo  {publisher} {John Wiley and Sons},\ \bibinfo {year} {1968})\BibitemShut {NoStop}%
\bibitem [{\citenamefont {Hamilton}(1975)}]{HamiltonNuclear}%
  \BibitemOpen
  \bibinfo {editor} {\bibfnamefont {W.~D.}\ \bibnamefont {Hamilton}},\ ed.,\ \href@noop {} {\emph {\bibinfo {title} {The Electromagnetic Interaction in Nuclear Spectroscopy}}}\ (\bibinfo  {publisher} {North-Holland Publishing Company},\ \bibinfo {year} {1975})\BibitemShut {NoStop}%
\bibitem [{\citenamefont {Weissbluth}(1978)}]{WeissbluthAtoms}%
  \BibitemOpen
  \bibfield  {author} {\bibinfo {author} {\bibfnamefont {M.}~\bibnamefont {Weissbluth}},\ }\href@noop {} {\emph {\bibinfo {title} {Atoms and Molecules}}}\ (\bibinfo  {publisher} {Academic Press},\ \bibinfo {year} {1978})\BibitemShut {NoStop}%
\bibitem [{\citenamefont {Berestetskii}\ \emph {et~al.}(1982)\citenamefont {Berestetskii}, \citenamefont {Lifshitz},\ and\ \citenamefont {Pitaevskii}}]{LandauQED}%
  \BibitemOpen
  \bibfield  {author} {\bibinfo {author} {\bibfnamefont {V.~B.}\ \bibnamefont {Berestetskii}}, \bibinfo {author} {\bibfnamefont {E.~M.}\ \bibnamefont {Lifshitz}},\ and\ \bibinfo {author} {\bibfnamefont {L.~P.}\ \bibnamefont {Pitaevskii}},\ }\href@noop {} {\emph {\bibinfo {title} {Quantum Electrodynamics}}},\ \bibinfo {edition} {2nd}\ ed.\ (\bibinfo  {publisher} {Pergamon Press},\ \bibinfo {year} {1982})\BibitemShut {NoStop}%
\bibitem [{\citenamefont {Biedenharn}\ and\ \citenamefont {Louck}(1982)}]{BiedenhornAndLouck}%
  \BibitemOpen
  \bibfield  {author} {\bibinfo {author} {\bibfnamefont {L.~C.}\ \bibnamefont {Biedenharn}}\ and\ \bibinfo {author} {\bibfnamefont {J.~D.}\ \bibnamefont {Louck}},\ }\href@noop {} {\emph {\bibinfo {title} {Angular Momentum in Quantum Physics}}}\ (\bibinfo  {publisher} {Cambridge University Press},\ \bibinfo {year} {1982})\BibitemShut {NoStop}%
\bibitem [{\citenamefont {Brink}\ and\ \citenamefont {Satchler}(1993)}]{BrinkAndSatchler}%
  \BibitemOpen
  \bibfield  {author} {\bibinfo {author} {\bibfnamefont {D.~M.}\ \bibnamefont {Brink}}\ and\ \bibinfo {author} {\bibfnamefont {G.~R.}\ \bibnamefont {Satchler}},\ }\href@noop {} {\emph {\bibinfo {title} {Angular Momentum}}},\ \bibinfo {edition} {3rd}\ ed.\ (\bibinfo  {publisher} {Oxford Science Publications},\ \bibinfo {year} {1993})\BibitemShut {NoStop}%
\bibitem [{\citenamefont {Johnson}(2007)}]{WalterRJohnson}%
  \BibitemOpen
  \bibfield  {author} {\bibinfo {author} {\bibfnamefont {W.~R.}\ \bibnamefont {Johnson}},\ }\href@noop {} {\emph {\bibinfo {title} {Atomic Structure Theory}}}\ (\bibinfo  {publisher} {Springer},\ \bibinfo {year} {2007})\BibitemShut {NoStop}%
\bibitem [{\citenamefont {Roberts}\ \emph {et~al.}(1997)\citenamefont {Roberts}, \citenamefont {Taylor}, \citenamefont {Barwood}, \citenamefont {Gill}, \citenamefont {Klein},\ and\ \citenamefont {Rowley}}]{Roberts1997Observation}%
  \BibitemOpen
  \bibfield  {author} {\bibinfo {author} {\bibfnamefont {M.}~\bibnamefont {Roberts}}, \bibinfo {author} {\bibfnamefont {P.}~\bibnamefont {Taylor}}, \bibinfo {author} {\bibfnamefont {G.~P.}\ \bibnamefont {Barwood}}, \bibinfo {author} {\bibfnamefont {P.}~\bibnamefont {Gill}}, \bibinfo {author} {\bibfnamefont {H.~A.}\ \bibnamefont {Klein}},\ and\ \bibinfo {author} {\bibfnamefont {W.~R.~C.}\ \bibnamefont {Rowley}},\ }\bibfield  {title} {\bibinfo {title} {Observation of an electric octupole transition in a single ion},\ }\href {https://doi.org/10.1103/PhysRevLett.78.1876} {\bibfield  {journal} {\bibinfo  {journal} {Phys. Rev. Lett.}\ }\textbf {\bibinfo {volume} {78}},\ \bibinfo {pages} {1876} (\bibinfo {year} {1997})}\BibitemShut {NoStop}%
\bibitem [{\citenamefont {Huntemann}\ \emph {et~al.}(2012)\citenamefont {Huntemann}, \citenamefont {Okhapkin}, \citenamefont {Lipphardt}, \citenamefont {Weyers}, \citenamefont {Tamm},\ and\ \citenamefont {Peik}}]{Huntemann2012HighAccuracy}%
  \BibitemOpen
  \bibfield  {author} {\bibinfo {author} {\bibfnamefont {N.}~\bibnamefont {Huntemann}}, \bibinfo {author} {\bibfnamefont {M.}~\bibnamefont {Okhapkin}}, \bibinfo {author} {\bibfnamefont {B.}~\bibnamefont {Lipphardt}}, \bibinfo {author} {\bibfnamefont {S.}~\bibnamefont {Weyers}}, \bibinfo {author} {\bibfnamefont {C.}~\bibnamefont {Tamm}},\ and\ \bibinfo {author} {\bibfnamefont {E.}~\bibnamefont {Peik}},\ }\bibfield  {title} {\bibinfo {title} {High-accuracy optical clock based on the octupole transition in $^{171}\mathrm{Yb}^{+}$},\ }\href {https://doi.org/10.1103/PhysRevLett.108.090801} {\bibfield  {journal} {\bibinfo  {journal} {Phys. Rev. Lett.}\ }\textbf {\bibinfo {volume} {108}},\ \bibinfo {pages} {090801} (\bibinfo {year} {2012})}\BibitemShut {NoStop}%
\bibitem [{\citenamefont {Tofful}\ \emph {et~al.}(2024)\citenamefont {Tofful}, \citenamefont {Baynham}, \citenamefont {Curtis}, \citenamefont {Parsons}, \citenamefont {Robertson}, \citenamefont {Schioppo}, \citenamefont {Tunesi}, \citenamefont {Margolis}, \citenamefont {Hendricks}, \citenamefont {Whale}, \citenamefont {Thompson},\ and\ \citenamefont {Godun}}]{Tofful2024Optical}%
  \BibitemOpen
  \bibfield  {author} {\bibinfo {author} {\bibfnamefont {A.}~\bibnamefont {Tofful}}, \bibinfo {author} {\bibfnamefont {C.~F.~A.}\ \bibnamefont {Baynham}}, \bibinfo {author} {\bibfnamefont {E.~A.}\ \bibnamefont {Curtis}}, \bibinfo {author} {\bibfnamefont {A.~O.}\ \bibnamefont {Parsons}}, \bibinfo {author} {\bibfnamefont {B.~I.}\ \bibnamefont {Robertson}}, \bibinfo {author} {\bibfnamefont {M.}~\bibnamefont {Schioppo}}, \bibinfo {author} {\bibfnamefont {J.}~\bibnamefont {Tunesi}}, \bibinfo {author} {\bibfnamefont {H.~S.}\ \bibnamefont {Margolis}}, \bibinfo {author} {\bibfnamefont {R.~J.}\ \bibnamefont {Hendricks}}, \bibinfo {author} {\bibfnamefont {J.}~\bibnamefont {Whale}}, \bibinfo {author} {\bibfnamefont {R.~C.}\ \bibnamefont {Thompson}},\ and\ \bibinfo {author} {\bibfnamefont {R.~M.}\ \bibnamefont {Godun}},\ }\bibfield  {title} {\bibinfo {title} {${}^{171}\mathrm{Yb}^+$ optical clock with $2.2\times 10^{-18}$ systematic uncertainty and absolute frequency measurements},\ }\href
  {https://doi.org/10.1088/1681-7575/ad53cd} {\bibfield  {journal} {\bibinfo  {journal} {Metrologia}\ }\textbf {\bibinfo {volume} {61}},\ \bibinfo {pages} {045001} (\bibinfo {year} {2024})}\BibitemShut {NoStop}%
\bibitem [{\citenamefont {Ludlow}\ \emph {et~al.}(2015)\citenamefont {Ludlow}, \citenamefont {Boyd}, \citenamefont {Ye}, \citenamefont {Peik},\ and\ \citenamefont {Schmidt}}]{Ludlow2015Optical}%
  \BibitemOpen
  \bibfield  {author} {\bibinfo {author} {\bibfnamefont {A.~D.}\ \bibnamefont {Ludlow}}, \bibinfo {author} {\bibfnamefont {M.~M.}\ \bibnamefont {Boyd}}, \bibinfo {author} {\bibfnamefont {J.}~\bibnamefont {Ye}}, \bibinfo {author} {\bibfnamefont {E.}~\bibnamefont {Peik}},\ and\ \bibinfo {author} {\bibfnamefont {P.~O.}\ \bibnamefont {Schmidt}},\ }\bibfield  {title} {\bibinfo {title} {Optical atomic clocks},\ }\href {https://doi.org/10.1103/RevModPhys.87.637} {\bibfield  {journal} {\bibinfo  {journal} {Rev. Mod. Phys.}\ }\textbf {\bibinfo {volume} {87}},\ \bibinfo {pages} {637} (\bibinfo {year} {2015})}\BibitemShut {NoStop}%
\bibitem [{\citenamefont {Schmiegelow}\ \emph {et~al.}(2016)\citenamefont {Schmiegelow}, \citenamefont {Schulz}, \citenamefont {Kaufmann}, \citenamefont {Ruster}, \citenamefont {Poschinger},\ and\ \citenamefont {Schmidt-Kaler}}]{Schmiegelow2016Transfer}%
  \BibitemOpen
  \bibfield  {author} {\bibinfo {author} {\bibfnamefont {C.~T.}\ \bibnamefont {Schmiegelow}}, \bibinfo {author} {\bibfnamefont {J.}~\bibnamefont {Schulz}}, \bibinfo {author} {\bibfnamefont {H.}~\bibnamefont {Kaufmann}}, \bibinfo {author} {\bibfnamefont {T.}~\bibnamefont {Ruster}}, \bibinfo {author} {\bibfnamefont {U.~G.}\ \bibnamefont {Poschinger}},\ and\ \bibinfo {author} {\bibfnamefont {F.}~\bibnamefont {Schmidt-Kaler}},\ }\bibfield  {title} {\bibinfo {title} {Transfer of optical orbital angular momentum to a bound electron},\ }\href {https://doi.org/10.1038/ncomms12998} {\bibfield  {journal} {\bibinfo  {journal} {Nature Communications}\ }\textbf {\bibinfo {volume} {7}},\ \bibinfo {pages} {12998} (\bibinfo {year} {2016})}\BibitemShut {NoStop}%
\bibitem [{\citenamefont {Sakurai}(1994)}]{Sakurai}%
  \BibitemOpen
  \bibfield  {author} {\bibinfo {author} {\bibfnamefont {J.~J.}\ \bibnamefont {Sakurai}},\ }\href@noop {} {\emph {\bibinfo {title} {Modern Quantum Mechanics}}},\ \bibinfo {edition} {revised}\ ed.\ (\bibinfo  {publisher} {Addison-Wesley},\ \bibinfo {year} {1994})\BibitemShut {NoStop}%
\bibitem [{\citenamefont {Jackson}(1999)}]{Jackson}%
  \BibitemOpen
  \bibfield  {author} {\bibinfo {author} {\bibfnamefont {J.~D.}\ \bibnamefont {Jackson}},\ }\href@noop {} {\emph {\bibinfo {title} {Classical Electrodynamics}}},\ \bibinfo {edition} {3rd}\ ed.\ (\bibinfo  {publisher} {John Wiley \& Sons},\ \bibinfo {year} {1999})\BibitemShut {NoStop}%
\bibitem [{\citenamefont {Varshalovich}\ \emph {et~al.}(1988)\citenamefont {Varshalovich}, \citenamefont {Moskalev},\ and\ \citenamefont {Khersonskii}}]{Varshalovich1988Quantum}%
  \BibitemOpen
  \bibfield  {author} {\bibinfo {author} {\bibfnamefont {D.~A.}\ \bibnamefont {Varshalovich}}, \bibinfo {author} {\bibfnamefont {A.~N.}\ \bibnamefont {Moskalev}},\ and\ \bibinfo {author} {\bibfnamefont {V.~K.}\ \bibnamefont {Khersonskii}},\ }\href {https://doi.org/10.1142/0270} {\emph {\bibinfo {title} {Quantum {Theory} of {Angular} {Momentum}}}}\ (\bibinfo  {publisher} {World Scientific},\ \bibinfo {year} {1988})\BibitemShut {NoStop}%
\bibitem [{\citenamefont {Zare}(1988)}]{Zare}%
  \BibitemOpen
  \bibfield  {author} {\bibinfo {author} {\bibfnamefont {R.~N.}\ \bibnamefont {Zare}},\ }\href@noop {} {\emph {\bibinfo {title} {Angular Momentum}}}\ (\bibinfo  {publisher} {Wiley-Interscience},\ \bibinfo {year} {1988})\BibitemShut {NoStop}%
\bibitem [{\citenamefont {James}(1998)}]{James1998Quantum}%
  \BibitemOpen
  \bibfield  {author} {\bibinfo {author} {\bibfnamefont {D.~F.~V.}\ \bibnamefont {James}},\ }\bibfield  {title} {\bibinfo {title} {Quantum dynamics of cold trapped ions with application to quantum computation},\ }\href {https://doi.org/10.1007/s003400050373} {\bibfield  {journal} {\bibinfo  {journal} {Applied Physics B}\ }\textbf {\bibinfo {volume} {66}},\ \bibinfo {pages} {181} (\bibinfo {year} {1998})}\BibitemShut {NoStop}%
\bibitem [{\citenamefont {Rosales-Guzm\'an}\ \emph {et~al.}(2018)\citenamefont {Rosales-Guzm\'an}, \citenamefont {Ndagano},\ and\ \citenamefont {Forbes}}]{Rosales-Guzman2018Review}%
  \BibitemOpen
  \bibfield  {author} {\bibinfo {author} {\bibfnamefont {C.}~\bibnamefont {Rosales-Guzm\'an}}, \bibinfo {author} {\bibfnamefont {B.}~\bibnamefont {Ndagano}},\ and\ \bibinfo {author} {\bibfnamefont {A.}~\bibnamefont {Forbes}},\ }\bibfield  {title} {\bibinfo {title} {A review of complex vector light fields and their applications},\ }\href {https://doi.org/10.1088/2040-8986/aaeb7d} {\bibfield  {journal} {\bibinfo  {journal} {Journal of Optics}\ }\textbf {\bibinfo {volume} {20}},\ \bibinfo {pages} {123001} (\bibinfo {year} {2018})}\BibitemShut {NoStop}%
\bibitem [{\citenamefont {Spreeuw}(1998)}]{Spreeuw1998Classical}%
  \BibitemOpen
  \bibfield  {author} {\bibinfo {author} {\bibfnamefont {R.~J.~C.}\ \bibnamefont {Spreeuw}},\ }\bibfield  {title} {\bibinfo {title} {A classical analogy of entanglement},\ }\href {https://doi.org/10.1023/A:1018703709245} {\bibfield  {journal} {\bibinfo  {journal} {Foundations of Physics}\ }\textbf {\bibinfo {volume} {28}},\ \bibinfo {pages} {361} (\bibinfo {year} {1998})}\BibitemShut {NoStop}%
\bibitem [{\citenamefont {Hasegawa}\ \emph {et~al.}(2003)\citenamefont {Hasegawa}, \citenamefont {Loidl}, \citenamefont {Badurek}, \citenamefont {Baron},\ and\ \citenamefont {Rauch}}]{Hasegawa2003Violation}%
  \BibitemOpen
  \bibfield  {author} {\bibinfo {author} {\bibfnamefont {Y.}~\bibnamefont {Hasegawa}}, \bibinfo {author} {\bibfnamefont {R.}~\bibnamefont {Loidl}}, \bibinfo {author} {\bibfnamefont {G.}~\bibnamefont {Badurek}}, \bibinfo {author} {\bibfnamefont {M.}~\bibnamefont {Baron}},\ and\ \bibinfo {author} {\bibfnamefont {H.}~\bibnamefont {Rauch}},\ }\bibfield  {title} {\bibinfo {title} {Violation of a bell-like inequality in single-neutron interferometry},\ }\href {https://doi.org/10.1038/nature01881} {\bibfield  {journal} {\bibinfo  {journal} {Nature}\ }\textbf {\bibinfo {volume} {425}},\ \bibinfo {pages} {45} (\bibinfo {year} {2003})}\BibitemShut {NoStop}%
\bibitem [{\citenamefont {Souza}\ \emph {et~al.}(2007)\citenamefont {Souza}, \citenamefont {Huguenin}, \citenamefont {Milman},\ and\ \citenamefont {Khoury}}]{Souza2007Topological}%
  \BibitemOpen
  \bibfield  {author} {\bibinfo {author} {\bibfnamefont {C.~E.~R.}\ \bibnamefont {Souza}}, \bibinfo {author} {\bibfnamefont {J.~A.~O.}\ \bibnamefont {Huguenin}}, \bibinfo {author} {\bibfnamefont {P.}~\bibnamefont {Milman}},\ and\ \bibinfo {author} {\bibfnamefont {A.~Z.}\ \bibnamefont {Khoury}},\ }\bibfield  {title} {\bibinfo {title} {Topological phase for spin-orbit transformations on a laser beam},\ }\href {https://doi.org/10.1103/PhysRevLett.99.160401} {\bibfield  {journal} {\bibinfo  {journal} {Phys. Rev. Lett.}\ }\textbf {\bibinfo {volume} {99}},\ \bibinfo {pages} {160401} (\bibinfo {year} {2007})}\BibitemShut {NoStop}%
\bibitem [{\citenamefont {Luis}(2009)}]{Luis2009Coherence}%
  \BibitemOpen
  \bibfield  {author} {\bibinfo {author} {\bibfnamefont {A.}~\bibnamefont {Luis}},\ }\bibfield  {title} {\bibinfo {title} {Coherence, polarization, and entanglement for classical light fields},\ }\href {https://doi.org/https://doi.org/10.1016/j.optcom.2009.06.024} {\bibfield  {journal} {\bibinfo  {journal} {Optics Communications}\ }\textbf {\bibinfo {volume} {282}},\ \bibinfo {pages} {3665} (\bibinfo {year} {2009})}\BibitemShut {NoStop}%
\bibitem [{\citenamefont {Chen}\ and\ \citenamefont {She}(2010)}]{Chen2010SinglePhoton}%
  \BibitemOpen
  \bibfield  {author} {\bibinfo {author} {\bibfnamefont {L.}~\bibnamefont {Chen}}\ and\ \bibinfo {author} {\bibfnamefont {W.}~\bibnamefont {She}},\ }\bibfield  {title} {\bibinfo {title} {Single-photon spin-orbit entanglement violating a {B}ell-like inequality},\ }\href {https://doi.org/10.1364/JOSAB.27.0000A7} {\bibfield  {journal} {\bibinfo  {journal} {J. Opt. Soc. Am. B}\ }\textbf {\bibinfo {volume} {27}},\ \bibinfo {pages} {A7} (\bibinfo {year} {2010})}\BibitemShut {NoStop}%
\bibitem [{\citenamefont {Borges}\ \emph {et~al.}(2010)\citenamefont {Borges}, \citenamefont {Hor-Meyll}, \citenamefont {Huguenin},\ and\ \citenamefont {Khoury}}]{Borges2010BellLike}%
  \BibitemOpen
  \bibfield  {author} {\bibinfo {author} {\bibfnamefont {C.~V.~S.}\ \bibnamefont {Borges}}, \bibinfo {author} {\bibfnamefont {M.}~\bibnamefont {Hor-Meyll}}, \bibinfo {author} {\bibfnamefont {J.~A.~O.}\ \bibnamefont {Huguenin}},\ and\ \bibinfo {author} {\bibfnamefont {A.~Z.}\ \bibnamefont {Khoury}},\ }\bibfield  {title} {\bibinfo {title} {Bell-like inequality for the spin-orbit separability of a laser beam},\ }\href {https://doi.org/10.1103/PhysRevA.82.033833} {\bibfield  {journal} {\bibinfo  {journal} {Phys. Rev. A}\ }\textbf {\bibinfo {volume} {82}},\ \bibinfo {pages} {033833} (\bibinfo {year} {2010})}\BibitemShut {NoStop}%
\bibitem [{\citenamefont {Qian}\ and\ \citenamefont {Eberly}(2011)}]{Qian2011Coherence}%
  \BibitemOpen
  \bibfield  {author} {\bibinfo {author} {\bibfnamefont {X.-F.}\ \bibnamefont {Qian}}\ and\ \bibinfo {author} {\bibfnamefont {J.~H.}\ \bibnamefont {Eberly}},\ }\bibfield  {title} {\bibinfo {title} {Entanglement and classical polarization states},\ }\href {https://doi.org/10.1364/OL.36.004110} {\bibfield  {journal} {\bibinfo  {journal} {Opt. Lett.}\ }\textbf {\bibinfo {volume} {36}},\ \bibinfo {pages} {4110} (\bibinfo {year} {2011})}\BibitemShut {NoStop}%
\bibitem [{\citenamefont {Goldin}\ \emph {et~al.}(2010)\citenamefont {Goldin}, \citenamefont {Francisco},\ and\ \citenamefont {Ledesma}}]{Goldin2010Simulating}%
  \BibitemOpen
  \bibfield  {author} {\bibinfo {author} {\bibfnamefont {M.~A.}\ \bibnamefont {Goldin}}, \bibinfo {author} {\bibfnamefont {D.}~\bibnamefont {Francisco}},\ and\ \bibinfo {author} {\bibfnamefont {S.}~\bibnamefont {Ledesma}},\ }\bibfield  {title} {\bibinfo {title} {Simulating {B}ell inequality violations with classical optics encoded qubits},\ }\href {https://doi.org/10.1364/JOSAB.27.000779} {\bibfield  {journal} {\bibinfo  {journal} {J. Opt. Soc. Am. B}\ }\textbf {\bibinfo {volume} {27}},\ \bibinfo {pages} {779} (\bibinfo {year} {2010})}\BibitemShut {NoStop}%
\bibitem [{\citenamefont {Karimi}\ and\ \citenamefont {Boyd}(2015)}]{Karimi2015Classical}%
  \BibitemOpen
  \bibfield  {author} {\bibinfo {author} {\bibfnamefont {E.}~\bibnamefont {Karimi}}\ and\ \bibinfo {author} {\bibfnamefont {R.~W.}\ \bibnamefont {Boyd}},\ }\bibfield  {title} {\bibinfo {title} {Classical entanglement?},\ }\href {https://doi.org/10.1126/science.aad7174} {\bibfield  {journal} {\bibinfo  {journal} {Science}\ }\textbf {\bibinfo {volume} {350}},\ \bibinfo {pages} {1172} (\bibinfo {year} {2015})}\BibitemShut {NoStop}%
\bibitem [{\citenamefont {T\"oppel}\ \emph {et~al.}(2014)\citenamefont {T\"oppel}, \citenamefont {Aiello}, \citenamefont {Marquardt}, \citenamefont {Giacobino},\ and\ \citenamefont {Leuchs}}]{Toppel2014Classical}%
  \BibitemOpen
  \bibfield  {author} {\bibinfo {author} {\bibfnamefont {F.}~\bibnamefont {T\"oppel}}, \bibinfo {author} {\bibfnamefont {A.}~\bibnamefont {Aiello}}, \bibinfo {author} {\bibfnamefont {C.}~\bibnamefont {Marquardt}}, \bibinfo {author} {\bibfnamefont {E.}~\bibnamefont {Giacobino}},\ and\ \bibinfo {author} {\bibfnamefont {G.}~\bibnamefont {Leuchs}},\ }\bibfield  {title} {\bibinfo {title} {Classical entanglement in polarization metrology},\ }\href {https://doi.org/10.1088/1367-2630/16/7/073019} {\bibfield  {journal} {\bibinfo  {journal} {New Journal of Physics}\ }\textbf {\bibinfo {volume} {16}},\ \bibinfo {pages} {073019} (\bibinfo {year} {2014})}\BibitemShut {NoStop}%
\bibitem [{\citenamefont {Saleh}\ and\ \citenamefont {Teich}(1991)}]{SalehAndTeich}%
  \BibitemOpen
  \bibfield  {author} {\bibinfo {author} {\bibfnamefont {B.~E.~A.}\ \bibnamefont {Saleh}}\ and\ \bibinfo {author} {\bibfnamefont {M.~C.}\ \bibnamefont {Teich}},\ }\href@noop {} {\emph {\bibinfo {title} {Fundamentals of Photonics}}}\ (\bibinfo  {publisher} {Wiley-Interscience},\ \bibinfo {year} {1991})\BibitemShut {NoStop}%
\bibitem [{\citenamefont {Hecht}(1998)}]{Hecht}%
  \BibitemOpen
  \bibfield  {author} {\bibinfo {author} {\bibfnamefont {E.}~\bibnamefont {Hecht}},\ }\href@noop {} {\emph {\bibinfo {title} {Optics}}},\ \bibinfo {edition} {3rd}\ ed.\ (\bibinfo  {publisher} {Addison Wesley},\ \bibinfo {year} {1998})\BibitemShut {NoStop}%
\bibitem [{\citenamefont {Yariv}(1997)}]{Yariv}%
  \BibitemOpen
  \bibfield  {author} {\bibinfo {author} {\bibfnamefont {A.}~\bibnamefont {Yariv}},\ }\href@noop {} {\emph {\bibinfo {title} {Optical Electronics in Modern Communications}}}\ (\bibinfo  {publisher} {Oxford University Press},\ \bibinfo {year} {1997})\BibitemShut {NoStop}%
\bibitem [{\citenamefont {Boyer}\ \emph {et~al.}(1975)\citenamefont {Boyer}, \citenamefont {Kalnins},\ and\ \citenamefont {Miller}}]{Boyer1975Lie}%
  \BibitemOpen
  \bibfield  {author} {\bibinfo {author} {\bibfnamefont {C.~P.}\ \bibnamefont {Boyer}}, \bibinfo {author} {\bibfnamefont {E.~G.}\ \bibnamefont {Kalnins}},\ and\ \bibinfo {author} {\bibfnamefont {J.}~\bibnamefont {Miller}, \bibfnamefont {W.}},\ }\bibfield  {title} {\bibinfo {title} {Lie theory and separation of variables. 7. {T}he harmonic oscillator in elliptic coordinates and {I}nce polynomials},\ }\href {https://doi.org/10.1063/1.522574} {\bibfield  {journal} {\bibinfo  {journal} {Journal of Mathematical Physics}\ }\textbf {\bibinfo {volume} {16}},\ \bibinfo {pages} {512} (\bibinfo {year} {1975})}\BibitemShut {NoStop}%
\bibitem [{\citenamefont {Bandres}\ and\ \citenamefont {Guti\'{e}rrez-Vega}(2004)}]{Bandres2004InceGaussian}%
  \BibitemOpen
  \bibfield  {author} {\bibinfo {author} {\bibfnamefont {M.~A.}\ \bibnamefont {Bandres}}\ and\ \bibinfo {author} {\bibfnamefont {J.~C.}\ \bibnamefont {Guti\'{e}rrez-Vega}},\ }\bibfield  {title} {\bibinfo {title} {Ince--{G}aussian modes of the paraxial wave equation and stable resonators},\ }\href {https://doi.org/10.1364/JOSAA.21.000873} {\bibfield  {journal} {\bibinfo  {journal} {J. Opt. Soc. Am. A}\ }\textbf {\bibinfo {volume} {21}},\ \bibinfo {pages} {873} (\bibinfo {year} {2004})}\BibitemShut {NoStop}%
\end{thebibliography}%

\onecolumngrid
\appendix
\renewcommand{\tocname}{Appendices}
\pagebreak

\section{Vector Spherical Harmonics}\label{app:VectorSphericalHarmonics}
Vector spherical harmonics are not typically introduced before substantial formalism of irreducible spherical tensors, but they can nonetheless be understood with little more than an understanding of scalar spherical harmonics and the spherical basis for vectors (the spherical basis is reviewed in Appendix \ref{app:HelicityFrame}).  Just as the scalar spherical harmonics ($Y_{\ell,m}(\mathbf{\hat{r}})$) are eigenfunctions of $\mathbf{L}^2$ and $L_z$,
\begin{align*}
    \mathbf{L}^2\, Y_{\ell,m}(\mathbf{\hat{r}}) = & \,\, \ell(\ell+1)\, Y_{\ell,m}(\mathbf{\hat{r}}) \\
    L_z\, Y_{\ell,m}(\mathbf{\hat{r}}) = & \,\, m\, Y_{\ell,m}(\mathbf{\hat{r}}),
\end{align*}
the vector spherical harmonics ($\mathbf{Y}^{(+1)}_{J,M}(\mathbf{\hat{r}})$ and $\mathbf{Y}^{(0)}_{J,M}(\mathbf{\hat{r}})$) are eigenfunctions of $\mathbf{J}^2$ and $J_z$ (where $\mathbf{J} = \mathbf{L} + \mathbf{S}$) as well as $\mathbf{S}^2$ (the square of the total spin operator for $S=1$),\footnote{When used with a vector field, this spin-1 operator $\mathbf{S}$ treats the three-dimensional vector it operates on as a three-component spinor.  The components of $\mathbf{S}$ in this case take the form $S_q \doteq \ii \mathbf{\hat{e}}_q \boldsymbol{\times}$ where $\boldsymbol{\times}$ is the vector cross product.  This gives $S_z \,\mathbf{\hat{e}}_q = q\,\mathbf{\hat{e}}_q$.} but \emph{not} necessarily $\mathbf{L}^2$, $L_z$, or $S_z$,
\begin{align*}
    \mathbf{J}^2\, \mathbf{Y}^{(\lambda)}_{J,M}(\mathbf{\hat{r}}) = & \,\, J(J+1)\, \mathbf{Y}^{(\lambda)}_{J,M}(\mathbf{\hat{r}})\\
    \mathbf{S}^2\, \mathbf{Y}^{(\lambda)}_{J,M}(\mathbf{\hat{r}}) = & \,\, 2\, \mathbf{Y}^{(\lambda)}_{J,M}(\mathbf{\hat{r}})\\
    J_z\, \mathbf{Y}^{(\lambda)}_{J,M}(\mathbf{\hat{r}}) = & \,\, M\, \mathbf{Y}^{(\lambda)}_{J,M}(\mathbf{\hat{r}}).
\end{align*}
The main idea at work is that these are the states of a spin $S=1$ particle that are \emph{total} angular momentum eigenstates.  $\mathbf{Y}^{(+1)}_{J,M}(\mathbf{\hat{r}})$, which is the one we're the most interested in here, contains both $L=J-1$ and $L=J+1$ character, whereas $\mathbf{Y}^{(0)}_{J,M}(\mathbf{\hat{r}})$ actually is an eigenfunction of $\mathbf{L}^2$ with associated quantum number $L=J$.  These vector spherical harmonics are also rank-$J$ irreducible spherical tensors and their components therefore transform the same way under rotations that the atomic angular momentum eigenstates do.

In this paper, I adopt the definitions of Varshalovich, Moskalev, and Khersonskii \cite{Varshalovich1988Quantum}, according to which the components of the two types ($\mathbf{Y}^{(\lambda)}$ with $\lambda \in \{0,+1\}$) of vector spherical harmonics that interest us\footnote{There is a third kind, $\mathbf{Y}^{(-1)}_{K,p}(\mathbf{\hat{r}}) \equiv \mathbf{\hat{r}}\,Y_{K,p}(\mathbf{\hat{r}})$, which is radial and therefore not applicable to the description of far-field (i.e., transverse) waves.  For our purposes, we will assume $\lambda$ can only take values $0$ or $+1$.} can be defined in terms of the gradient of the corresponding scalar spherical harmonic,
\begin{align}
    \mathbf{Y}^{(+1)}_{K,M}(\mathbf{\hat{r}}) \equiv &\,\, \frac{1}{\sqrt{K(K+1)}}\, (r\,\boldsymbol{\nabla}) Y_{K,M}(\mathbf{\hat{r}}) \\
    \mathbf{Y}^{(0)}_{K,M}(\mathbf{\hat{r}}) \equiv &\,\, - \ii\, \mathbf{\hat{r}} \boldsymbol{\times} \mathbf{Y}^{(+1)}_{K,M}(\mathbf{\hat{r}}).\label{eq:Y0def}
\end{align}
Here, since $Y_{K,M}(\mathbf{\hat{r}})$ has no dependence on the magnitude ($r$) of $\mathbf{r}$, the object
\begin{align}
    (r\,\boldsymbol{\nabla}) Y_{K,M}(\mathbf{\hat{r}}) = & \,\, \left(\boldgreek{\hat{\vartheta}}\frac{\partial}{\partial \vartheta} + \boldgreek{\hat{\varphi}}\frac{1}{\sin(\vartheta)}\frac{\partial}{\partial \varphi} \right)Y_{K,M}(\vartheta, \varphi) \label{eq:gradYKM}
\end{align}
is a quantity that is independent of $r$, despite the appearance of $r$ in the expression that may at first glance suggest the contrary.

The explicit form (\ref{eq:gradYKM}) with Eq.\ (\ref{eq:Y0def}) shows that both $\mathbf{Y}^{(+1)}_{K,M}(\mathbf{\hat{r}})$ and $\mathbf{Y}^{(0)}_{K,M}(\mathbf{\hat{r}})$ are vector fields that are everywhere perpendicular\footnote{\label{footnotePerp}Here the term \emph{perpendicular} is used loosely only in reference the vanishing dot products that follow.  This author unaware of a globally satisfying definition of what is means for complex vectors to be ``perpendicular'' to one another.} to $\mathbf{\hat{r}}$, 
\begin{align*}
    \mathbf{\hat{r}} \bigcdot \mathbf{Y}^{(+1)}_{K,M}(\mathbf{\hat{r}}) = \mathbf{\hat{r}} \bigcdot \mathbf{Y}^{(0)}_{K,M}(\mathbf{\hat{r}}) = &\,\, 0
\end{align*}
and their dot product likewise vanishes everywhere,
\begin{align*}
    \mathbf{Y}^{(+1)}_{K,M}(\mathbf{\hat{r}}) \bigcdot \mathbf{Y}^{(0)}_{K,M}(\mathbf{\hat{r}}) = &\,\, 0.
\end{align*}
This makes them convenient for describing vector transverse waves such as the electric or magnetic fields of light in the far field by replacing the argument $\mathbf{\hat{r}} \doteq (\vartheta, \varphi)$ with $\mathbf{\hat{k}} \doteq (\vartheta_k, \varphi_k)$.

The vector spherical harmonics are also orthonormal under integration of the dot product of the conjugate of one with another over all solid angle ($\mathrm{d}^2\Omega \equiv \mathrm{d}\varphi \,\mathrm{d}\vartheta \, \sin(\vartheta)$),
\begin{align}
    \int \mathrm{d}^2\Omega\,\,\, \mathbf{Y}^{(\lambda') \ast}_{K',M'}(\vartheta, \varphi)\bigcdot \mathbf{Y}^{(\lambda)}_{K,M}(\vartheta, \varphi) = \delta_{K,K'}\delta_{M,M'}\delta_{\lambda, \lambda'}. 
\end{align}

Given a particular value for $\vartheta_k$ and $\varphi_k$, the quantity $\mathbf{Y}^{(\lambda)}_{K,M}(\mathbf{\hat{k}})$ is a 3-space vector, and will have three components.  While those components can certainly be expressed in the quantization frame (see, e.g., \cite{Varshalovich1988Quantum}), it is convenient to take advantage of the orthogonality of these vectors to $\mathbf{\hat{k}}$ and express them in the helicity frame (Appendix \ref{app:HelicityFrame}), which reduces the number of nonzero components to two.  In particular, I will start with the helicity-frame \emph{components} ($q'$) of the vector spherical harmonics in terms of helicity unit vectors, for which the only potentially-nonzero components are $q'=\pm 1$.  The helicity-frame vector spherical harmonic components take a rather simple form:
\begin{align}
    \mathbf{Y}^{(+1)}_{K,q'}(\vartheta_k^\prime \!=\! 0) = &\,\, |q'| \,\,\sqrt{\frac{2K + 1}{8 \pp}}\,\, \mathbf{\hat{e}}'_{q'} (\vartheta_k,\varphi_k)\\
    \mathbf{Y}^{(0)}_{K,q'}(\vartheta_k^\prime \!=\! 0) = &\,\, -q' \,\, \sqrt{\frac{2K + 1}{8 \pp}}\,\, \mathbf{\hat{e}}'_{q'}(\vartheta_k,\varphi_k),
\end{align}
where I use the helicity-frame components $\mathbf{\hat{k}} \doteq (\vartheta_k^\prime \!=\! 0,\varphi_k^\prime \equiv 0)$ on the left and $\mathbf{\hat{e}}'_{q'} = 0$ for $|q'| \ge 2$.
Written this way, it is suggestive that the vector spherical harmonics form a convenient basis for describing the polarization of electromagnetic waves in the far field, as they appear to (and indeed do) span the space of allowed transverse vectors.  To obtain the components in the quantization frame, we can use the definition of the transformation of irreducible spherical tensor components under rotation,
\begin{align*}
    T^{(K)}_p(\vartheta_k, \varphi_k) = \sum_{q'}D^{(K)}_{-q',-p}(0,\vartheta_k,\varphi_k) \,\,T^{(K)}_{q'}(\vartheta^\prime_k \! = \!0),
\end{align*}
to get
\begin{align}
    \mathbf{Y}^{(+1)}_{K,p}(\vartheta_k, \varphi_k) = & \,\, \sqrt{\frac{2K + 1}{8 \pp}}\,\,\left[ D^{(K)}_{-1,-p}(0,\vartheta_k,\varphi_k)\,\, \mathbf{\hat{e}}'_{+1}(\vartheta_k,\varphi_k) + D^{(K)}_{+1,-p}(0,\vartheta_k,\varphi_k)\,\, \mathbf{\hat{e}}'_{-1}(\vartheta_k,\varphi_k)\right] \label{eq:Yp1circ}\\
    \mathbf{Y}^{(0)}_{K,p}(\vartheta_k, \varphi_k) = & \,\, \sqrt{\frac{2K + 1}{8 \pp}}\,\,\left[ -D^{(K)}_{-1,-p}(0,\vartheta_k,\varphi_k)\,\, \mathbf{\hat{e}}'_{+1}(\vartheta_k,\varphi_k) + D^{(K)}_{+1,-p}(0,\vartheta_k,\varphi_k)\,\, \mathbf{\hat{e}}'_{-1}(\vartheta_k,\varphi_k)\right] \label{eq:Y0circ}
\end{align}
where I adopt the definition of the Wigner rotation matrix elements of Ref.\ \cite{Varshalovich1988Quantum}: 
$D^{(J)}_{M_1,M_2}(\alpha, \beta, \gamma) \equiv \bra{J\,M_1} \ee^{-\ii \alpha \hat{J}_z} \ee^{-\ii \beta \hat{J}_y} \ee^{-\ii \gamma \hat{J}_z} \ket{J\, M_2}$.  The Wigner rotation matrix elements needed to calculate the vector spherical harmonics for ranks $K=1-3$ are provided in Table \ref{table:WignerD}, and can be calculated directly for arbitrary rank using \cite{Varshalovich1988Quantum}
\begin{align}
    D^{(K)}_{\pm 1,M_2}(0,\vartheta,\varphi) = & \,\, \mp \sqrt{\frac{4 \pp}{K(K+1)(2K+1)}} \Bigg\{ \frac{1}{2}\sqrt{(K-M_2)(K+M_2+1)}\,\, (1 \mp \cos(\vartheta))\, Y_{K,-M_2-1}(\vartheta,\varphi)\;\ee^{\ii \varphi} \nonumber \\
    & \,\, \mp M_2\,\sin(\vartheta)\, Y_{K,-M_2}(\vartheta,\varphi) \label{eq:WignerD} \\
    & \,\, - \frac{1}{2}\sqrt{(K+M_2)(K-M_2+1)}\,\,(1 \pm \cos(\vartheta))\, Y_{K,-M_2+1}(\vartheta,\varphi)\;\ee^{-\ii \varphi} \Bigg\} \nonumber\\
    = & \,\, (-)^{K-M_2}\sqrt{(K \pm 1)!\,(K \mp 1)!\,(K+M_2)!\,(K-M_2)!}\,\,\ee^{-\ii M_2 \varphi} \nonumber \\
    & \,\, \times \sum_n(-1)^n \frac{\left(\cos(\shrinkify{\frac{\vartheta}{2}})\right)^{2n + M_2 \pm 1} \left(\sin(\shrinkify{\frac{\vartheta}{2}})\right)^{2K - 2n - M_2 \mp 1} }{n!\,(K  - n \mp 1)!\,(K-M_2-n)!\,(M_2 + n \pm 1)!},\label{eq:WignerDalt}
\end{align}
where in the second form (\ref{eq:WignerDalt}), the sum over $n$ extends over all non-negative integers $n$ for which the factorials in the denominator all have non-negative arguments.

\begin{figure}[t]
\centering
\includegraphics[scale=1.0]{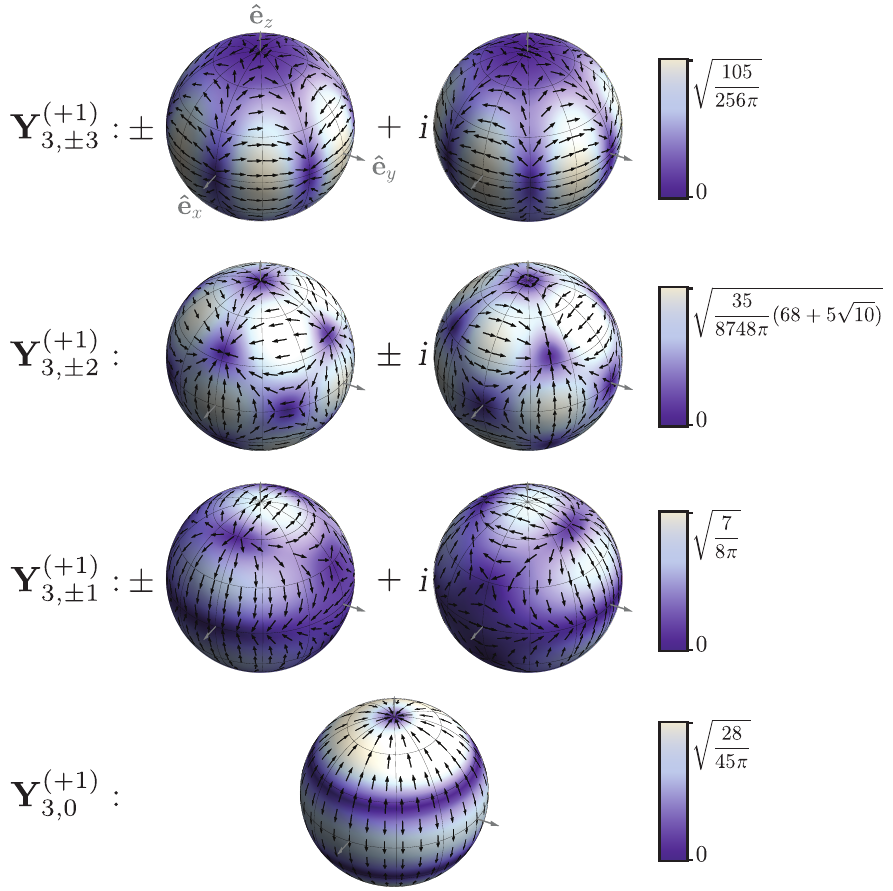}
\caption{Rank-3 vector spherical harmonics of type $\lambda = +1$.}\label{fig:Y3s}
\end{figure}

\begin{table}
\begin{center}
\begin{tabular}{|l||c|c|c|} 
 \hline
 Wigner $D$ Matrix Element & $K=1$ & $K=2$ & $K=3$ \\ [0.5ex] 
 \hline\hline
 \hspace{7mm} $D^{(K)}_{-1,-3}(0,\vartheta,\varphi)^{\strut}$ &  &  & $\ee^{\ii 3 \varphi}\, \sqrt{15}\, \cos^4(\vartheta/2)\sin^2(\vartheta/2)$ \\[1ex]
 \hline
 \hspace{7mm} $D^{(K)}_{+1,-3}(0,\vartheta,\varphi)^{\strut}$ &  &  & $\ee^{\ii 3 \varphi}\, \sqrt{15}\, \sin^4(\vartheta/2)\cos^2(\vartheta/2)$ \\[1ex]
 \hline\hline
 \hspace{7mm} $D^{(K)}_{-1,-2}(0,\vartheta,\varphi)^{\strut}$ &  & $-\ee^{\ii 2 \varphi} \sin(\vartheta)\cos^2(\vartheta/2)$ & $\ee^{\ii 2 \varphi} \frac{1}{16}\sqrt{\frac{5}{2}}\, [ \sin(\vartheta) - 4 \sin(2 \vartheta) - 3 \sin(3 \vartheta) ]$ \\[1ex]
 \hline
 \hspace{7mm} $D^{(K)}_{+1,-2}(0,\vartheta,\varphi)^{\strut}$ &  & $-\ee^{\ii 2 \varphi} \sin(\vartheta) \sin^2(\vartheta/2)$ & \hspace{1mm} $-\ee^{\ii 2 \varphi} \frac{1}{16}\sqrt{\frac{5}{2}}\, [ \sin(\vartheta) + 4 \sin(2 \vartheta) - 3 \sin(3 \vartheta) ]$ \hspace{1mm} \\[1ex]
 \hline\hline
 \hspace{7mm} $D^{(K)}_{-1,-1}(0,\vartheta,\varphi)^{\strut}$ & $\ee^{\ii \varphi} \cos^2(\vartheta/2)$ & \hspace{1mm} $\ee^{\ii \varphi} \frac{1}{2} [\cos(\vartheta) + \cos(2\vartheta)]$ \hspace{1mm} & $\ee^{\ii \varphi} \frac{1}{32} [6 + \cos(\vartheta) + 10 \cos(2 \vartheta) + 15 \cos(3 \vartheta) ]$ \\[1ex]
 \hline
 \hspace{7mm} $D^{(K)}_{+1,-1}(0,\vartheta,\varphi)^{\strut}$ & $\ee^{\ii \varphi} \sin^2(\vartheta/2)$ & $\ee^{\ii \varphi} \frac{1}{2} [\cos(\vartheta) - \cos(2\vartheta)]$ & $\ee^{\ii \varphi} \frac{1}{32} [6 - \cos(\vartheta) + 10 \cos(2 \vartheta) - 15 \cos(3 \vartheta) ]$ \\[1ex]
 \hline\hline
 \hspace{7mm} $D^{(K)}_{-1,0}(0,\vartheta,\varphi)^{\strut}$ & $\frac{1}{\sqrt{2}} \sin(\vartheta)$ & $\sqrt{\frac{3}{8}}\, \sin(2\vartheta)$ & $\frac{1}{8}\sqrt{3}\, \sin(\vartheta)[3 + 5 \cos(2 \vartheta)]$ \\[1ex]
 \hline
 \hspace{7mm} $D^{(K)}_{+1,0}(0,\vartheta,\varphi)^{\strut}$ & $-\frac{1}{\sqrt{2}} \sin(\vartheta)$ & $-\sqrt{\frac{3}{8}}\, \sin(2\vartheta)$ & $-\frac{1}{8}\sqrt{3}\, \sin(\vartheta)[3 + 5 \cos(2 \vartheta)]$ \\[1ex]
 \hline\hline
 \hspace{7mm} $D^{(K)}_{-1,+1}(0,\vartheta,\varphi)^{\strut}$ & \hspace{1mm} $\ee^{-\ii \varphi} \sin^2(\vartheta/2)$ \hspace{1mm}& $\ee^{-\ii \varphi} \frac{1}{2} [\cos(\vartheta) - \cos(2\vartheta)]$ & $\ee^{-\ii \varphi} \frac{1}{32} [6 - \cos(\vartheta) + 10 \cos(2 \vartheta) - 15 \cos(3 \vartheta) ]$ \\[1ex]
 \hline
 \hspace{7mm} $D^{(K)}_{+1,+1}(0,\vartheta,\varphi)^{\strut}$ & $\ee^{-\ii \varphi} \cos^2(\vartheta/2)$ & $\ee^{-\ii \varphi} \frac{1}{2} [\cos(\vartheta) + \cos(2\vartheta)]$ & $\ee^{-\ii \varphi} \frac{1}{32} [6 + \cos(\vartheta) + 10 \cos(2 \vartheta) + 15 \cos(3 \vartheta) ]$ \\[1ex]
 \hline\hline
 \hspace{7mm} $D^{(K)}_{-1,+2}(0,\vartheta,\varphi)^{\strut}$ &  & $\ee^{-\ii 2 \varphi} \sin(\vartheta) \sin^2(\vartheta/2)$ & $\ee^{-\ii 2 \varphi} \frac{1}{16}\sqrt{\frac{5}{2}}\, [ \sin(\vartheta) + 4 \sin(2 \vartheta) - 3 \sin(3 \vartheta) ]$ \\[1ex]
 \hline
 \hspace{7mm} $D^{(K)}_{+1,+2}(0,\vartheta,\varphi)^{\strut}$ & & $\ee^{-\ii 2 \varphi} \sin(\vartheta) \cos^2(\vartheta/2)$ & $-\ee^{-\ii 2 \varphi} \frac{1}{16}\sqrt{\frac{5}{2}}\, [ \sin(\vartheta) - 4 \sin(2 \vartheta) - 3 \sin(3 \vartheta) ]$ \\[1ex]
 \hline\hline
 \hspace{7mm} $D^{(K)}_{-1,+3}(0,\vartheta,\varphi)^{\strut}$ &  &  & $\ee^{-\ii 3 \varphi}\, \sqrt{15}\, \sin^4(\vartheta/2)\cos^2(\vartheta/2)$ \\[1ex]
 \hline
 \hspace{7mm} $D^{(K)}_{+1,+3}(0,\vartheta,\varphi)^{\strut}$ & &  & $\ee^{-\ii 3 \varphi}\, \sqrt{15}\, \cos^4(\vartheta/2)\sin^2(\vartheta/2)$ \\[1ex]
 \hline
\end{tabular}
\caption{Selected low-rank matrix elements of the Wigner rotation matrix with the first angle set to $0$, $D^{(K)}_{\pm 1,M'}(0,\vartheta,\varphi) \equiv  \bra{K,\pm 1}\, \ee^{-\ii \vartheta \hat{J}_y}\,\ee^{-\ii \varphi \hat{J}_z} \, \ket{K\,M'}$.}\label{table:WignerD}
\end{center}
\end{table}

\begin{figure}[t]
\centering
\includegraphics[scale=1.0]{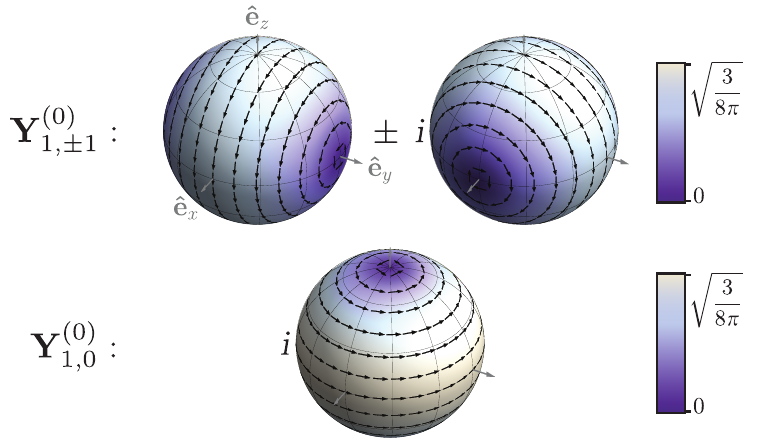}
\caption{Rank-1 vector spherical harmonics of type $\lambda = 0$, which can be used to describe the quadrature field to the field interacting with a multipole moment.}\label{fig:Y1Quad}
\end{figure}

Equations (\ref{eq:Yp1circ}) and (\ref{eq:Y0circ}) can also be transformed from the circular-polarization basis in the helicity frame ($\mathbf{\hat{e}}'_{\pm 1}$) to the linear-polarization basis in the helicity frame ($\mathbf{\hat{e}}'_{x'} = \boldgreek{\hat{\vartheta}}(\vartheta_k,\varphi_k)$ and $\mathbf{\hat{e}}'_{y'} = \boldgreek{\hat{\varphi}}(\vartheta_k,\varphi_k)$) for convenience when working with linearly-polarized light,
\begin{align}
    \mathbf{Y}^{(+1)}_{K,p}(\vartheta_k, \varphi_k) = & \,\, \frac{1}{2\sqrt{K(K+1)}} \Bigg[ \ii \boldgreek{\hat{\varphi}}(\vartheta_k,\varphi_k) \, \frac{ 2 p}{\sin(\vartheta_k)}Y_{K,p}(\vartheta_k,\varphi_k)  \nonumber\\
    & \hspace{-15mm} + \boldgreek{\hat{\vartheta}}(\vartheta_k,\varphi_k) \, \left( \sqrt{K(K+1) - p(p+1)}\,\,Y_{K,p+1}(\vartheta_k,\varphi_k)\,\ee^{-\ii \varphi_k} - \sqrt{K(K+1) -p(p-1)}\,\, Y_{K,p-1}(\vartheta_k, \varphi_k)\,\ee^{\ii \varphi_k}\right)\Bigg] \\
    \mathbf{Y}^{(0)}_{K,p}(\vartheta_k, \varphi_k) = & \,\, -\frac{1}{2\sqrt{K(K+1)}} \Bigg[ \boldgreek{\hat{\vartheta}}(\vartheta_k,\varphi_k) \, \frac{ 2 p}{\sin(\vartheta_k)}Y_{K,p}(\vartheta_k,\varphi_k)  \nonumber\\
    & \hspace{-15mm}+ \ii \boldgreek{\hat{\varphi}}(\vartheta_k,\varphi_k) \, \left( \sqrt{K(K+1) - p(p+1)}\,\,Y_{K,p+1}(\vartheta_k,\varphi_k)\,\ee^{-\ii \varphi_k} - \sqrt{K(K+1) -p(p-1)}\,\, Y_{K,p-1}(\vartheta_k, \varphi_k)\,\ee^{\ii \varphi_k}\right)\Bigg].
\end{align}
The helicity-frame unit vectors $\mathbf{\hat{e}}'_{x'}$ and $\mathbf{\hat{e}}'_{y'}$ are related to the quantization-frame basis vectors $\mathbf{\hat{e}}_x$, $\mathbf{\hat{e}}_y$, and $\mathbf{\hat{e}}_z$ via
\begin{align*}
    \mathbf{\hat{e}}'_{x'} = & \,\, \boldgreek{\hat{\vartheta}}(\vartheta_k,\varphi_k) \nonumber \\
    =&\,\, \mathbf{\hat{e}}_x\, \cos(\vartheta_k)\cos(\varphi_k) + \mathbf{\hat{e}}_y \, \cos(\vartheta_k) \sin(\varphi_k) - \mathbf{\hat{e}}_z \, \sin(\vartheta_k) \\
    \mathbf{\hat{e}}'_{y'} = & \,\, \boldgreek{\hat{\varphi}}(\vartheta_k,\varphi_k) \nonumber \\
    =&\,\, -\mathbf{\hat{e}}_x\, \sin(\varphi_k) + \mathbf{\hat{e}}_y \,  \cos(\varphi_k).
\end{align*}

For some calculations, it may be convenient to separate a vector spherical harmonic into the product of its magnitude (denoted by the square root of $W_{K,|p|}(\vartheta_k)$ (defined below), which has no dependence on $\lambda \in \{+1,0 \}$, $\varphi_k$, or the sign of $p$) and a unit vector for its direction ($\mathbf{\widehat{Y}}^{(\lambda)}_{K,p}(\mathbf{\hat{k}})$) according to \cite{Varshalovich1988Quantum}
\begin{align}
    W_{K,|p|}(\vartheta_k) \equiv &\,\, \left| \mathbf{Y}^{(\lambda)}_{K,p}(\mathbf{\hat{k}}) \right|^2 \nonumber \\
    = & \,\, \frac{1}{2K(K+1)} \Bigg\{ (K+p)(K-p+1) \left| Y_{K,p-1}(\mathbf{\hat{k}}) \right|^2 +2 p^2\left| Y_{K,p}(\mathbf{\hat{k}}) \right|^2 +(K-p)(K+p+1) \left| Y_{K,p+1}(\mathbf{\hat{k}}) \right|^2 \Bigg\} \label{eq:Wperp}
\end{align}
and
\begin{align}
    \mathbf{\widehat{Y}}^{(\lambda)}_{K,p}(\mathbf{\hat{k}}) \equiv &\,\, \frac{\mathbf{Y}^{(\lambda)}_{K,p}(\mathbf{\hat{k}})}{\sqrt{W_{K,|p|}(\vartheta_k)}}.\label{eq:VSHhat}
\end{align}
Equation (\ref{eq:Wperp}) gives the well-known angular distribution of spontaneously emitted \emph{power} for a $2^K$-pole transition with $p = M_\mathrm{e} - M_\mathrm{g}$, while $\mathbf{\widehat{Y}}^{(+1)}_{K,-p}(\mathbf{\hat{k}})$ gives the \emph{polarization vector} of the emitted light. 

\begin{table}
\begin{center}
\begin{tabular}{|l||c|c|c|} 
 \hline
 & Magnitude, $\sqrt{W(\vartheta)}^{\strut}$ & Direction, $\mathbf{\widehat{Y}}^{(+1)}/n_i$, circular basis & Direction, $\mathbf{\widehat{Y}}^{(+1)}/n_i$, linear basis\\ [1ex] 
 \hline\hline
 $\mathbf{Y}^{(+1) \strut}_{1,0}(\mathbf{\hat{r}})$ & $\sqrt{\frac{3}{8 \pp} \sin^2(\vartheta)}$ & $\mathbf{\hat{e}}'_{+1} - \mathbf{\hat{e}}'_{-1}$ & $-\boldgreek{\hat{\vartheta}}$ \\[1.5ex]
 \hline
 $\mathbf{Y}^{(+1) \strut}_{1,\pm 1}(\mathbf{\hat{r}})$ & $\sqrt{\frac{3}{16 \pp} \left[1 + \cos^2(\vartheta)\right]}$ & $\ee^{\pm \ii \varphi}\!\left[ \cos^2\!\big(\frac{\vartheta}{2}\big) \, \mathbf{\hat{e}}'_{\pm 1} + \sin^2 \!\big(\frac{\vartheta}{2}\big)\, \mathbf{\hat{e}}'_{\mp 1} \right]$ & $\mp \ee^{\pm \ii \varphi}\!\left[ \cos(\vartheta) \,\boldgreek{\hat{\vartheta}} \pm \ii \boldgreek{\hat{\varphi}}\right]$ \\[1.5ex]
 \hline\hline
 $\mathbf{Y}^{(+1) \strut}_{2,0}(\mathbf{\hat{r}})$ & $\sqrt{\frac{15}{8 \pp} \sin^2(\vartheta)\cos^2(\vartheta)}$ & $\cos(\vartheta)\left(\mathbf{\hat{e}}'_{+1} - \mathbf{\hat{e}}'_{-1}\right)$ & $-\cos(\vartheta)\, \boldgreek{\hat{\vartheta}}$ \\[1.5ex]
 \hline
 $\mathbf{Y}^{(+1) \strut}_{2,\pm 1}(\mathbf{\hat{r}})$ & $\sqrt{\frac{5}{16 \pp}\!\mbox{\scriptsize$\left[1 \!- \!3 \cos^2(\vartheta)\! +\! 4 \cos^4(\vartheta) \right]$}}$ & $\ee^{\pm \ii \varphi}\!\mbox{\scriptsize$\cos(\vartheta)$}\!\shrinkify{\left(\left[1 \!+ \!\frac{\cos(2 \vartheta)}{\cos(\vartheta)} \right]\!\mathbf{\hat{e}}'_{\pm 1}\! + \!\left[1 \!-\! \frac{\cos(2 \vartheta)}{\cos(\vartheta)} \!\right]\mathbf{\hat{e}}'_{\mp 1} \right)}$ & $\mp \ee^{\pm \ii \varphi}\left[ \cos(2\vartheta) \, \boldgreek{\hat{\vartheta}} \pm \ii \cos(\vartheta)\, \boldgreek{\hat{\varphi}}\right]$ \\[1ex]
 \hline
 $\mathbf{Y}^{(+1) \strut}_{2,\pm 2}(\mathbf{\hat{r}})$ & $\sqrt{\frac{5}{16 \pp}\left[1 - \cos^4(\vartheta) \right]}$ & $\mp \ee^{\pm \ii 2 \varphi}\!\left[ \cos^2\!\big(\frac{\vartheta}{2}\big) \, \mathbf{\hat{e}}'_{\pm 1} + \sin^2 \!\big(\frac{\vartheta}{2}\big)\, \mathbf{\hat{e}}'_{\mp 1} \right]$ & $\ee^{\pm \ii 2 \varphi}\!\left[ \cos(\vartheta) \,\boldgreek{\hat{\vartheta}} \pm \ii \boldgreek{\hat{\varphi}}\right]$ \\[1ex]
 \hline\hline
 $\mathbf{Y}^{(+1) \strut}_{3,0}(\mathbf{\hat{r}})$ & $\sqrt{\frac{21}{64 \pp}\mbox{\scriptsize$\sin^2(\vartheta)\left[ 1\!-\!5 \cos^2(\vartheta)\right]^2$}}$ & $\left[3+5\cos(2 \vartheta)\right]\left(\mathbf{\hat{e}}'_{+1} - \mathbf{\hat{e}}'_{-1}\right)$ & $-\left[3+5\cos(2 \vartheta)\right]\, \boldgreek{\hat{\vartheta}}$ \\[1ex]
 \hline
 $\mathbf{Y}^{(+1) \strut}_{3,\pm 1}(\mathbf{\hat{r}})$ & $\mbox{\tiny$\sqrt{\frac{1 + 111\cos^2(\vartheta) - 305\cos^4(\vartheta)+225\cos^6(\vartheta)}{256 \pp/7}}$}$ &$\begin{array}{l}\ee^{\pm \ii \varphi}\big\{ \left[6 + \cos(\vartheta) + 10\cos(2 \vartheta) + 15\cos(3 \vartheta)\right] \mathbf{\hat{e}}'_{\pm 1} \\ \hspace{3.5mm}+\left[6 - \cos(\vartheta) + 10\cos(2 \vartheta) - 15\cos(3 \vartheta)\right] \mathbf{\hat{e}}'_{\mp 1} \big\}\end{array}$&$\begin{array}{l} \mp \ee^{\pm \ii \varphi} \big\{ \left[ \cos(\vartheta) + 15 \cos(3 \vartheta)\right] \boldgreek{\hat{\vartheta}} \\\hspace{12mm} \pm \ii \left[ 6 + 10\cos(2 \vartheta)\right]\boldgreek{\hat{\varphi}} \big\} \end{array}$\\[1ex]
 \hline
 $\mathbf{Y}^{(+1) \strut}_{3,\pm 2}(\mathbf{\hat{r}})$ & $\sqrt{\mbox{$\frac{35 \sin^2(\vartheta)\left[1-2\cos^2(\vartheta) + 9 \cos^4(\vartheta) \right]}{128 \pp}$}}$ &$\begin{array}{l} \pm \ee^{\pm \ii 2 \varphi} \big\{ \left[\sin(\vartheta) - 4 \sin(2\vartheta) - 3\sin(3 \vartheta) \right]\mathbf{\hat{e}}'_{\pm 1} \\\hspace{12mm} - \left[\sin(\vartheta) + 4 \sin(2\vartheta) - 3\sin(3 \vartheta) \right]\mathbf{\hat{e}}'_{\mp 1} \big\}\end{array}$&$\begin{array}{l} -\ee^{\pm \ii 2 \varphi} \big\{ \left[ \sin(\vartheta) - 3 \sin(3 \vartheta)\right] \boldgreek{\hat{\vartheta}} \\ \hspace{22mm}\mp \ii  4\sin(2 \vartheta)\,\boldgreek{\hat{\varphi}} \big\} \end{array}$\\[1ex]
 \hline
 $\mathbf{Y}^{(+1) \strut}_{3,\pm 3}(\mathbf{\hat{r}})$ & $\sqrt{\frac{105}{256 \pp}\sin^4(\vartheta)\left[ 1+\cos^2(\vartheta)\right]}$ & $\ee^{\pm \ii 3 \varphi}\!\left[ \cos^2\!\big(\frac{\vartheta}{2}\big) \, \mathbf{\hat{e}}'_{\pm 1} + \sin^2 \!\big(\frac{\vartheta}{2}\big)\, \mathbf{\hat{e}}'_{\mp 1} \right]$ & $\mp \ee^{\pm \ii 3 \varphi}\!\left[ \cos(\vartheta) \,\boldgreek{\hat{\vartheta}} \pm \ii \boldgreek{\hat{\varphi}}\right]$ \\[1ex]
 \hline
\end{tabular}
\caption{Selected low-rank, $\lambda=+1$ vector spherical harmonics in terms of magnitude and direction.  For the (complex) direction, $n_i$ represents an entry-specific multiplicative normalization factor (which may be a function of $\vartheta$, but is defined to be positive and real) required to make each entry a unit vector; entries are un-normalized to allow for space, but are hopefully straightforward to normalize for use.  Entries in the direction columns are written so that, when normalized according to $\mathbf{\hat{a}} \equiv \mathbf{a}/\sqrt{\mathbf{a}^\ast \bigcdot \mathbf{a}}$ and multiplied by the entry in the magnitude column, that product yields $\mathbf{Y}^{(+1)}_{K,p}(\vartheta,\varphi)$.}\label{table:WYhat}
\end{center}
\end{table}

Table \ref{table:WYhat} provides explicit expressions for the $\lambda=+1$ vector spherical harmonics up to rank 3, separated into a magnitude and a direction.  The direction entries in Table \ref{table:WYhat} are not normalized due to space constraints, but the full vector spherical harmonic can be reconstructed from the table by normalizing the direction vector and multiplying it by the magnitude.

For example, to use Table \ref{table:WYhat} to write $\mathbf{Y}^{(+1)}_{2,0}(\mathbf{\hat{k}})$ using circular-polarization basis vectors, the first and second columns of the third row of the table give
\begin{align*}
    \mathbf{Y}^{(+1)}_{2,0}(\vartheta_k,\varphi_k) = & \,\, \sqrt{\frac{15}{8 \pp} \sin^2(\vartheta_k)\cos^2(\vartheta_k)}\,\,\frac{\cos(\vartheta_k)\left(\mathbf{\hat{e}}'_{+1} - \mathbf{\hat{e}}'_{-1}\right)}{\sqrt{\cos^2(\vartheta_k)\left(\mathbf{\hat{e}}^{\prime}_{+1} - \mathbf{\hat{e}}^{\prime}_{-1}\right)^\ast \bigcdot \left(\mathbf{\hat{e}}'_{+1} - \mathbf{\hat{e}}'_{-1}\right)}} \\
    = & \,\, \sqrt{\frac{15}{8 \pp}}\,\,\sin(\vartheta_k)\,|\!\cos(\vartheta_k)|\,\frac{\cos(\vartheta_k)}{|\!\cos(\vartheta_k)|}\, \frac{1}{\sqrt{2}}\left(\mathbf{\hat{e}}'_{+1} - \mathbf{\hat{e}}'_{-1}\right) \\
    = & \,\, \sqrt{\frac{15}{32 \pp}}\,\,\sin(2\vartheta_k)\,\frac{1}{\sqrt{2}}\left(\mathbf{\hat{e}}'_{+1} - \mathbf{\hat{e}}'_{-1}\right).
\end{align*}

As described in the main body of this paper, the $\lambda = +1$ vector spherical harmonics describe the polarization of the field interacting with the multipole in the far field, and are therefore the main type considered when calculating single-moment processes.  The $\lambda=0$ vector spherical harmonics, being uniformly perpendicular\footnote{see footnote \ref{footnotePerp}} to their $\lambda=+1$ counterparts under the dot product in space and also purely transverse, therefore describe the complimentary (or quadrature) field of the light for that process.  For example, the electric field polarization for $E2$ transitions will be given by $\mathbf{Y}^{(+1)}_{(2)}$ and the magnetic field polarization will be given by $\ii \mathbf{Y}^{(0)}_{(2)}$, whereas for $M2$ transitions, it is the magnetic field polarization that points in the $\mathbf{Y}^{(+1)}_{(2)}$ direction and the electric field polarization is along $-\ii \mathbf{Y}^{(0)}_{(2)}$.  The flow of energy, momentum, and angular momentum is described by the (complex) Poynting unit vector $\mathbf{\hat{s}} = \boldgreek{\hat{\epsilon}} \!\boldsymbol{\times}\! \boldgreek{\hat{\beta}}^\ast$ \cite{BiedenhornAndLouck}.  The three tensor components of $\mathbf{Y}^{(0)}_{(1)}$ are shown in Fig.\ \ref{fig:Y1Quad}.

In the linear-polarization basis, the first few $\lambda = +1$ vector spherical harmonics take the form
\begin{align*}
    \mathbf{Y}^{(+1)}_{1,-1}(\vartheta, \varphi) = & \,\, \ee^{-\ii \varphi}\sqrt{\shrinkify{\frac{3}{16 \pp}}}\, \left[ \cos(\vartheta) \,\boldgreek{\hat{\vartheta}} - \ii \boldgreek{\hat{\varphi}}\right] \\
    \mathbf{Y}^{(+1)}_{1,0}(\vartheta, \varphi) = & \,\, -\sqrt{\shrinkify{\frac{3}{8 \pp}}}\, \, \sin(\vartheta)\, \boldgreek{\hat{\vartheta}} \\
    \mathbf{Y}^{(+1)}_{1,+1}(\vartheta, \varphi) = & \,\, -\ee^{\ii \varphi}\sqrt{\shrinkify{\frac{3}{16 \pp}}}\, \left[ \cos(\vartheta) \,\boldgreek{\hat{\vartheta}} + \ii \boldgreek{\hat{\varphi}}\right]
\end{align*}
\begin{align*}
    \mathbf{Y}^{(+1)}_{2,-2}(\vartheta, \varphi) = & \,\, \ee^{-\ii 2\varphi}\sqrt{\shrinkify{\frac{5}{16 \pp}}}\,\sin(\vartheta) \left[ \cos(\vartheta) \,\boldgreek{\hat{\vartheta}} - \ii \boldgreek{\hat{\varphi}}\right] \\
    \mathbf{Y}^{(+1)}_{2,-1}(\vartheta, \varphi) = & \,\, \ee^{-\ii \varphi} \sqrt{\shrinkify{\frac{5}{16 \pp}}}\,\left[ \cos(2\vartheta) \,\boldgreek{\hat{\vartheta}} - \ii \cos(\vartheta)\,\boldgreek{\hat{\varphi}}\right]\\
    \mathbf{Y}^{(+1)}_{2,0}(\vartheta, \varphi) = & \,\, -\sqrt{\shrinkify{\frac{15}{32 \pp}}}\, \,\sin(2\vartheta) \,\boldgreek{\hat{\vartheta}} \\
    \mathbf{Y}^{(+1)}_{2,+1}(\vartheta, \varphi) = & \,\, -\ee^{\ii \varphi} \sqrt{\shrinkify{\frac{5}{16 \pp}}}\,\left[ \cos(2\vartheta) \,\boldgreek{\hat{\vartheta}} + \ii \cos(\vartheta)\,\boldgreek{\hat{\varphi}}\right]\\
    \mathbf{Y}^{(+1)}_{2,+2}(\vartheta, \varphi) = & \,\, \ee^{\ii 2\varphi}\sqrt{\shrinkify{\frac{5}{16 \pp}}}\,\sin(\vartheta) \left[ \cos(\vartheta) \,\boldgreek{\hat{\vartheta}} + \ii \boldgreek{\hat{\varphi}}\right]
\end{align*}
\begin{align*}
    \mathbf{Y}^{(+1)}_{3,-3}(\vartheta, \varphi) = & \,\, \ee^{-\ii 3\varphi}\sqrt{\shrinkify{\frac{105}{256 \pp}}}\,\sin^2(\vartheta) \left[ \cos(\vartheta) \,\boldgreek{\hat{\vartheta}} - \ii \boldgreek{\hat{\varphi}}\right] \\
    \mathbf{Y}^{(+1)}_{3,-2}(\vartheta, \varphi) = & \,\, -\ee^{-\ii 2\varphi}\sqrt{\shrinkify{\frac{35}{2048 \pp}}}\,\left\{ \left[\sin(\vartheta) - 3\sin(3\vartheta) \right]\boldgreek{\hat{\vartheta}} + \ii 4 \sin(2\vartheta)\,\boldgreek{\hat{\varphi}}\right\} \\
    \mathbf{Y}^{(+1)}_{3,-1}(\vartheta, \varphi) = & \,\, \ee^{-\ii \varphi} \sqrt{\shrinkify{\frac{7}{4096 \pp}}}\,\left\{ \left[ \cos(\vartheta) + 15\cos(3\vartheta) \right]\boldgreek{\hat{\vartheta}} - \ii \left[6 + 10\cos(2\vartheta) \right]\boldgreek{\hat{\varphi}}\right\}\\
    \mathbf{Y}^{(+1)}_{3,0}(\vartheta, \varphi) = & \,\, -\sqrt{\shrinkify{\frac{21}{256 \pp}}}\, \,\sin(\vartheta) \left[3 + 5 \cos(2\vartheta) \right]\boldgreek{\hat{\vartheta}} \\
    \mathbf{Y}^{(+1)}_{3,+1}(\vartheta, \varphi) = & \,\, -\ee^{\ii \varphi} \sqrt{\shrinkify{\frac{7}{4096 \pp}}}\,\left\{ \left[ \cos(\vartheta) + 15\cos(3\vartheta) \right]\boldgreek{\hat{\vartheta}} + \ii \left[6 + 10\cos(2\vartheta) \right]\boldgreek{\hat{\varphi}}\right\}\\
    \mathbf{Y}^{(+1)}_{3,+2}(\vartheta, \varphi) = & \,\, -\ee^{\ii 2\varphi}\sqrt{\shrinkify{\frac{35}{2048 \pp}}}\,\left\{ \left[\sin(\vartheta) - 3\sin(3\vartheta) \right]\boldgreek{\hat{\vartheta}} - \ii 4 \sin(2\vartheta)\,\boldgreek{\hat{\varphi}}\right\} \\
    \mathbf{Y}^{(+1)}_{3,+3}(\vartheta, \varphi) = & \,\, -\ee^{\ii 3\varphi}\sqrt{\shrinkify{\frac{105}{256 \pp}}}\,\sin^2(\vartheta) \left[ \cos(\vartheta) \,\boldgreek{\hat{\vartheta}} + \ii \boldgreek{\hat{\varphi}}\right].
\end{align*}

In the circular-polarization basis, the first few $\lambda = +1$ vector spherical harmonics take the form
\begin{align*}
    \mathbf{Y}^{(+1)}_{1,-1}(\vartheta, \varphi) = & \,\, \ee^{-\ii \varphi}\sqrt{\shrinkify{\frac{3}{8 \pp}}}\, \left[ \sin^2(\shrinkify{\frac{\vartheta}{2}})\,\mathbf{\hat{e}}'_{+1} + \cos^2(\shrinkify{\frac{\vartheta}{2}})\,\mathbf{\hat{e}}'_{-1}\right] \\
    \mathbf{Y}^{(+1)}_{1,0}(\vartheta, \varphi) = & \,\, \sqrt{\shrinkify{\frac{3}{16 \pp}}}\, \, \sin(\vartheta)\left[ \mathbf{\hat{e}}'_{+1} - \mathbf{\hat{e}}'_{-1}\right] \\
    \mathbf{Y}^{(+1)}_{1,+1}(\vartheta, \varphi) = & \,\, \ee^{\ii \varphi}\sqrt{\shrinkify{\frac{3}{8 \pp}}}\, \left[ \cos^2(\shrinkify{\frac{\vartheta}{2}})\,\mathbf{\hat{e}}'_{+1} + \sin^2(\shrinkify{\frac{\vartheta}{2}})\,\mathbf{\hat{e}}'_{-1}\right]
\end{align*}
\begin{align*}
    \mathbf{Y}^{(+1)}_{2,-2}(\vartheta, \varphi) = & \,\, \ee^{-\ii 2\varphi}\sqrt{\shrinkify{\frac{5}{8 \pp}}}\,\sin(\vartheta) \left[ \sin^2(\shrinkify{\frac{\vartheta}{2}})\,\mathbf{\hat{e}}'_{+1} + \cos^2(\shrinkify{\frac{\vartheta}{2}})\,\mathbf{\hat{e}}'_{-1}\right] \\
    \mathbf{Y}^{(+1)}_{2,-1}(\vartheta, \varphi) = & \,\, \ee^{-\ii \varphi} \sqrt{\shrinkify{\frac{5}{32 \pp}}} \,\left\{ \left[ \cos(\vartheta) - \cos(2\vartheta) \right] \,\mathbf{\hat{e}}'_{+1} + \left[ \cos(\vartheta) + \cos(2\vartheta) \right] \,\mathbf{\hat{e}}'_{-1}  \right\}\\
    \mathbf{Y}^{(+1)}_{2,0}(\vartheta, \varphi) = & \,\, \sqrt{\shrinkify{\frac{15}{64 \pp}}}\,\sin(2\vartheta) \left[\mathbf{\hat{e}}'_{+1} - \mathbf{\hat{e}}'_{-1} \right]  \\
    \mathbf{Y}^{(+1)}_{2,+1}(\vartheta, \varphi) = & \,\, \ee^{\ii \varphi} \sqrt{\shrinkify{\frac{5}{32 \pp}}} \,\left\{ \left[ \cos(\vartheta) + \cos(2\vartheta) \right] \,\mathbf{\hat{e}}'_{+1} + \left[ \cos(\vartheta) - \cos(2\vartheta) \right] \,\mathbf{\hat{e}}'_{-1}  \right\}\\
    \mathbf{Y}^{(+1)}_{2,+2}(\vartheta, \varphi) = & \,\, -\ee^{\ii 2\varphi}\sqrt{\shrinkify{\frac{5}{8 \pp}}}\,\sin(\vartheta) \left[ \cos^2(\shrinkify{\frac{\vartheta}{2}})\,\mathbf{\hat{e}}'_{+1} + \sin^2(\shrinkify{\frac{\vartheta}{2}})\,\mathbf{\hat{e}}'_{-1}\right]
\end{align*}
\begin{align*}
    \mathbf{Y}^{(+1)}_{3,-3}(\vartheta, \varphi) = & \,\, \ee^{-\ii 3\varphi}\sqrt{\shrinkify{\frac{105}{128 \pp}}}\,\sin^2(\vartheta)\left[ \sin^2(\shrinkify{\frac{\vartheta}{2}})\,\mathbf{\hat{e}}'_{+1} + \cos^2(\shrinkify{\frac{\vartheta}{2}})\,\mathbf{\hat{e}}'_{-1}\right] \\
    \mathbf{Y}^{(+1)}_{3,-2}(\vartheta, \varphi) = & \,\, \ee^{-\ii 2\varphi}\sqrt{\shrinkify{\frac{35}{4096 \pp}}}\,\left\{ \left[ \sin(\vartheta) + 4 \sin(2 \vartheta) - 3 \sin(3 \vartheta) \right] \mathbf{\hat{e}}'_{+1} - \left[ \sin(\vartheta) - 4 \sin(2 \vartheta) - 3 \sin(3 \vartheta) \right] \mathbf{\hat{e}}'_{-1} \right\} \\
    \mathbf{Y}^{(+1)}_{3,-1}(\vartheta, \varphi) = & \,\, \ee^{-\ii \varphi} \sqrt{\shrinkify{\frac{7}{8192 \pp}}}\,\left\{ \left[ 6 - \cos(\vartheta) + 10 \cos(2 \vartheta) - 15 \cos(3 \vartheta) \right] \mathbf{\hat{e}}'_{+1} + \left[ 6 + \cos(\vartheta) + 10 \cos(2 \vartheta) + 15 \cos(3 \vartheta) \right] \mathbf{\hat{e}}'_{-1} \right\} \\
    \mathbf{Y}^{(+1)}_{3,0}(\vartheta, \varphi) = & \,\, \sqrt{\shrinkify{\frac{21}{512 \pp}}}\,\sin(\vartheta)\left[ 3 + 5 \cos(2 \vartheta)\right] \, (\mathbf{\hat{e}}'_{+1} - \mathbf{\hat{e}}'_{-1}) \\
    \mathbf{Y}^{(+1)}_{3,+1}(\vartheta, \varphi) = & \,\, \ee^{\ii \varphi} \sqrt{\shrinkify{\frac{7}{8192 \pp}}}\,\left\{ \left[ 6 + \cos(\vartheta) + 10 \cos(2 \vartheta) + 15 \cos(3 \vartheta) \right] \mathbf{\hat{e}}'_{+1} + \left[ 6 - \cos(\vartheta) + 10 \cos(2 \vartheta) - 15 \cos(3 \vartheta) \right] \mathbf{\hat{e}}'_{-1} \right\} \\
    \mathbf{Y}^{(+1)}_{3,+2}(\vartheta, \varphi) = & \,\, \ee^{\ii 2\varphi}\sqrt{\shrinkify{\frac{35}{4096 \pp}}}\,\left\{ \left[ \sin(\vartheta) - 4 \sin(2 \vartheta) - 3 \sin(3 \vartheta) \right] \mathbf{\hat{e}}'_{+1} - \left[ \sin(\vartheta) + 4 \sin(2 \vartheta) - 3 \sin(3 \vartheta) \right] \mathbf{\hat{e}}'_{-1} \right\} \\
    \mathbf{Y}^{(+1)}_{3,+3}(\vartheta, \varphi) = & \,\, \ee^{\ii 3\varphi}\sqrt{\shrinkify{\frac{105}{128 \pp}}}\,\sin^2(\vartheta)\left[ \cos^2(\shrinkify{\frac{\vartheta}{2}})\,\mathbf{\hat{e}}'_{+1} + \sin^2(\shrinkify{\frac{\vartheta}{2}})\,\mathbf{\hat{e}}'_{-1}\right].
\end{align*}
Expressions for the first few $\lambda = 0$ vector spherical harmonic components with the same $K$ and $p$ can be obtained from these with the substitutions of $\boldgreek{\hat{\varphi}} \rightarrow \ii \boldgreek{\hat{\vartheta}}$ and $\boldgreek{\hat{\vartheta}} \rightarrow -\ii \boldgreek{\hat{\varphi}}$ for the linear-polarization basis, or $\mathbf{\hat{e}}'_{+1} \rightarrow -\mathbf{\hat{e}}'_{+1}$ for the circular-polarization basis.

\section{The complex spherical basis and the helicity frame}\label{app:HelicityFrame}

As is standard practice when using atomic states in the basis of angular momentum eigenstates, I find it convenient (and hopefully not overly-cumbersome) to use the complex, spherical basis for describing vectors. The spherical basis vectors of the quantization frame are related to the Cartesian basis vectors of the quantization frame via the standard definitions
\begin{align*}
    \mathbf{\hat{e}}_{+1} = & \,\, - \frac{1}{\sqrt{2}}(\mathbf{\hat{e}}_x + \ii \mathbf{\hat{e}}_y) \\
    \mathbf{\hat{e}}_{0} = & \,\, \mathbf{\hat{e}}_z \\
    \mathbf{\hat{e}}_{-1} = & \,\, \frac{1}{\sqrt{2}}(\mathbf{\hat{e}}_x - \ii \mathbf{\hat{e}}_y),
\end{align*}
and all spherical basis vectors obey
\begin{align}
    \mathbf{\hat{e}}_p^\ast \equiv (\mathbf{\hat{e}}_p^{})^\ast = (-1)^p\, \mathbf{\hat{e}}_{-p}^{} \label{eq:estar}
\end{align}
and
\begin{align}
    \mathbf{\hat{e}}_p^\ast \bigcdot \mathbf{\hat{e}}_M^{} = \delta_{p,M} \label{eq:edote}
\end{align}
where $p,M \in \{-1,0,+1\}$.

When working with complex vectors, it is important to keep in mind that the definition of the magnitude (or norm) of any vector $\mathbf{A}$ is \emph{defined} to be given by
\begin{align*}
    |\mathbf{A}| \equiv \sqrt{\mathbf{A}^\ast \bigcdot \mathbf{A}},
\end{align*}
much like the magnitude (or norm) of a complex number $A$ is given by $|A| = \sqrt{A^\ast A}$.  This definition is entirely consistent with what may be more familiar with real vectors ($|\mathbf{x}| = \sqrt{\mathbf{x} \bigcdot \mathbf{x}}$) since $\mathbf{x}^\ast = \mathbf{x}$ for all real vectors $\mathbf{x}$, and so applies to both.

Another subtlety is that the definition of the (covariant) vector \emph{components}, which obey an identical rule to real vectors,
\begin{align*}
    A_p \equiv \mathbf{A} \bigcdot \mathbf{\hat{e}}_p,
\end{align*}
has the consequence that the expansion of any vector in the spherical basis takes the form
\begin{align*}
    \mathbf{A} = \sum_p  A^{}_p\, \mathbf{\hat{e}}^\ast_p
\end{align*}
owing to (\ref{eq:estar}) and (\ref{eq:edote}).

The quantization frame is typically chosen such that any static magnetic field $\mathbf{B}_0$ points along $\mathbf{\hat{e}}_0 = \mathbf{\hat{e}}_z$, and is the frame whose $z$ axis is used when stipulating the eigenvalues $M$ of $J_z$, for example.  The atomic eigenstates are convenient to work with in this frame, but this may not be the most convenient frame to use for describing, say, the polarization of an incoming laser beam. 

\begin{figure}
\centering
\includegraphics[width=\textwidth]{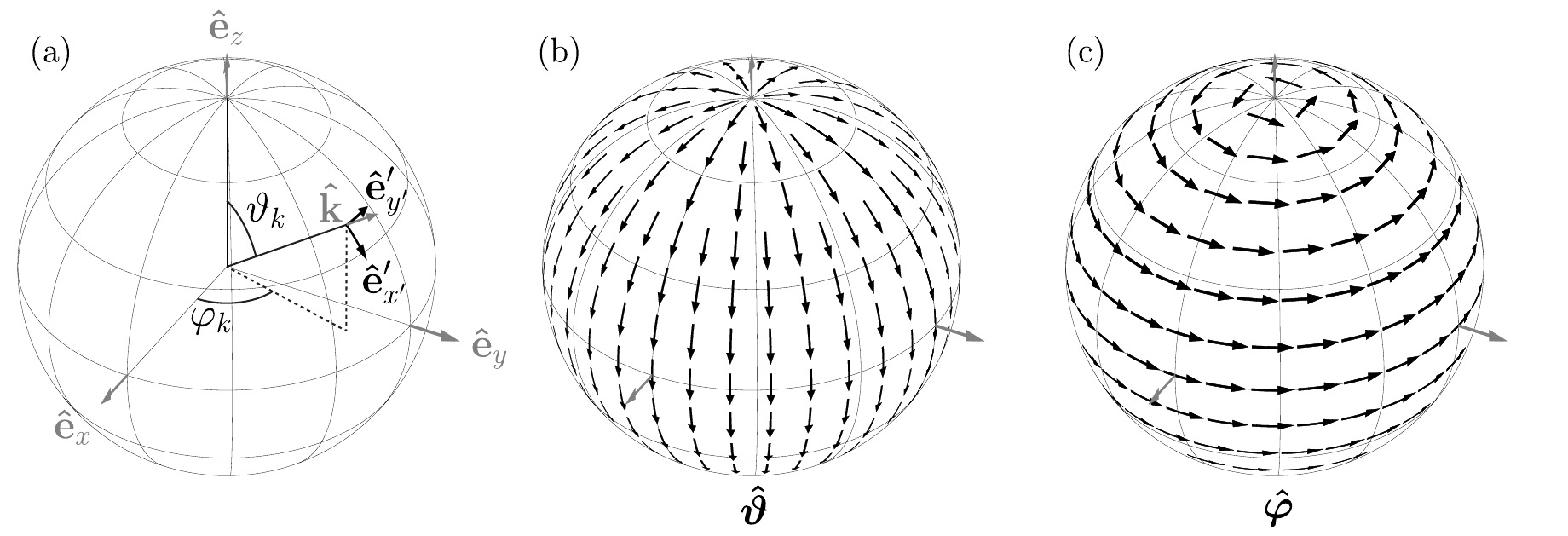}
\caption{Basis vectors for describing transverse vector fields in the far-field limit.  The atom is at the origin.  (a) The $\mathbf{k}$ vector of the light points in the radial direction $\mathbf{\hat{k}} = \mathbf{\hat{e}}'_{z'} \doteq (\vartheta_k,\varphi_k)$.  I choose the helicity-frame basis vectors such that: (b) $\mathbf{\hat{e}}'_{x'} = \boldgreek{\hat{\vartheta}}$ and (c) $\mathbf{\hat{e}}'_{y'} = \boldgreek{\hat{\varphi}}$.}\label{fig:BasisVectors}
\end{figure}

The \emph{helicity frame} (see Fig.\ \ref{fig:BasisVectors}) is used here to designate the reference frame in which $\mathbf{\hat{k}}$ is defined to be along the frame's positive $z$ axis (hereafter called $z'$ to distinguish it from $z$), and the helicity basis vectors (denoted by primes) are functions of $\mathbf{\hat{k}}$ even when written without that dependence explicitly indicated.  The helicity basis vectors are related to the quantization-frame basis vectors by the direction of $\mathbf{k} \doteq (k,\vartheta_k,\varphi_k)$,
\begin{align*}
    \mathbf{\hat{e}}'_{+1} = & \,\, \shrinkify{\mathbf{\hat{e}}_{+1} \frac{1 + \cos(\vartheta_k)}{2}\ee^{-\ii \varphi_k}} + \shrinkify{\mathbf{\hat{e}}_0\, \frac{\sin(\vartheta_k)}{\sqrt{2}}} + \shrinkify{\mathbf{\hat{e}}_{-1} \frac{1 - \cos(\vartheta_k)}{2}\ee^{\ii \varphi_k}}\\
    \mathbf{\hat{e}}'_{0} = & \,\, -\shrinkify{\mathbf{\hat{e}}_{+1} \frac{\sin(\vartheta_k)}{\sqrt{2}}\ee^{-\ii \varphi_k}} + \shrinkify{\mathbf{\hat{e}}_0\, \cos(\vartheta_k)} + \shrinkify{\mathbf{\hat{e}}_{-1} \frac{\sin(\vartheta_k)}{\sqrt{2}}\ee^{\ii \varphi_k}}\\
    \mathbf{\hat{e}}'_{-1} = & \,\, \shrinkify{\mathbf{\hat{e}}_{+1} \frac{1 - \cos(\vartheta_k)}{2}\ee^{-\ii \varphi_k}} - \shrinkify{\mathbf{\hat{e}}_0\, \frac{\sin(\vartheta_k)}{\sqrt{2}}} + \shrinkify{\mathbf{\hat{e}}_{-1} \frac{1 + \cos(\vartheta_k)}{2}\ee^{\ii \varphi_k}},
\end{align*}
where $\vartheta_k$ and $\varphi_k$ are the polar coordinates of $\mathbf{\hat{k}}$ as measured in the quantization frame.
I adopt the notation that indices for components in the helicity frame are primed (usually $q'$) while those in the quantization frame are not (usually $p$ or $M$).  In the helicity frame, the only nonzero component of $\mathbf{k}$ is the $q'=0$ component,
\begin{align*}
    k_{q'}^{}
    = k\, \,\delta_{q',0}.
\end{align*}
Accordingly, since the angular coordinate of $\mathbf{\hat{k}}$ in the helicity frame is $\vartheta^{\prime}_k = 0$, we see that spherical harmonics of $\mathbf{\hat{k}}$ also have only one nonzero component, $q'=0$, in the helicity frame,
\begin{align}
    Y_{K,q'}(\mathbf{\hat{k}}) = \delta_{q',0} \sqrt{\frac{2K+1}{4 \pp}}, \label{eq:YKq'Helicity}
\end{align}
where we adopt the Condon-Shortley phase convention.

Further, since the polarization $\boldgreek{\hat{\epsilon}}$ is always perpendicular to $\mathbf{\hat{k}}$, it only has two potentially nonzero components in the helicity frame, which I will denote in the circular-polarization basis as $\varepsilon'_{+1}$ and $\varepsilon'_{-1}$ according to
\begin{align}
    \boldgreek{\hat{\epsilon}} = \varepsilon'_{+1} \mathbf{\hat{e}}_{+1}^{\prime \ast} + \varepsilon'_{-1} \mathbf{\hat{e}}_{-1}^{\prime \ast}.\label{eq:PolCircularHelicityBasis}
\end{align}
The two complex components $\varepsilon'_{\pm 1} \equiv \boldgreek{\hat{\epsilon}} \bigcdot \mathbf{\hat{e}}'_{\pm 1}$ can be determined from (or used to make) a Jones-vector representation of the polarization.
The transformation of the polarization of a laser beam by optical elements is convenient to calculate using the Jones formalism in the helicity frame, whereas the same transformations tend to be less obvious when expressed in the quantization frame.  An explanation of Jones vectors with some principal examples can be found in Appendix \ref{app:JonesCalculus}. 

\section{Polarization and Jones Calculus}\label{app:JonesCalculus}
For electric multipole transitions, I express the classical electric field of infinite traveling plane waves in free space in terms of a complex polarization unit vector $\boldsymbol{\hat{\epsilon}}$ in the form
\begin{align}
    \mathbf{E}(\mathbf{r},t) = \frac{\mathcal{E}_0}{2}\left( \boldgreek{\hat{\epsilon}}\, \ee^{\ii(\mathbf{k} \boldsymbol{\cdot} \mathbf{r}- \omega t)} + \boldgreek{\hat{\epsilon}}^\ast\, \ee^{-\ii(\mathbf{k} \boldsymbol{\cdot} \mathbf{r}- \omega t)}\right).
\end{align}
The complimentary magnetic field in this case would be
\begin{align}
    \mathbf{B}(\mathbf{r},t) = \frac{\mathcal{B}_0}{2}\left( \boldgreek{\hat{\beta}}\, \ee^{\ii(\mathbf{k} \boldsymbol{\cdot} \mathbf{r}- \omega t)} + \boldgreek{\hat{\beta}}^\ast\, \ee^{-\ii(\mathbf{k} \boldsymbol{\cdot} \mathbf{r}- \omega t)}\right)
\end{align}
where $\boldgreek{\hat{\beta}} \equiv \mathbf{\hat{k}} \boldsymbol{\times} \boldgreek{\hat{\epsilon}}$ is the polarization unit vector for the magnetic field and $\mathcal{B}_0 = \mathcal{E}_0/c$ for a travelling wave in free space.  Most of the results in this manuscript are discussed in terms of electric fields and $EK$ transitions for simplicity of wording, but can be applied equally well to magnetic fields for $MK$ transitions by making the substitutions $\boldgreek{\hat{\epsilon}} \rightarrow \boldgreek{\hat{\beta}}$, $\mathcal{E}_0 \rightarrow c \,\mathcal{B}_0$ and utilizing the magnetic atomic multipole operator.

\begin{table}
\begin{center}
\begin{tabular}{|c||c|c|c|c|c|} 
 \hline
 Geometry & $\varepsilon'_{+1}$ & $\varepsilon'_{-1}$ & $\varepsilon'_{x'}$ & $\varepsilon'_{y'}$ & $\mathbf{J}$ \\[1.0ex] 
 \hline\hline &&&&&\\[-8pt]
 any $\mathbf{\hat{k}}$, LCP & $0$ & $1$ & $\frac{1}{\sqrt{2}}$ & $\frac{\ii}{\sqrt{2}}$ & $\frac{1}{\sqrt{2}} \left( \begin{array}{c} 1 \\ \ii \end{array}\right)$ \\[15pt]
 any $\mathbf{\hat{k}}$, RCP & $1$ & $0$ & $\frac{1}{\sqrt{2}}$ & $-\frac{\ii}{\sqrt{2}}$ & $\frac{1}{\sqrt{2}} \left( \begin{array}{c} 1 \\ -\ii \end{array}\right)$\\[15pt]
 $\mathbf{\hat{k}}=+\mathbf{\hat{e}}_z$, $\sigma^+$ & $0$ & $1$ & $\frac{1}{\sqrt{2}}$ & $\frac{\ii}{\sqrt{2}}$ & $\frac{1}{\sqrt{2}} \left( \begin{array}{c} 1 \\ \ii \end{array}\right)$ \\[15pt]
 $\mathbf{\hat{k}}=-\mathbf{\hat{e}}_z$, $\sigma^+$ & $1$ & $0$ & $\frac{1}{\sqrt{2}}$ & $-\frac{\ii}{\sqrt{2}}$ & $\frac{1}{\sqrt{2}} \left( \begin{array}{c} 1 \\ -\ii \end{array}\right)$\\[15pt]
 $\mathbf{\hat{k}}\perp \mathbf{\hat{e}}_z$, $\boldgreek{\hat{\epsilon}}=\mathbf{\hat{e}}_z$ & $\frac{1}{\sqrt{2}}$ & $-\frac{1}{\sqrt{2}}$ & $-1$ & $0$ & $-\left( \begin{array}{c} 1 \\ 0 \end{array}\right)$\\[15pt]
 $\mathbf{\hat{k}}\perp \mathbf{\hat{e}}_z$, linear pol., $\boldgreek{\hat{\epsilon}}\bigcdot\mathbf{\hat{e}}_z = \cos(\gamma)$ & $\frac{1}{\sqrt{2}}\,\ee^{\ii \gamma}$ & $-\frac{1}{\sqrt{2}}\,\ee^{-\ii \gamma}$ & $-\cos(\gamma)$ & $-\sin(\gamma)$ & $-\left( \begin{array}{c} \cos(\gamma) \\ \sin(\gamma) \end{array}\right)$\\[15pt]
 $\mathbf{\hat{k}} = -\mathbf{\hat{e}}_y$, linear pol., $\boldgreek{\hat{\epsilon}} = \mathbf{\hat{e}}_x$ & $-\frac{\ii}{\sqrt{2}}$ & $-\frac{\ii}{\sqrt{2}}$ & $0$ & $1$ & $\left( \begin{array}{c} 0 \\ 1 \end{array}\right)$ \\[15pt]
 $\mathbf{\hat{k}} = -\mathbf{\hat{e}}_x$, linear pol., $\boldgreek{\hat{\epsilon}} = \mathbf{\hat{e}}_y$ & $\frac{\ii}{\sqrt{2}}$ & $\frac{\ii}{\sqrt{2}}$ & $0$ & $-1$ & $-\left( \begin{array}{c} 0 \\ 1 \end{array}\right)$ \\[15pt]
 $\mathbf{\hat{k}}\bigcdot \mathbf{\hat{e}}_z = \cos(\vartheta_k)$, linear pol., $\boldgreek{\hat{\epsilon}}\bigcdot(\mathbf{\hat{k}}\boldsymbol{\times}(\mathbf{\hat{e}}_z\boldsymbol{\times} \mathbf{\hat{k}})) = \cos(\gamma)$ & $\frac{1}{\sqrt{2}}\,\ee^{\ii \gamma}$ & $-\frac{1}{\sqrt{2}}\,\ee^{-\ii \gamma}$ & $-\cos(\gamma)$ & $-\sin(\gamma)$ & $-\left( \begin{array}{c} \cos(\gamma) \\ \sin(\gamma) \end{array}\right)$\\[15pt]
 $\mathbf{\hat{k}}=+\mathbf{\hat{e}}_z$, linear pol., $\boldgreek{\hat{\epsilon}} = \mathbf{\hat{e}}_x$, (define $\varphi_k=0$) & $-\frac{1}{\sqrt{2}}$ & $\frac{1}{\sqrt{2}}$ & $1$ & $0$ & $\left( \begin{array}{c} 1 \\ 0 \end{array}\right)$\\[15pt]
 $\mathbf{\hat{k}}=+\mathbf{\hat{e}}_z$, linear pol., $\boldgreek{\hat{\epsilon}}\bigcdot\mathbf{\hat{e}}_x = \cos(\beta)$, (define $\varphi_k=0$) & $-\frac{1}{\sqrt{2}}\,\ee^{\ii \beta}$ & $\frac{1}{\sqrt{2}}\,\ee^{-\ii \beta}$ & $\cos(\beta)$ & $\sin(\beta)$ & $\left( \begin{array}{c} \cos(\beta) \\ \sin(\beta) \end{array}\right)$\\[15pt]
 \hline
\end{tabular}
\caption{Components of the polarization unit vector $\boldgreek{\hat{\epsilon}}$ (to within a multiplicative global phase term of the form $\ee^{\ii \phi}$). The vector $(\mathbf{\hat{k}}\boldsymbol{\times}(\mathbf{\hat{e}}_z\boldsymbol{\times} \mathbf{\hat{k}}))$ is purely transverse to $\mathbf{\hat{k}}$ and is in the plane that contains both $\mathbf{\hat{k}}$ and $\mathbf{\hat{e}}_z$ so that $\gamma$ is the angle between the linear polarization and this plane.}\label{table:Polarizations}
\end{center}
\end{table}

Equation (\ref{eq:PolCircularHelicityBasis}) gives the polarization vector expanded in the spherical helicity basis, which is the basis of circular polarization states.  I adopt the convention of Jackson \cite{Jackson} in designating the handedness of right-handed circular polarization (RCP) and left-handed circular polarization (LCP) according to the handedness of the helix traced out by the tip of the polarization vector in space at a fixed instant of time.  As such, pure RCP light would have $\boldgreek{\hat{\epsilon}}_\mathrm{RCP} \propto \mathbf{\hat{e}}'_{-1} = -\mathbf{\hat{e}}^{\prime \ast}_{+1}$, so $\varepsilon'_{+1} \propto 1$ and the light would be said to have \emph{negative helicity}, indicating that the projection of its spin angular momentum on $\mathbf{k}$ is negative.  Likewise, pure LCP light would have $\boldgreek{\hat{\epsilon}}_\mathrm{LCP} \propto \mathbf{\hat{e}}'_{+1} = - \mathbf{\hat{e}}^{\prime \ast}_{-1}$, so $\varepsilon'_{-1} \propto 1$ and it would be said to have \emph{positive helicity}.

In order to adhere to the prevailing convention in the optics community \cite{SalehAndTeich,Hecht,Yariv}, I adopt the linear basis for defining a two-dimensional Jones-vector representation of polarization, which can be related to the circular components in the helicity frame ($\varepsilon'_{q'} \equiv \boldgreek{\hat{\epsilon}} \bigcdot \mathbf{\hat{e}}'_{q'}$) according to
\begin{align}
    \mathbf{J}(\boldgreek{\hat{\epsilon}}) \doteq  \left( \begin{array}{c} \boldgreek{\hat{\epsilon}} \bigcdot \mathbf{\hat{e}}'_{x'} \\ \boldgreek{\hat{\epsilon}} \bigcdot \mathbf{\hat{e}}'_{y'} \end{array} \right) = \frac{1}{\sqrt{2}} \left(  \begin{array}{c} - \varepsilon'_{+1} + \varepsilon'_{-1}  \\ \ii \varepsilon'_{+1} + \ii \varepsilon'_{-1} \end{array} \right).\label{eq:JonesVec}
\end{align}
Alternatively, if it is more convenient to express the polarization in terms of linear components in the helicity frame ($\varepsilon'_{i'} \equiv \boldgreek{\hat{\epsilon}} \bigcdot \mathbf{\hat{e}}'_{i'}$ with $i \in \{x,y,z \}$),
\begin{align}
    \boldgreek{\hat{\epsilon}} = \varepsilon'_{x'}\, \mathbf{\hat{e}}'_{x'} + \varepsilon'_{y'}\, \mathbf{\hat{e}}'_{y'},
\end{align}
the Jones-vector representation takes the form
\begin{align}
    \mathbf{J}(\boldgreek{\hat{\epsilon}}) \doteq  \left(  \begin{array}{c} \varepsilon'_{x'}  \\ \varepsilon'_{y'} \end{array} \right).
\end{align}
The $\mathbf{\hat{e}}'_{x'}$ and $\mathbf{\hat{e}}'_{y'}$ helicity-frame directions are related to the quantization frame via
\begin{align}
    \mathbf{\hat{e}}'_{x'} = -\frac{1}{\sqrt{2}}(\mathbf{\hat{e}}_{+1}' - \mathbf{\hat{e}}_{-1}') = \boldgreek{\hat{\vartheta}}(\vartheta_k,\varphi_k) \\
    \mathbf{\hat{e}}'_{y'} = \frac{\ii}{\sqrt{2}}(\mathbf{\hat{e}}_{+1}' + \mathbf{\hat{e}}_{-1}') = \boldgreek{\hat{\varphi}}(\vartheta_k,\varphi_k)
\end{align}
where the $\mathbf{k}$ vector's spherical polar coordinates in the quantization frame are represented by $\mathbf{k} \doteq (k,\vartheta_k,\varphi_k)$.

We see from this that the helicity frame's $y'$ axis is always perpendicular to the quantization frame's $z$ axis, but the same cannot be said of the $x'$ axis (see Fig.\ \ref{fig:BasisVectors}).  Components and Jones-vector representations for some principal states of polarization are listed in Table \ref{table:Polarizations}.

One of the strengths of the Jones-vector representation is that it provides an easy way to calculate the effect of cascaded polarization optics, which transformations are representations of elements of SU(2).  For example, when a laser beam travels through a half-wave plate whose fast axis is along $y'$, this transforms the Jones vector according to
\begin{align*}
    \mathbf{J}'  = \sigma_Z\, \mathbf{J}
\end{align*}
where $\sigma_Z$ is the Pauli $Z$ operator.  If this optic is rotated about $+\mathbf{\hat{k}}$ through an angle $\theta_\mathrm{h}$, the transformation operator for the Jones vector is given by
\begin{align*}
    \ee^{-\ii \theta_\mathrm{h} \sigma_Y}\,\sigma_Z \,\ee^{\ii \theta_\mathrm{h} \sigma_Y} 
    \doteq & \,\, \left( \begin{array}{cc} \cos(2 \theta_\mathrm{h}) & \sin(2 \theta_\mathrm{h}) \\
     \sin(2 \theta_\mathrm{h}) & -\cos(2 \theta_\mathrm{h}) \end{array} \right).
\end{align*}
For the same scenario but with a quarter-wave plate, the transformation operator is
\begin{align*}
    \ee^{-\ii \theta_\mathrm{q} \sigma_Y}\,\ee^{\ii \frac{\pp}{4}\sigma_Z} \,\ee^{\ii \theta_\mathrm{q} \sigma_Y} \doteq & \,\, \left( \begin{array}{cc} \cos^2(\theta_\mathrm{q}) - \ii \sin^2(\theta_\mathrm{q}) & (1+\ii)\cos(\theta_\mathrm{q})\sin(\theta_\mathrm{q}) \\
    (1+\ii)\cos(\theta_\mathrm{q})\sin(\theta_\mathrm{q}) & \sin^2(\theta_\mathrm{q}) - \ii \cos^2(\theta_\mathrm{q}) \end{array} \right).
\end{align*}
Here I ignore global phases, which can be done as long as we're not comparing the optical phase to another reference.

These two types of wave plates are enough to create any arbitrary pure polarization state from any other pure polarization state.  This framework provides us with a convenient means to parameterize the polarization state of the light (in the helicity frame) in terms of, for example, the angles of wave plates in the lab.

\section{Spontaneous emission rate}\label{app:Spon}
Spontaneous emission can be connected to the Hamiltonian (\ref{eq:HEK}) via Fermi's golden rule,
\begin{align}
    A_{EK} = &\,\, \frac{2 \pp}{\hbar} \, \frac{\mathrm{d}N}{\mathrm{d} E}\, \sum_\mathrm{channels} \big| \bra{\mathrm{e}}\, H_{EK}^{(\mathrm{vac})} \, \ket{\mathrm{g}}\big|^2 \label{eq:FermisGoldenRule}
\end{align}
where $\mathrm{d}N/\mathrm{d}E$ is the number-density of energy states (or \emph{modes}) at the resonant frequency and $H_{EK}^{(\mathrm{vac})}$ is the Hamiltonian (\ref{eq:HEK}) on resonance and with the semi-classical electric field amplitude set to
\begin{align}
    \frac{\mathcal{E}_0^{(\mathrm{vac})}}{2} =&\,\,  \sqrt{\frac{\hbar \omega}{2 \epsilon_0 V}} \label{eq:Evac}
\end{align}
where $V$ is the volume of the field mode.  The sum over decay channels will be described below.

Since the form we have for $H_{EK}$ (Eq.\ (\ref{eq:HEK})) applies to an infinite-plane-wave field of traveling waves, it is convenient to calculate the density of states for three, orthogonal Cartesian directions in a cube with side length $L$ with periodic boundary conditions.  Summing over all three directions and two helicities for each gives
\begin{align}
    \frac{\mathrm{d}N}{\mathrm{d}E} = &\,\, \frac{L^3}{\hbar \pp^2\, c^3}\omega^2. \label{eq:DOS}
\end{align}

For the sum over decay channels (which term is not yet precisely defined), we essentially need to consider any process through which the excited state can spontaneously decay.  In most cases, this means a sum over the components ($p = M_\mathrm{e}-M_\mathrm{g}$) and lower-state projection quantum number ($M_\mathrm{g}$), and an average over the spontaneously-emitted photon's direction ($\mathbf{\hat{k}}$) and helicity ($\sigma \in \{+1,-1 \}$).  The reason we average over the last two instead of summing is that $H_{EK}$ was written with the assumption that $\mathbf{\hat{k}}$ and $\boldgreek{\hat{\epsilon}}$ were known, and we need to include the probability that the emitted photon takes these values in the first place.  All of this amounts to the replacement
\begin{align*}
    \sum_\mathrm{channels} \rightarrow \sum_{p,M_\mathrm{g}}\,\,\int \frac{\mathrm{d}^2\Omega_k}{4 \pp}\,\, \sum_\sigma \frac{1}{2}.
\end{align*}
The sum over emission directions and polarizations is particularly easy with the vector spherical harmonic form of (\ref{eq:HEK}),
\begin{align}
    \int \frac{\mathrm{d}^2\Omega_k}{4 \pp}\, \sum_\sigma \frac{1}{2} \left| \mathbf{\hat{e}}'_\sigma \bigcdot \mathbf{Y}^{(+1)}_{K,-p}(\mathbf{\hat{k}})\right|^2 = & \,\, \frac{1}{8 \pp} \int \mathrm{d}^2\Omega_k\, \left| \mathbf{Y}^{(+1)}_{K,-p}(\mathbf{\hat{k}})\right|^2 \nonumber \\
    = & \,\, \frac{1}{8 \pp}.
\end{align}
The sum over $M_\mathrm{g}$ is simplified with the application of the Wigner-Eckart theorem, for which I adopt the Racah convention for defining the reduced matrix element:
\begin{align}
    \bra{J'\,M'}\, T^{(K)}_p \,\ket{J\, M} = (-1)^{J'-M'} \WignerThreej{J'}{K}{J}{-M'}{p}{M} \bra{J'}\!|\, T^{(K)}_\cdot \, | \!\ket{J}.\label{eq:WignerEckart}
\end{align}
This allows us to rewrite the matrix element in terms of the orientation-independent reduced matrix element,
\begin{align}
    \sum_{p,M_\mathrm{g}} \left| \bra{J_\mathrm{e}\, M_\mathrm{e}} \, T^{(K)}_p[Q^{(E)}]\, \ket{J_\mathrm{g} \, M_\mathrm{g} }\right|^2 \nonumber = & \,\, \big| \bra{J_\mathrm{e}}\!| \, T^{(K)}_\cdot [Q^{(E)}]\, |\!\ket{J_\mathrm{g}}\big|^2\, \sum_{p,M_\mathrm{g}} \left| \WignerThreej{J_\mathrm{e}}{K}{J_\mathrm{g}}{-M_\mathrm{e}}{p}{M_\mathrm{g}} \right|^2 \nonumber \\
    = & \,\, \frac{\big| \bra{J_\mathrm{e}}\!| \, T^{(K)}_\cdot [Q^{(E)}]\, |\!\ket{J_\mathrm{g}}\big|^2}{2J_\mathrm{e} + 1}.\label{eq:SumRuleResult}
\end{align}
Results (\ref{eq:Evac}) through (\ref{eq:SumRuleResult}) can now be combined and used in (\ref{eq:FermisGoldenRule}) to give the Einstein $A$ coefficient for a $2^K$-pole transition in Eq.\ (\ref{eq:AEK}).

The magnetic counterpart of Eq.\ (\ref{eq:AEK}) is \cite{LandauQED}
\begin{align}
    A_{MK} = &\,\, \frac{2(K+1)}{K\,(2K-1)!!\, (2K+1)!!} \, \left( \frac{\omega}{c} \right)^{2K+1} \frac{\alpha}{c\,e^2} \,\, \frac{\big|\bra{J_\mathrm{e}}\! | \, T^{(K)}_{\cdot}[Q^{(M)}]\, |\!\ket{J_\mathrm{g}} \big|^2}{2J_\mathrm{e} + 1},\label{eq:AMK}
\end{align}
where the $p$ component of the atomic magnetic $2^K$-pole moment is \cite{ShoreAndMenzel}
\begin{align}
    T^{(K)}_p[Q^{(M)}] = & \,\, -\sqrt{K\, 4 \pp}\, \sum_i \mu_\mathrm{B} \,r_i^{K-1} \, T^{(K)}_p[Y_{(K-1)}(\mathbf{\hat{r}}_i), \shrinkify{\frac{2}{K+1}}\mathbf{l}_i + g_s\mathbf{s}_i] \label{eq:QM}
\end{align}
with $g_s$ defined to be positive. For nuclear transitions, the $p$ component of the nuclear magnetic $2^K$-pole moment is \cite{HamiltonNuclear}
\begin{align}
    T^{(K)}_p[Q^{(M)}] = & \,\, \sqrt{K\, 4 \pp}\, \sum_i \,r_i^{K-1} \, T^{(K)}_p[Y_{(K-1)}(\mathbf{\hat{r}}_i), \shrinkify{\frac{2}{K+1}}\mathbf{m}_{\mathrm{c},i} + \mathbf{m}_{\mathrm{s},i}] \label{eq:QM}
\end{align}
where $\mathbf{m}_\mathrm{c}$ is the ``convection'' magnetic moment operator from orbit and $\mathbf{m}_\mathrm{s}$ is the ``intrinsic'' magnetic moment operator from nucleon spin.

The relative magnitudes of $A_{EK}$ and $A_{MK}$ for atomic transitions can be highlighted by normalizing the transition moment to an ``atomic scale'' $2^K$-pole moment using the moment for a hydrogenic orbital with nuclear charge $+eZ$ ($e(a_0/Z)^K$ for electric and $\mu_\mathrm{B}(a_0/Z)^{K-1}$ for magnetic):
\begin{align*}
    A_{EK} = & \,\,  \frac{2(K+1)}{K\,(2K-1)!!\, (2K+1)!!} \, \alpha \omega \, \left( \frac{\hbar \omega}{\alpha Z \,m_ec^2}\right)^{2K} \,\, \frac{\big|\bra{J_\mathrm{e}}\! | \, T^{(K)}_{\cdot}[Q^{(E)}]\, |\!\ket{J_\mathrm{g}}/(e\,a_0^KZ^{-K}) \big|^2}{2J_\mathrm{e} + 1} \\
    A_{MK} = & \,\,  \frac{2(K+1)}{K\,(2K-1)!!\, (2K+1)!!} \, \frac{\alpha^2 Z^2}{4} \,\alpha \omega \, \left( \frac{\hbar \omega}{\alpha Z \,m_ec^2}\right)^{2K} \,\, \frac{\big|\bra{J_\mathrm{e}}\! | \, T^{(K)}_{\cdot}[Q^{(M)}]\, |\!\ket{J_\mathrm{g}}/(\mu_\mathrm{B}\, a_0^{K-1}Z^{1-K}) \big|^2}{2J_\mathrm{e} + 1}.
\end{align*}

\section{Including hyperfine structure}\label{app:Hyperfine}
The adaptation of the results in section \ref{sec:MultipoleExpansion} to include hyperfine structure can be accomplished by using $\{ F_i \}$ as the total angular momentum quantum numbers in place of $\{ J_i \}$ in those expressions.  The Einstein $A$ coefficient in Eqs.\ (\ref{eq:AEK}) and (\ref{eq:RabiFreq}) (which we may call $A_{EK}(F_\mathrm{e},F_\mathrm{g})$) would in this case need to apply for a specific $F_\mathrm{e} \rightarrow F_\mathrm{g}$ line.

However, it is probably more convenient to use the Einstein $A$ coefficient specified for a specific $J_\mathrm{e} \rightarrow J_\mathrm{g}$ multiplet of $\mathbf{J} = \mathbf{L} + \mathbf{S}$ ($A_{EK}(J_\mathrm{e},J_\mathrm{g})$), which would apply to all isotopes, whether they have hyperfine structure or not.  The fact that a single Einstein $A$ coefficient can be used for all of the $\ket{F_\mathrm{e}\,M_\mathrm{e}}$ states within a manifold of total electron angular momentum $J_\mathrm{e}$ relies on the assumptions that the radial wavefunctions are independent of $F_\mathrm{e}$ and that the splittings between states with different $F$ are small compared to the splitting between $J_\mathrm{e}$ and $J_\mathrm{g}$.

When calculating a resonant Rabi frequency in an atom with hyperfine structure (according to $\mathbf{F} = \mathbf{J}+ \mathbf{I}$ with $I_\mathrm{e}= I_\mathrm{g} \equiv I$), we could start by applying the Wigner-Eckart theorem (\ref{eq:WignerEckart}) to write the desired matrix element in terms of the reduced matrix element at the level of $F$:
\begin{align*}
    \bra{J_\mathrm{e}; F_\mathrm{e}\,M_\mathrm{e}}\, T^{(K)}_p[Q^{(E)}] \,\ket{J_\mathrm{g};F_\mathrm{g}\, M_\mathrm{g}} = (-1)^{F_\mathrm{e}-M_\mathrm{e}} \WignerThreej{F_\mathrm{e}}{K}{F_\mathrm{g}}{-M_\mathrm{e}}{p}{M_\mathrm{g}} \bra{F_\mathrm{e}}\!|\, T^{(K)}_\cdot[Q^{(E)}] \, | \!\ket{F_\mathrm{g}}.
\end{align*}
Next, we apply the repeated-reduction formula,
\begin{align}
    \bra{F_\mathrm{e}}\!| \, T^{(K)}_\cdot \, | \! \ket{F_\mathrm{g}} = & \,\, (-1)^{J_\mathrm{e} +I+F_\mathrm{g} + K} \sqrt{(2F_\mathrm{e} + 1)(2F_\mathrm{g} + 1)} \WignerSixj{J_\mathrm{e}}{F_\mathrm{e}}{I}{F_\mathrm{g}}{J_\mathrm{g}}{K} \bra{J_\mathrm{e}}\!| \, T^{(K)}_\cdot \, | \! \ket{J_\mathrm{g}},
\end{align}
which allows us to write $\bra{F_\mathrm{e}}\!| \, (\shrinkify{-\frac{1}{e}})\,T^{(K)}_\cdot[Q^{(E)}] \, | \! \ket{F_\mathrm{g}}$ in terms of
\begin{align*}
    \big|\bra{J_\mathrm{e}}\!| \, (\shrinkify{-\frac{1}{e}})\,T^{(K)}_\cdot[Q^{(E)}] \, | \! \ket{J_\mathrm{g}}\big| = & \,\, \sqrt{\frac{A_{EK}(J_\mathrm{e},J_\mathrm{g})}{\alpha c}(2J_\mathrm{e} + 1)\left(\frac{c}{\omega}\right)^{2K+1}}\, \sqrt{\frac{K(2K-1)!!(2K+1)!!}{2(K+1)}}.
\end{align*}
Putting this together gives the resonant Rabi frequency 
\begin{align}
    \Omega_\mathrm{eg} = &\,\, s_J (-1)^{F_\mathrm{e}-M_\mathrm{g}} \frac{e\mathcal{E}_0}{\hbar}\, \sqrt{\frac{2 \pp \,A_{EK}(J_\mathrm{e},J_\mathrm{g})}{\alpha \, c} \, \,(2 J_\mathrm{e} + 1) \left( \frac{c}{\omega}\right)^{3}} \WignerThreej{F_\mathrm{e}}{K}{F_\mathrm{g}}{-M_\mathrm{e}}{p}{M_\mathrm{g}}  \,\, \left( \boldgreek{\hat{\epsilon}} \bigcdot \mathbf{Y}^{(+1)}_{K,-p}(\mathbf{\hat{k}}) \right)\nonumber \\
    & \times (-1)^{J_\mathrm{e}+I+F_\mathrm{g}+K} \sqrt{(2 F_\mathrm{g} + 1)(2F_\mathrm{e}+1)} \WignerSixj{J_\mathrm{e}}{F_\mathrm{e}}{I}{F_\mathrm{g}}{J_\mathrm{g}}{K}.
\end{align}

\section{Paraxial Beams}\label{app:Beams}
For our purposes in this paper, the required machinery of geometric optics sits somewhere in between the overly-simplistic treatment of ray optics and the full, 3D treatment using Maxwell's equations, with the former being too inaccurate to capture the essential physics and the latter too unwieldy to furnish much intuition.  However, laser light tends to be produced in a \emph{beam}, by which I mean most of the light is close to a particular line in space (called the \emph{optical axis}, which I will designate as coinciding with the axis $z'$ of the helicity frame) and the field falls to zero quickly with the distance from that line.  As such, the local $\mathbf{k}$ vector for the light cannot be steeply inclined away from $\mathbf{\hat{e}}'_{z'}$, and this affords us a simplification known as the \emph{paraxial approximation} (essentially, $k_\perp \ll k_{z'}$).

In the paraxial approximation, the most basic solution to the wave equation for the electric field of a traveling light beam is the lowest-order Gaussian beam, which can be written in the form
\begin{align}
    \mathbf{E}(\mathbf{r},t) = & \,\, \frac{\mathcal{E}_0}{2}\left( \boldgreek{\hat{\epsilon}} \, \frac{w_0}{w(z')}\ee^{- \ii \phi_\mathrm{G}(z')} \exp \left( - \frac{\boldgreek{\rho}^2}{w^2(z')} + \ii k\frac{\boldgreek{\rho}^2}{2R(z')}\right) \ee^{\ii(k z' - \omega t)}+ \text{c.c.}\right).\label{eq:TEM00}
\end{align}
Here, $\boldgreek{\rho} \equiv x' \mathbf{\hat{e}}'_{x'} + y' \mathbf{\hat{e}}'_{y'}$ is the transverse vector position from the optical axis and $w_0$ is the minimum beam waist (the $1/\ee^2$ intensity radius of the beam at the plane of the minimum waist, $z'=0$).
The waist as a function of axial position $z'$ is
\begin{align*}
    w(z') = w_0^2 \left[1 + \left( \shrinkify{\frac{\lambda z'}{\pp w_0^2}}\right)^2 \right].
\end{align*}
The function $R(z')$ is the radius of curvature of the wave fronts, set by
\begin{align*}
    \frac{1}{R(z')} \equiv \frac{z'}{z'^2 + \left( \shrinkify{\frac{\pp w_0^2}{\lambda}}\right)^2}
\end{align*}
and $\phi_\mathrm{G}$ is known as the \emph{Gouy phase},
\begin{align*}
    \phi_\mathrm{G}(z') \equiv \mathrm{arctan}\left( \shrinkify{\frac{\lambda z'}{\pp w_0^2}}\right).
\end{align*}

The solution (\ref{eq:TEM00}) is a stable solution for optical cavities (or resonators) commonly employed in lasers, and is also approximately equal to the stable solution for light confined in a single-mode optical fiber \cite{Yariv}.  Despite its wide adoption, it is only the lowest-order solution in a whole family of stable resonator solutions known as Ince-Gauss modes \cite{Boyer1975Lie,Bandres2004InceGaussian}, of which the more-familiar Hermite-Gauss (\S \ref{sec:HG}) and Laguerre-Gauss (\S \ref{sec:LG}) modes are limiting cases.  In particular, the $\mathrm{HG}_{mn}$ mode (with $m$ and $n$ non-negative integers) can be described by
\begin{align*}
    \mathbf{E}^{(\mathrm{HG})}_{mn}(\mathbf{r},t) = & \,\, \frac{\mathcal{E}_0}{2}\left( \boldgreek{\hat{\epsilon}} \, \frac{w_0}{w(z')}H_m\!\left[\shrinkify{\frac{\sqrt{2}\,x'}{w(z')}} \right]H_n\!\left[\shrinkify{\frac{\sqrt{2}\,y'}{w(z')}} \right]\ee^{- \ii(m+n+1) \phi_\mathrm{G}(z')} \exp \left( - \frac{\boldgreek{\rho}^2}{w^2(z')} + \ii k\frac{\boldgreek{\rho}^2}{2R(z')}\right) \ee^{\ii(k z' - \omega t)}+ \text{c.c.}\right)
\end{align*}
and the helical $\mathrm{LG}_{n,\ell}$ mode (with $n$ a non-negative integer and $\ell$ an integer) can be written
\begin{align*}
    \mathbf{E}^{(\mathrm{LG})}_{n,\ell}(\mathbf{r},t) = & \,\, \frac{\mathcal{E}_0}{2}\left( \boldgreek{\hat{\epsilon}} \, \shrinkify{\frac{w_0}{w(z')}}\left(\shrinkify{\frac{\sqrt{2}\,\rho}{w(z')}} \right)^{|\ell|} \, L^{(|\ell|)}_n\!\left[\shrinkify{\frac{2 \rho^2}{w^2(z')}} \right]\ee^{- \ii(2n + |\ell| + 1) \phi_\mathrm{G}(z')} \ee^{\ii \ell \varphi_\rho}\exp \left( - \frac{\boldgreek{\rho}^2}{w^2(z')} + \ii k\frac{\boldgreek{\rho}^2}{2R(z')}\right) \ee^{\ii(k z' - \omega t)}+ \text{c.c.}\right)
\end{align*}
where $H_k[x]$ is a Hermite polynomial (using the physics convention such that $H_1[x] = 2x$) and $L^{(\alpha)}_{k}[x]$ is a generalized Laguerre polynomial.

\section{Nested tensor operators}\label{app:PolIdty} 
The multipole expansion involves stretched, nested tensor products of the form
\begin{align}
    T^{(K)}[\underbrace{\mathbf{a},\mathbf{a}, \ldots , \mathbf{a}}_{K}] \equiv T^{(K)}[\cdots T^{(2)}[T^{(1)}[T^{(0)}[1],\mathbf{a}],\mathbf{a}] \cdots ,\mathbf{a}]\label{eq:nested}
\end{align}
where
\begin{align*}
    T^{(K)}_p[T^{(k_A)}[\mathbf{A}],T^{(k_B)}[\mathbf{B}]] = \sum_{m_A,m_B} (-1)^{k_A-k_B+p}\, \sqrt{2K+1}\, \WignerThreej{k_A}{k_B}{K}{m_A}{m_B}{-p}\,\,T^{(k_A)}_{m_A}[\mathbf{A}]\,\,T^{(k_B)}_{m_B}[\mathbf{B}]
\end{align*}
is the $p$ component of the rank-$K$ tensor product of a rank-$k_A$ tensor with a rank-$k_B$ tensor (with the $T^{(k)}[\cdot]$ notation often omitted for $k=1$ in favor of bold font denoting vectors) and $T^{(0)}[1] \equiv 1$.
This can be rewritten recursively as
\begin{align}
    T^{(K)}[\underbrace{\mathbf{a},\mathbf{a}, \ldots , \mathbf{a}}_{K}] = \left\{ \begin{array}{ll} 1 & \,\, K=0 \\ T^{(K)}[T^{(K-1)}[\underbrace{\mathbf{a},\mathbf{a}, \ldots , \mathbf{a}}_{K-1}],\mathbf{a}] & \,\, K \ge 1\end{array} \right. .\label{eq:recursivedef}
\end{align}
This can be simplified with the help of the following lemma:
\begin{lemma} \label{lma:NestedLemma}
    A nested, rank-$K$ irreducible tensor product $T^{(K)}[\mathbf{a},\mathbf{a}, \ldots , \mathbf{a}]$ of a vector $\mathbf{a}$ with itself $K$ times is proportional to a solid harmonic and given by
    \begin{align}
        T^{(K)}[\underbrace{\mathbf{a},\mathbf{a}, \ldots , \mathbf{a}}_{K}] = \sqrt{\frac{K!\,\, 4\pp}{(2K+1)!!}}\,\,|\mathbf{a}|^K Y_{(K)}(\mathbf{\hat{a}}) \label{eq:NestedLemma}
    \end{align}
    where $Y_{(K)}(\mathbf{\hat{a}})$ is the rank-$K$ irreducible tensor whose $p$ component is given by the spherical harmonic $Y_{K,p}(\vartheta_a,\varphi_a)$.
\end{lemma}

This is not a new result, and appears as Eq.\ 3.2.(23) in \cite{Varshalovich1988Quantum}.  The proof introduces helpful concepts and is provided below due to its importance for converting the Hamiltonian into familiar spherical tensor operators.

\vspace{3mm}
\textit{Proof.} This can be shown by induction.  If this is true for $K=J-1$, we have
\begin{align*}
    T^{(J)}[\underbrace{\mathbf{a},\mathbf{a}, \ldots , \mathbf{a}}_{J}] = & \,\,
    \sqrt{\frac{(J-1)! \, 4 \pp}{(2J-1)!!}}\,\, |\mathbf{a}|^{J-1}\,T^{(J)}[Y_{(J-1)}(\mathbf{\hat{a}}), \mathbf{a}].
\end{align*}
In a coordinate frame in which the $q=0$ axis points along $\mathbf{\hat{a}}$, both $Y_{J-1,q}(\mathbf{\hat{a}})$ and $a_q$ are only nonzero for $q=0$.  Computing all components allows us to identify
\begin{align*}
    T^{(J)}[Y_{(J-1)}(\mathbf{\hat{a}}), \mathbf{a}] = |\mathbf{a}|\, \sqrt{\frac{J}{2J+1}}\,\, Y_{(J)}(\mathbf{\hat{a}}),
\end{align*}
which shows that the $K=J-1$ case implies the $K=J$ case.  Since (\ref{eq:NestedLemma})  can be computed directly for the $K=0$ base case and gives 1 in agreement with (\ref{eq:recursivedef}), Lemma \ref{lma:NestedLemma} is shown to hold in general.

\vspace{3mm}

The field-derivative part of the Hamiltonian can now be written as
\begin{align*}
    T^{(K)}[T^{(K-1)}[\underbrace{\mathbf{k},\mathbf{k},\ldots ,\mathbf{k}}_{K-1}], \mathbf{E}] = & \,\, \sqrt{\frac{(K-1)!\,\, 4\pp}{(2K-1)!!}}\,\,|\mathbf{k}|^{K-1} \,\,T^{(K)}[Y_{(K-1)}(\mathbf{\hat{k}}), \mathbf{E}].
\end{align*}
We now have a need to connect this to the desired form, for which I will make use of the following lemma.
\begin{lemma} \label{lma:Kplus1}
    Given two, orthogonal unit vectors $\mathbf{\hat{k}}$ and $\boldgreek{\hat{\epsilon}}$ in 3D, the irreducible rank-$K$ tensor product of $Y_{(K+1)}(\mathbf{\hat{k}})$ and $\boldgreek{\hat{\epsilon}}$ can be related to the irreducible rank-$K$ tensor product of $Y_{(K-1)}(\mathbf{\hat{k}})$ and $\boldgreek{\hat{\epsilon}}$ via
    \begin{align}
        T^{(K)}[Y_{(K+1)}(\mathbf{\hat{k}}),\boldgreek{\hat{\epsilon}}] = \sqrt{\frac{K}{K+1}}\,\, T^{(K)}[Y_{(K-1)}(\mathbf{\hat{k}}),\boldgreek{\hat{\epsilon}}].
    \end{align}
\end{lemma}
\textit{Proof.}
Since $\mathbf{\hat{k}} \perp \boldgreek{\hat{\epsilon}}$, we see that the only nonzero components $\epsilon'_{q'}$ of $\boldgreek{\hat{\epsilon}}$ in the helicity frame are $q'=\pm 1$.  Likewise, there is only one nonzero component of the spherical harmonic $Y_{(K)}(\mathbf{\hat{k}})$ in the helicity frame, given by Eq. (\ref{eq:YKq'Helicity}).
From the definition of the irreducible tensor product components, we have
\begin{align*}
    T^{(K)}_{q'}[Y_{(K+1)}(\mathbf{\hat{k}}),\mathbf{\hat{\epsilon}}] 
    = & \,\, \delta_{|q'|,1}\, \sqrt{\frac{K}{8 \pp }} \,\,\epsilon'_{q'}.
\end{align*}
Likewise,
\begin{align*}
    T^{(K)}_{q'}[Y_{(K-1)}(\mathbf{\hat{k}}),\mathbf{\hat{\epsilon}}] = & \,\, \delta_{|q'|,1}\, \sqrt{\frac{(K+1)}{8 \pp}} \,\,\epsilon'_{q'}.
\end{align*}
Since all components are proportional to one another with this same proportionality constant, this proves Lemma \ref{lma:Kplus1}.

\vspace{3mm}

I now proceed to the polarization identity:

\begin{theorem} \label{thm:PolarizationIdentity} Given two, orthogonal unit vectors $\mathbf{\hat{k}}$ and $\boldgreek{\hat{\epsilon}}$ in 3D,
the rank-$K$ irreducible tensor product of $Y_{(K-1)}(\mathbf{\hat{k}})$ and $\boldgreek{\hat{\epsilon}}$ can be written in terms of the vector dot product of $\boldgreek{\hat{\epsilon}}$ with the vector spherical harmonic $\mathbf{Y}^{(+1)}_{(K)}$ according to
\begin{align*}
    T^{(K)}[Y_{(K-1)}(\mathbf{\hat{k}}),\boldgreek{\hat{\epsilon}}] = \sqrt{\frac{K+1}{2K+1}}\left( \boldgreek{\hat{\epsilon}} \bigcdot \mathbf{Y}^{(+1)}_{(K)}(\mathbf{\hat{k}})\right).
\end{align*}
\end{theorem}
\textit{Proof.} We can start with the definition of the vector spherical harmonic components in terms of components of tensor products of scalar spherical harmonics with the rank-1 unit orthonormalized irreducible \emph{tensor} $\mathbf{e}_{(1)}$ whose (covariant) \emph{components} are the spherical basis \emph{vectors} $\mathbf{e}_{1,q}= \mathbf{e}_{(1)}\bigcdot \mathbf{\hat{e}}_q = \mathbf{\hat{e}}_q$,
\begin{align}
    \mathbf{Y}^{(+1)}_{K,M}(\mathbf{\hat{k}}) =&\,\, \sqrt{\frac{K+1}{2K+1}}\,\,T^{(K)}_M[Y_{(K-1)}(\mathbf{\hat{k}}),\mathbf{e}_{(1)}] + \sqrt{\frac{K}{2K+1}}\,\,T^{(K)}_M[Y_{(K+1)}(\mathbf{\hat{k}}),\mathbf{e}_{(1)}].
\end{align}
The vector dot product of this with $\boldgreek{\hat{\epsilon}}$ gives
\begin{align*}
    \boldgreek{\hat{\epsilon}} \bigcdot \mathbf{Y}^{(+1)}_{K,M}(\mathbf{\hat{k}}) =&\,\, \sqrt{\frac{K+1}{2K+1}}\,\,T^{(K)}_M[Y_{(K-1)}(\mathbf{\hat{k}}),\boldgreek{\hat{\epsilon}}] + \sqrt{\frac{K}{2K+1}}\,\,T^{(K)}_M[Y_{(K+1)}(\mathbf{\hat{k}}),\boldgreek{\hat{\epsilon}}].
\end{align*}
The two tensor products above can be related by lemma \ref{lma:Kplus1} to prove the identity \ref{thm:PolarizationIdentity},
\begin{align*}
    \boldgreek{\hat{\epsilon}} \bigcdot \mathbf{Y}^{(+1)}_{K,M}(\mathbf{\hat{k}}) =&\,\, \left\{ \sqrt{\frac{K+1}{2K+1}} + \sqrt{\frac{K}{2K+1}\frac{K}{K+1}}\right\}\,\,T^{(K)}_M[Y_{(K-1)}(\mathbf{\hat{k}}),\boldgreek{\hat{\epsilon}}] \\
    = & \,\, \sqrt{\frac{2K+1}{K+1}}\,\,T^{(K)}_M[Y_{(K-1)}(\mathbf{\hat{k}}),\boldgreek{\hat{\epsilon}}].
\end{align*}
Theorem \ref{thm:PolarizationIdentity} can be combined with Lemma \ref{lma:NestedLemma} to arrive at Eq.\ (\ref{eq:PolIdty}).

\end{document}